\documentclass[aps,preprint,epsfig,rotate]{revtex4}
\usepackage{graphicx}
\usepackage{bm}
\usepackage{epsfig}
\input epsf
\begin{document}
\title{On matrix elements of the vector physical quantities}

%\title{On matrix elements of the vector physical quantities and methods of angular momentum 
%       in quantum mechanics}

\author{Alexei M. Frolov}
 \email[E--mail address: ]{alex1975frol@gmail.com} 

%\affiliation{ITAMP, Harvard-Smithonian Center for Astrophysics, \\
%         MS 14, 60 Garden Street, Cambridge MA 02138-1516, USA}  

\affiliation{Department of Applied Mathematics \\
 University of Western Ontario, London, Ontario N6H 5B7, Canada}

%\date{October 30, 2024}
\date{\today}

\begin{abstract}

Methods of angular momenta are modified and used to solve some actual problems in quantum mechanics.  
In particular, we re-derive some known formulas for analytical and numerical calculations of matrix 
elements of the vector physical quantities. These formulas are applied to a large number of quantum 
systems which have an explicit spherical symmetry. Multiple commutators of different powers of the 
angular momenta $\hat{\bf J}^{2}$ and vector-operator $\hat{\bf A}$ are determined in the general 
form. Calculations of the expectation values averaged over orbital angular momenta are also described 
in detail. This effective and elegant old technique, which was successfully used by E. Fermi and A. 
Bohr, is almost forgotten in modern times. We also discuss quantum systems with additional relations
(or constraints) between some vector-operators and orbital angular momentum. For similar systems such 
relations allow one to obtain some valuable additional information about their properties, including 
the bound state spectra, correct asymptotics of actual wave functions, etc. As an example of unsolved 
problems we consider applications of the algebras of angular momenta to investigation of the 
one-electron, two-center (Coulomb) problem $(Q_1, Q_2)$. For this problem it is possible to obtain the 
closed analytical solutions which are written as the `correct' linear combinations of products of the 
two one-electron wave functions of the hydrogen-like ions with the nuclear charges $Q_1 + Q_2$ and $Q_1 
- Q_2$, respectively. However, in contrast with the usual hydrogen-like ions such hydrogenic wave 
functions must be constructed in three-dimensional pseudo-Euclidean space with the metric (-1,-1,1). \\
  
% \noindent 
% PACS number(s): 31.15.xh, 03.65.-w and 31.30.Gs \\
% doi: 10.13140/RG.2.2.16168.17925

\end{abstract}

\maketitle

%\noindent \vspace{0.2in}

\section{Introduction} 

In this communication we discuss analytical and numerical calculations of the matrix elements of arbitrary 
vector physical quantities (or vector-operators, for short) and applications of similar matrix elements to 
a number of actual problems in Quantum Mechanics. Soon we will mark the centenary since publication of the 
famous article \cite{BHJ}, in which Quantum Mechanics was applied (for the very first time) to describe the 
general transformations of vector physical quantities. At that time there were many questions and discussions 
about the laws of transformations and actual calculations of different products and various powers of vector 
physical quantities. For instance, there were certain confusions related with the instantaneous use (often 
in the same equation) of the quantum Poisson bracket(s) between two (or more) vector-operators and vector 
product(s) of the same operators. In this study we reconsider the same and few other closely related problems, 
carefully investigate the methods developed in \cite{BHJ} and apply them to solve a number of actual problems. 

First of all, following \cite{BHJ} we introduce some basic notations. Everywhere below in this study the 
notation $\hat{\bf A} = (\hat{A}_1, \hat{A}_2, \hat{A}_3)$ stands for a vector-operator which represents 
some vector physical quantity and satisfies the following commutation rules $[ \hat{J}_{i}, \hat{A}_{j}] 
= \imath \varepsilon_{ijk} \hat{A}_{k} = \imath \varepsilon_{kij} \hat{A}_{k}$ \cite{BHJ}, where $(i,j,k) 
= (1,2,3) = (x,y,z)$, while the vector $\hat{\bf J} = (\hat{J}_1, \hat{J}_2, \hat{J}_3)$ is the angular 
momentum of some closed physical system and $\varepsilon_{ijk}$ is the complete antisymmetric tensor of 
the third rank (see, e.g., \cite{Kochin}) which has only six non-zero components (of twenty seven) and 
$\varepsilon_{123} = 1$. The well known properties of this tensor are: $\varepsilon_{ijk} \varepsilon_{ijm} 
= 2 \delta_{km}$ and $\varepsilon_{ijk} \varepsilon_{ijk} = 6$ (see,e.g., \cite{KS}). From antisymmetry of 
$\varepsilon_{ijk}$ it follows that this tensor has only those non-vanishing components which have three 
different indices. Note also that in this study we apply the double suffix summation convention, which is 
widely known as the dummy suffix notation (see, e.g., \cite{Kochin}, \cite{LLFT} and \cite{VHein}). This 
convention means that we shall always omit (unless otherwise specified) the summation signs over every 
suffix that occurs twice in one expression, e.g., in one term, or in some product. 

Below, the equality $[ \hat{J}_{i}, \hat{A}_{j}] = \imath \varepsilon_{ijk} \hat{A}_{k}$ is called the main 
(or fundamental) relation for vector-operators in Quantum Mechanics. The vector-operator ${\bf A}$ in this 
relation can, in principle, be arbitrary. At this point it is important to note an interesting fact that the 
same complete antisymmetric tensor of third rank $\varepsilon_{ijk}$ is used to define the cross product (or 
vector product) of the two vectors. Indeed, in the component form the cross-product (or vector-product) is 
written in the form $Z_a = ({\bf X} \times {\bf Y})_{a} = \varepsilon_{abc} X_{b} Y_{c}$, where $(a,b,c)$ =
(1,2,3). This vector ${\bf Z}$ is called the cross product (or vector product) of the two vectors ${\bf X}$ 
and ${\bf Y}$. From here one finds that if ${\bf X} \parallel {\bf Y}$, then $({\bf X} \times {\bf Y}) = 0$, 
i.e., their vector product equals zero identically. A close relation between the quantum Poisson bracket and 
usual cross product of two vectors is crucial for our analysis in this study. The tensor notations used here 
are very compact and convenient for analytical calculations. 

Now, we can generalize these definitions to the scalar and cross products to the vector-operators. First, the 
dot product (or scalar product) of any two vector-operators $\hat{\bf X}$ and $\hat{\bf Y}$ their dot product 
is is defined as follows: $\hat{s} = \hat{\bf X} \cdot \hat{\bf Y} = \hat{X}_{a} \hat{Y}_{a}$, where $\hat{s}$ 
is a scalar-operator. It is clear that in general case $\hat{\bf X} \cdot \hat{\bf Y} \ne \hat{\bf Y} \cdot 
\hat{\bf X}$, since these two vector-operators do not commute with each other. However, if one of these 
operators is ${\bf J}$ (or ${\bf L}$), then we can write (from the main relation): $[ \hat{J}_{i}, \hat{A}_{i}] 
= 0 = [ \hat{A}_{i}, \hat{J}_{i}]$ and $[ \hat{J}_{i}, \hat{J}_{i}] = 0 = [ \hat{A}_{i}, \hat{A}_{i}]$ (here we 
do not mean summation over the two repeated indexes). In other words, the $i-$th component of any vector-operator 
$\hat{\bf A}$, including the angular momentum $\hat{\bf J}$ itself, commutes with the $i-$th component of 
$\hat{\bf J}$. From here one immediately obtains that $\hat{A}_{a} \hat{J}_{a} = \hat{J}_{a} \hat{A}_{a}$ (here 
the summation rule is applied), or in other words, $\hat{\bf J} \cdot \hat{\bf A} = \hat{\bf A} \cdot \hat{\bf J}$. 
More details about the properties of cross products defined for the two vector-operators  can be found in \$\$ 35 
- 36 from \cite{Dir59}. 

Analogously, for any three non-coplanar vectors ${\bf X}, {\bf Y}$ and ${\bf Z}$ it is possible to introduce the 
mixed product $v = {\bf X} \cdot ({\bf Y} \times {\bf Z}) = \varepsilon_{abc} X_{a} Y_{b} Z_{c}$, where $v$ is a 
scalar. This mixed product is a complete antisymmetric scalar form in our three-dimensional space which can be 
formed only from three different and non-coplanar vectors (or `objects' in the general case \cite{VIA}). In our 
three-dimensional Euclidean space any two vectors are always coplanar. Therefore, the mixed product, which 
includes a pair of identical vectors, equals zero identically. As follows from this definition $v = {\bf X} \cdot 
({\bf Y} \times {\bf Z}) = ({\bf X} \times {\bf Y}) \cdot {\bf Z}$. Also, if ${\bf X} = c_1 {\bf Y}$, or ${\bf Z} 
= c_2 {\bf X}$, or ${\bf Y} = c_3 {\bf Z}$, where $c_i$ are some real numbers, then $v = 0$. These formulas for 
the mixed product(s) can be generalized to the vector-operators too, but one has to be extra careful by performing 
similar operations, since here it is easy to make a large number of serious mistakes and obtain contradictory 
results. The source of such  troubles is obvious, since we are trying to combine the quantum commutation relations 
for vector-operators and their components with the traditional multi-vector construction which is relatively 
complex by itself. In other words, in contrast with the dot and cross products, the definition of mixed product 
for any three non-coplanar vectors cannot be easily generalized to the three vector-operator, since this often 
leads to various mistakes. One such an example is discussed in the Appendix A. 

This paper has the following structure. In the next two Sections we discuss some basic relations, which play a 
central role in quantum theory of angular momentum, and perform analytical calculations of the quantum Poisson 
brackets of different powers of angular momenta $\hat{\bf J}$ with an arbitrary vector-operator $\hat{\bf A}$. 
Formulas for the matrix elements of arbitrary vector-operator $\hat{\bf A}$ in the basis of spherical $|\lambda 
J M \rangle$ states are re-derived and discussed in the fourth Section. Here $J$ and $M$ are the two conserving 
quantum numbers which play a great role in numerous applications. The so-called `selection rules' for permitted 
variations of theses $J$ and $M$ quantum numbers are also discussed in this Section. In Section V we consider an 
effective old technique which was used for calculations of a large number of expectation values averaged over 
partial angular momenta. Unfortunately, this elegant technique is almost forgotten in modern times. Quantum 
systems with additional relations (or constraints) between some vector-operators and angular momentum are 
discussed in Section VI. Here we describe the one-electron, two-center (Coulomb) problem for which there is a 
closed analytical (or algebraic) solution. The final development of the corresponding procedure to determine 
the closed analytical solutions of this complex problem will be a great success for the both modern theoretical 
physics and theory of group representations. Concluding remarks can be found in the last Section. The paper also 
contains three Appendixes where some important technical details of our calculations are explained. In Appendix 
C we discuss a short version of one-electron atomic theory which is investigated in three-dimensional 
pseudo-Euclidean space. 

\section{Basic relations between components of angular momenta in quantum mechanics}

In this Section we briefly discuss a number of useful relations which are extensively used in quantum theory 
of angular momentum ${\bf L}$ (or ${\bf J}$) (see, e.g., \cite{LLQ} - \cite{Rose}). Formulas and results from 
these   trusted books are used below without any additional reference. The angular momentum ${\bf L}$ in 
classical mechanics is defined as follows ${\bf L} = {\bf r} \times {\bf p}$ for a single particle and ${\bf L} 
= \sum^{N}_{i=1} {\bf r}_{i} \times {\bf p}_{i}$ ($i = 1, \ldots, N$) for the few- and many-particle systems. 
The corresponding counterparts of these cross products in quantum mechanics are the operators $\hat{\bf L} = 
\hat{\bf r} \times \hat{\bf p}$ (for single particle) and $\hat{\bf L} = \sum^{N}_{i=1} \hat{\bf r}_{i} \times 
\hat{\bf p}_{i}$ (for $N-$particle system), where $\hat{\bf r}_{i}$ and $\hat{\bf p}_{i}$ are the position and 
momentum operator of the particle $i$. If we deal with spinless particles and in the absence of any electric 
field, then the operator momenta of a single particle takes the following form $\hat{\bf p} = -\imath \hbar 
\nabla$, where $\hbar$ is the reduced Planck constant which is also often called the Dirac constant. In respect 
to this, the operator of angular momentum (also called the operator of orbital angular momentum) is written in 
the form $\hat{\bf L} = -\imath \hbar (\hat{\bf r} \times \nabla)$. For its components we can write   
\begin{equation}
 \hat{L}_{x} = \imath \hbar \Bigl( y \frac{\partial}{\partial z} - z \frac{\partial}{\partial y} \Bigr) 
 \; \; , \; \;  \hat{L}_{y} = \imath \hbar \Bigl( z \frac{\partial}{\partial x} - x 
 \frac{\partial}{\partial z} \Bigr) \; \; , \; \; \hat{L}_{z} = \imath \hbar 
 \Bigl( x \frac{\partial}{\partial y} - y \frac{\partial}{\partial x} \Bigr) \; \; . \; \label{equant1}
\end{equation}
It is straightforward to check that the fundamental commutation relations $[ \hat{L}_{i}, \hat{L}_{j}] = 
\imath \varepsilon_{ijk} \hat{L}_{k}$ do obey for these three operators $\hat{L}_{x}, \hat{L}_{y}$ and  
$\hat{L}_{z}$. In other words, the three operators $\hat{L}_{x}, \hat{L}_{y}$ and $\hat{L}_{z}$ form the 
closed compact operator algebra $SO(3)$. Furthermore, each of these three operators is self-conjugate, or 
Hermitian. 

The invariant operator of second order, or Casimir operator $C_2$, for this $SO(3)$ algebra is $C_2 = 
\hat{L}^{2} = \hat{L}^{2}_{x} + \hat{L}^{2}_{y} + \hat{L}^{2}_{z}$. For bound states with angular momentum 
$L$ this $C_2$ operator takes a diagonal form and can be considered as a numerical constant which equals 
$L(L + 1)$. The operator $\hat{L}^{2}$ is called the square of angular momentum. It is clear that the $C_2
= \hat{L}^{2}$ operator always commutes with any of its operator-components $\hat{L}_{x}, \hat{L}_{y}, 
\hat{L}_{z}$. Now, let us introduce the two non-Hermitian operators $\hat{L}_{+}$ and $\hat{L}_{-}$ which 
are defined as follows: $\hat{L}_{+} = \hat{L}_{x} + \imath \hat{L}_{y}$ and $\hat{L}_{-} = \hat{L}_{x} 
- \imath \hat{L}_{y}$. They obey the following commutation relations 
\begin{equation}
 [ \hat{L}^{2}, \hat{L}_{\pm}] = 0 \; \; \; , \; \; \; [ \hat{L}_{z}, \hat{L}_{+}] = \hat{L}_{+} \; 
 \; \; , \; \; \; [ \hat{L}_{z}, \hat{L}_{-}] = - \hat{L}_{-} \; \; \; , \; \; \; [ \hat{L}_{+}, 
 \hat{L}_{-}] = 2 \hat{L}_{z} \; \; . \; \; \label{equant2}
\end{equation}
The three following expressions of the $\hat{L}^{2}$ operator written in terms of the $\hat{L}_{+}$ and  
$\hat{L}_{-}$ operators are useful for the explicit constructions of all irreducible representations of 
the compact $SO(3)-$group: 
\begin{equation}
 \hat{L}^{2} = \frac12 \Bigl( \hat{L}_{+} \hat{L}_{-} + \hat{L}_{-} \hat{L}_{+} \Bigr) + 
 \hat{L}^{2}_{z} = \hat{L}_{+} \hat{L}_{-} + \hat{L}^{2}_{z} - \hat{L}_{z} = \hat{L}_{-} 
 \hat{L}_{+} + \hat{L}^{2}_{z} + \hat{L}_{z} \; \; . \; \; \label{equant3}
\end{equation}

Note that all these equations, Eqs.(\ref{equant1}) - (\ref{equant3}), mentioned above are easily generalized 
to the few- and many-body quantum systems. Moreover, by using these equations one can explicitly construct 
all irreducible representations of the rotation $SO(3)$-group. This process is well described in details and 
at very good level in various books (see, e.g., \cite{GelfMS}, \cite{Rose} and \cite{VandWard}). Here we do 
not want to repeat any of these descriptions here. Instead, let us discuss the two alternative approaches 
which are less known, but have the same high efficiency for obtaining all irreducible representations of the 
rotation group.  

\subsection{Second quantization for the angular momentum}

Here we want to describe the two effective approaches which are often used to construct the irreducible 
representations of the rotation $SO(3)$ group. The both these approaches which are based on the second 
quantization technique. First, we shall assume that we have two bosonic operators $a$ and $a^{+}$ for 
which the following commutation relations are obeyed $[ a, a^{+}] = 1, [ a, a] = 0$ and $[ a^{+}, a^{+}] 
= 0$. For a given positive integer and/or semi-integer number $j$ we can construct the two following 
operators 
\begin{equation}
 \hat{j}_{-} = \sqrt{2 j - a^{+} a} \; a \; \; \; , \; \; {\rm and} \; \; \; \hat{j}_{+} = a^{+} \;
 \sqrt{2 j - a^{+} a} \; . \; \label{secquant1}
\end{equation}
As follows from these definitions the products of these two operators are 
\begin{equation}
 \hat{j}_{-} \hat{j}_{+} = (2 j - a^{+} a) a^{+} a + 2 j - a^{+} a \; \; , \; {\rm and} \; \; \; 
 \hat{j}_{+} \hat{j}_{-} = (2 j - a^{+} a) a^{+} a + 2 j + a^{+} a \; , \; \label{secquant2}
\end{equation}
respectively. Here we used the fact that the operator $a^{+} a$ commute with any analytical function (or
operator-function) of the $a^{+} a$ operator, i.e., $[a^{+} a, \hat{f}(a^{+} a)] = 0$. From here it is 
easy to find the explicit form of the $j_z = \frac12 \; [ j_{+}, j_{-}]$ operator:  
\begin{equation}
 2 j_{z} = [ \hat{j}_{+}, \hat{j}_{-}] = \hat{j}_{+} \hat{j}_{-} - \hat{j}_{-} \hat{j}_{+} = - 2 j 
 + 2 a^{+} a \; \; \; , \; {\rm or} \; \; \; \hat{j}_z = - j + a^{+} a \; \; . \; \label{secquant3}
\end{equation} 
This formula allows one to obtain the following expression for the $\hat{j}^{2}_z$ operator 
\begin{equation}
 \hat{j}^{2}_{z} = j^{2} - 2 j a^{+} a + a^{+} a a^{+} a \; \; . \; \label{secquant4}
\end{equation} 
Finally, we obtain the explicit formula for the $\hat{j}^{2}$ operator
\begin{eqnarray}
 \hat{j}^{2} &=& \hat{j}_{+} \hat{j}_{-} + \hat{j}^{2}_{z} - \hat{j}_{z} = \hat{j}_{-}\hat{j}_{+} + 
 \hat{j}^{2}_{z} + \hat{j}_{z} = (2 j - a^{+} a)\; a^{+} 
 a \; \; + \; a^{+} a \; + \; a^{+} a a^{+} a \nonumber \\
 &-& 2 \; j \;a^{+} a \; + \; j^{2} \; + \; j \; - \; a^{+} a = j^{2} + j = j (j + 1) \; . \; 
 \label{secquant5}
\end{eqnarray} 
This result is crucial to test the overall correctness of our method.  

Let us discuss briefly the construction of basis set. The lowest $\mid 0 \rangle$ vector (or state) is defined 
by the equation $a \mid 0 \rangle = 0$. For this vector one finds $\hat{j}_z \mid 0 \rangle = (- j + a^{+} a) 
\mid 0 \rangle = - j \mid 0 \rangle$. The next vector is $\mid 1 \rangle = a^{+} \mid 0 \rangle$, for which we 
can write 
\begin{eqnarray}
 \hat{j}_z \mid 1 \rangle &=& (- j + a^{+} a) a^{+} \mid 0 \rangle = (- j a^{+} + a^{+} a a^{+}) \mid 0 \rangle 
 = a^{+} (- j + 1 + a^{+} a) \mid 0 \rangle \nonumber \\
 &=& a^{+} (- j + 1) \mid 0 \rangle =  (- j + 1) \mid 1 \rangle \; \; . \; \label{secquant6}
\end{eqnarray}
Analogously, the vector $\mid 2 \rangle = a^{+} \mid 1 \rangle = (a^{+})^{2} \mid 0 \rangle$ obeys the following 
equation $\hat{j}_z \mid 2 \rangle = (-j + 2) \mid 2 \rangle$. In general, for the $\mid k \rangle = (a^{+})^{k} 
\mid 0 \rangle$ vector one finds: $\hat{j}_z \mid k \rangle = (-j + k) \mid k \rangle$, where $k \le 2 j$. The 
corresponding unit-norm vectors are written in the form $\mid k \rangle = \frac{1}{\sqrt{k!}} (a^{+})^{k} \mid 0 
\rangle$, where $0! = 1$ and $k$ is positive integer. For $k = 2 j$ we obtain: $\hat{j}_z \mid 2 j \rangle = j 
\mid 2 j \rangle$. Furthermore, as follows from here that $a^{+} \mid 2 j \rangle = 0$, since otherwise it would 
be possible to obtain the equality $\hat{j}_z a^{+} \mid 2 j \rangle = (j + 1) a^{+} \mid 2 j \rangle$ which means
that the actual dimension of our $(2 j + 1)-$dimensional irreducible representation exceeds $(2 j + 1)$. To avoid 
this problem we have to assume that $a^{+} \mid 2 j \rangle = 0$. 

Another approach which is often used to construct the irreducible representations of the rotation $SO(3)$ group 
is slightly more complicated, but it is more transparent, simple and advanced in applications. It is also based 
on the second quantization technique for the two combined (or conjugate) bosonic fields. In this cases the 
unit-norm vectors of state are written in the form 
\begin{eqnarray}
  \mid j \; m \rangle = \frac{1}{\sqrt{(j + m)! (j - m)!}} \; (a^{+}_{1})^{j+m} \; (a^{+}_{2})^{j-m} 
   = N_{j;m} \; (a^{+}_{1})^{j+m} \; (a^{+}_{2})^{j-m} \; \; , \; \; \label{secquant7}
\end{eqnarray}
where $N_{j;m} = \frac{1}{\sqrt{(j + m)! (j - m)!}}$ is the normalization constant. The three operators $\hat{j}_{x}, 
\hat{j}_{y}$ and $\hat{j}_{z}$ take the form 
\begin{eqnarray}
  \hat{j}_{x} &=& \frac12 {\bf a}^{+} \hat{\sigma}_{x} {\bf a} = \frac12 \Bigl( a^{+}_{1} a_{2} + a^{+}_{2} a_{1} \Bigr) 
  \; \; , \; \; 
  \hat{j}_{y} = \frac12 {\bf a}^{+} \hat{\sigma}_{y} {\bf a} = \frac{\imath}{2} \Bigl( a^{+}_{2} a_{1} - a^{+}_{1} a_{2} 
  \Bigr)\; \; , \; \;  \nonumber \\
  \hat{j}_{z} &=& \frac12 {\bf a}^{+} \hat{\sigma}_{z} {\bf a} = \frac12 \Bigl( a^{+}_{1} a_{1} - a^{+}_{2} a_{2} 
  \Bigr)\; \; , \; 
\end{eqnarray}
where $\sigma_{i}$ ($i = x, y, z$), are the three Pauli matrices
\begin{eqnarray}
 \hat{\sigma}_x = \left( \begin{array}{cc} 0 \; & \; 1 \\
                                     1 \; & \; 0 \\
  \end{array} \right) \; \; , \; \; 
  \hat{\sigma}_y = \left( \begin{array}{cc} 0 \; & \; - \imath \\
                                     \imath \; & \; 0 \\
  \end{array} \right) \; \; , \; \; 
   \hat{\sigma}_z = \left( \begin{array}{cc} 1 \; & \; 0 \\
                                     0 \; & \; -1 \\
   \end{array} \right) \; \; , \; \;  \label{PauliM}
\end{eqnarray} 
while the two-dimensional (or two-component) vectors ${\bf a}^{+}$ and ${\bf a}$ are $(a^{+}_1, a^{+}_2)$ 
and $\left( \begin{array}{c} a_1 \\ a_2 \end{array} \right)$, respectively. This method was originally 
developed for the compact $SO(3)$ group by J. Schwinger in 1952 (his report (1952) from Oak Ridge 
Laboratory is mentioned, e.g., in \cite{VHein}). The complete form of this method is openly described in 
Chapter 3 (entitled `\textit{Angular Momentum}') of his book \cite{Schwing}. Similar approaches have 
later been developed for many other groups (see, e.g., \cite{BarR} and references therein), including the 
non-compact groups such as $SO(2,1), SO(3,1)$, etc. In particular, our original method developed in 
\cite{Fro2002} and used for obtaining analytical solutions of the Coulomb three-body problems in Quantum 
Mechanics is based on a rigorous procedure of second quantization performed for the three independent 
non-compact $SO(2,1)-$groups.  

\section{Analytical calculation of the quantum Poisson brackets of momenta with vector-operators}

In this Section we derive some important formulas for the quantum Poisson bracket (or commutator, for short) 
$[ \hat{\bf J}^{2}, \hat{\bf A}]$, where $\hat{\bf J}^{2} = J_{x} J_{x} + J_{y} J_{y} + J_{z} J_{z} = J_{a} 
J_{a}$. By using the `product law', which is one of the fundamental properties of the Poisson brackets (see, 
Eq.(1-12) on p.10 in \cite{Dir64} and \$ 21 in \cite{Dir59}), we can write the following identity
\begin{equation}
 [ \hat{\bf J}^{2}, \hat{\bf A}] = \hat{\bf J} \; [ \hat{\bf J}, \hat{\bf A}] + [ \hat{\bf J}, \hat{\bf A}] 
 \; \hat{\bf J} \; . \; \label{eq1}
\end{equation}
The same equality written in Cartesian components of the $\hat{\bf J}$ and $\hat{\bf A}$ vector-operators 
takes the form 
\begin{equation}
 [ \hat{\bf J}^{2}, \hat{\bf A}] = \hat{J}_{a} [ \hat{J}_{a}, \hat{A}_{b}] + [\hat{J}_{a}, \hat{A}_{b}]  
 \hat{J}_{a} = \hat{J}_{a} \imath \varepsilon_{abc} \hat{A}_{c} + \imath \varepsilon_{abc} \hat{A}_{c} 
 \hat{J}_{a} = - \imath \varepsilon_{bac} \hat{J}_{a} \hat{A}_{c} + \imath \varepsilon_{bca} \hat{A}_{c} 
 \hat{J}_{a} \; , \; \label{eq2}
\end{equation}
where we have applied the fundamental identity. In Eq.(\ref{eq2}) we can introduce the notation $({\bf a} 
\times {\bf b})$ for the vector product of the vectors ${\bf a}$ and ${\bf b}$ (see above). In components 
we can write $\varepsilon_{bac} \hat{J}_{a} \hat{A}_{c} = - (\hat{\bf J} \times \hat{\bf A})_{b}$ and 
$\varepsilon_{bca} \hat{A}_{c} \hat{J}_{a} = (\hat{\bf A} \times \hat{\bf J})_{b}$. With this notation the 
last equation is re-written to its final form 
\begin{equation}
 [ \hat{\bf J}^{2}, \hat{\bf A} ] = \imath \Bigl((\hat{\bf A} \times \hat{\bf J}) - (\hat{\bf J} \times 
 \hat{\bf A})\Bigr) \; . \; \label{eq3}
\end{equation}
From this formula one easily obtains that $[ \hat{\bf J}^{2}, \hat{\bf J} ] = 0$. 

Now, consider the quantum Poisson bracket $[ \hat{\bf J}^{2}, [ \hat{\bf J}^{2}, \hat{\bf A}]]$ which is 
slightly more complicated than the Poisson bracket, Eq.(\ref{eq3}), considered above. By introducing the 
new vector ${\bf Z} = [ \hat{\bf J}^{2}, \hat{\bf A}]$ we can reduce the double commutator $[ \hat{\bf 
J}^{2}, [ \hat{\bf J}^{2}, \hat{\bf A}]]$ to the form of Eq.(\ref{eq3}). After a few additional and 
obvious steps of transformations one finds  
\begin{eqnarray}
 [ \hat{\bf J}^{2}, [ \hat{\bf J}^{2}, \hat{\bf A} ] ] &=& [ \hat{\bf J}^{2}, \hat{\bf Z}] =  \imath 
 \Bigl((\hat{\bf Z} \times \hat{\bf J}) - (\hat{\bf J} \times \hat{\bf Z})\Bigr) \nonumber \\
 &=& - (\hat{\bf A} \times \hat{\bf J}) \times \hat{\bf J} + (\hat{\bf J} \times \hat{\bf A}) \times 
 \hat{\bf J} + \hat{\bf J} \times (\hat{\bf A} \times \hat{\bf J}) - \hat{\bf J} \times (\hat{\bf J} 
 \times \hat{\bf A}) \; . \; \label{eq4}
\end{eqnarray}
The last formula is transformed by using the expression(s) for the double vector product 
\begin{equation}
 {\bf a} \times ({\bf b} \times {\bf c}) = {\bf b} ({\bf a} \cdot {\bf c}) - {\bf c} ({\bf a} \cdot 
 {\bf b}) \; \; , \; {\rm and} \; \; ({\bf a} \times {\bf b}) \times {\bf c} = {\bf b} ({\bf a} 
 \cdot {\bf c}) - {\bf a} ({\bf b} \cdot {\bf c}) \; , \; \label{eq5} 
\end{equation}
where the first formula is the well known `a b c = b a c minus c a b' formula (see, e.g., Eqs.(15) and 
(16) in \$ 7 from \cite{Kochin}), while the second formula is easily derived from the first one. By 
applying these formulas one finds the following expressions for the four terms $T_i$ in Eq.(\ref{eq5}):
\begin{eqnarray}
 &T_1& = -(\hat{\bf A} \times \hat{\bf J}) \times \hat{\bf J} = - (\hat{\bf A} \cdot \hat{\bf J}) \; 
 \hat{\bf J} + \hat{\bf A} \; \hat{\bf J}^{2} \; \; , \; T_2 = (\hat{\bf J} \times \hat{\bf A}) \times 
 \hat{\bf J} = \hat{\bf A} \; \hat{\bf J}^{2} - \hat{\bf J} \; (\hat{\bf A} \cdot \hat{\bf J}) \; , 
 \; \label{eq6a} \\
 &T_3& = \hat{\bf J} \times (\hat{\bf A} \times \hat{\bf J}) = \hat{\bf J}^{2} \; \hat{\bf A} - 
 \hat{\bf J} \; (\hat{\bf A} \cdot \hat{\bf J}) \; \; , \; \; \; T_4 = - \hat{\bf J} \times 
 (\hat{\bf J} \times \hat{\bf A}) = - \hat{\bf J} \; (\hat{\bf J} \cdot \hat{\bf A}) + \hat{\bf J}^{2} \; 
 \hat{\bf A} \; . \; \label{eq6b} 
\end{eqnarray}  
The sum of these four terms takes the form 
\begin{eqnarray}
 T_1 + T_2 + T_3 + T_4 = [ \hat{\bf J}^{2}, [ \hat{\bf J}^{2}, \hat{\bf A}]] = 2 \hat{\bf A} \; 
 \hat{\bf J}^{2} + 2 \hat{\bf J}^{2} \; \hat{\bf A} - 4 \hat{\bf J} \; (\hat{\bf J} \cdot \hat{\bf A}) 
 \; . \; \; \label{eq7} 
\end{eqnarray} 
For the first time this result was obtained by Dirac and its `classical', derivation can be found, e.g., 
in \cite{ConSho}. Note that the two last formulas can be re-written in a different forms, since the 
operator $\hat{\bf J}$ always commute with the scalar $(\hat{\bf J} \cdot \hat{\bf A})$. This allows one 
to write the third term in Eq.(\ref{eq7}) in a few different (but equivalent!) forms, e.g., $\hat{\bf J} 
\; (\hat{\bf J} \cdot \hat{\bf A}) = (\hat{\bf J} \cdot \hat{\bf A}) \; \hat{\bf J} = \hat{\bf J} \; 
(\hat{\bf A} \cdot \hat{\bf J}) = (\hat{\bf A} \cdot \hat{\bf J}) \; \hat{\bf J}$, etc. 

To finish this part let us obtain an important quantum formula which is a direct generalization of the 
well known classical formula from analytical geometry. Suppose that we have some given non-zero vector 
${\bf a}$ and an arbitrary vector ${\bf b}$. Our goal is to decompose this vector ${\bf b}$ into two 
components, one of which is parallel and the other perpendicular to the given vector ${\bf a}$. Solution 
of this problem is described by the formula (see, p. 64 in \cite{Kochin})  
\begin{eqnarray}
   {\bf b} = \frac{{\bf a} \cdot {\bf b}}{a^{2}} \; {\bf a} \; +  \; \frac{1}{a^{2}} \; {\bf a} \times 
   ({\bf b} \times {\bf a}) \; \; , \; \; \label{ParalPerp}
\end{eqnarray} 
where $a^{2} = {\bf a} \cdot {\bf a} \ne 0$. This formula plays a very important role in many actual 
problems known from analytical geometry, classical mechanics and physics. Here we want to obtain a similar 
formula in quantum mechanics for two non-commuting operators. For the angular momentum this means that we 
have to represent an arbitrary vector-operator $\hat{\bf A}$ as a sum of two components, one of which is 
parallel to a given vector-operator of angular momentum $\hat{\bf J}$, while the second is perpendicular to 
it. As follows from Eq.(\ref{eq7}) and other formulas derived in this Section the exact solution of this 
problem takes the form   
\begin{eqnarray}
 \hat{\bf A} = \frac{(\hat{\bf J} \cdot \hat{\bf A})}{J (J + 1)} \; \hat{\bf J} + \frac{1}{4 \; 
  J (J + 1)} \; [ \hat{\bf J}^{2}, [ \hat{\bf J}^{2}, \hat{\bf A} ] ] \; \; , \; \; \label{ParalPerpVO}
\end{eqnarray} 
where the first term on the right side is also called the longitudinal (or irrotational) component of 
vector-operator $\hat{\bf A}$, while the second term is its transverse (or solenoidal) component. This 
relation plays a great role in atomic physics (see, e.g., \cite{ConSho}), and it will be used in the next
Section. 

\subsection{More complex Poisson brackets}

The method developed above allows one to obtain the explicit expressions for the quantum Poisson brackets 
which include higher powers of angular momenta such as $[ \hat{\bf J}^{2}, [ \hat{\bf J}^{2}, [ \hat{\bf J}^{2}, 
\hat{\bf A}]]], [ \hat{\bf J}^{2}, [ \hat{\bf J}^{2}, [ \hat{\bf J}^{2}, [ \hat{\bf J}^{2}, \hat{\bf A}]]]]$, 
etc. Furthermore, based on our experience in analytical calculations of similar commutators we could formulate 
a few general rules which can be useful in this procedure. First, let us determine the following Poisson 
bracket 
\begin{eqnarray}
 & &[ \hat{\bf J}^{2}, [ \hat{\bf J}^{2}, [ \hat{\bf J}^{2}, \hat{\bf A}]]] = 2 \hat{\bf Z} \; \hat{\bf J}^{2} 
 + 2 \hat{\bf J}^{2} \; \hat{\bf Z} - 4 \hat{\bf J} \; (\hat{\bf J} \cdot \hat{\bf Z}) \nonumber \\
  & & = 2 \imath \Bigl\{ \hat{\bf J}^{2} \; \Bigl((\hat{\bf A} \times \hat{\bf J}) - (\hat{\bf J}  \times 
  \hat{\bf A})\Bigr) + \Bigl((\hat{\bf A} \times \hat{\bf J}) - (\hat{\bf J} \times \hat{\bf A}) \Bigr) 
  \; \hat{\bf J}^{2} \Bigr\} + 8 \; \hat{\bf J} (\hat{\bf J} \cdot \hat{\bf A}) \; , \; \label{eq7a} 
\end{eqnarray}  
where in the second equation $\hat{\bf Z} = [ \hat{\bf J}^{2}, \hat{\bf A}]$. To obtain the result in 
Eq.(\ref{eq7a}), we have used the fact that $(\hat{\bf J} \cdot \hat{\bf Z}) = - 2 (\hat{\bf J} \cdot 
\hat{\bf A})$. The next Poisson bracket which we want to consider here is slightly more complicated, since it 
contains one additional $\hat{\bf J}^{2}$ operator. By applying the formula, Eq.(\ref{eq7}), one obtains the 
following formula
\begin{eqnarray}
 [ \hat{\bf J}^{2}, [ \hat{\bf J}^{2}, [ \hat{\bf J}^{2}, [ \hat{\bf J}^{2}, \hat{\bf A}]]]] &=& 4 \;  
 \hat{\bf J}^{4} \; \hat{\bf A} + 8 \; \hat{\bf J}^{2} \hat{\bf A} \hat{\bf J}^{2} + 4 \; \hat{\bf A} 
 \hat{\bf J}^{4} - 8 \Bigl(\hat{\bf J}^{3} \; (\hat{\bf J} \cdot \hat{\bf A}) + (\hat{\bf J} \cdot 
 \hat{\bf A}) \hat{\bf J}^{3} \Bigr) \; , \; \label{eq8} 
\end{eqnarray}  
which can also be written in a number of equivalent forms, since it is clear that $\hat{\bf J}^{3} \; 
(\hat{\bf J} \cdot \hat{\bf A}) = \hat{\bf J}^{2} \; (\hat{\bf J} \cdot \hat{\bf A}) \hat{\bf J} = 
\hat{\bf J} \; (\hat{\bf J} \cdot \hat{\bf A}) \hat{\bf J}^{2} = (\hat{\bf J} \cdot \hat{\bf A}) 
\hat{\bf J}^{3}$, etc. In general, for the commutators with large number(s) of operators $\hat{\bf J}^{2}$ 
the scalar terms which contain the scalar product $(\hat{\bf J} \cdot \hat{\bf A})$ may present a problem. 
To avoid all related headaches it is better to re-define the vector-operator(s) $\hat{\bf A}^{\prime} = 
\hat{\bf A} - \Bigl( \hat{\bf J}^{2} \Bigr)^{-1} \hat{\bf J} (\hat{\bf J} \cdot \hat{\bf A})$. This `new'
vector-operator $\hat{\bf A}^{\prime}$ is transverse in respect to the vector-operator of angular momentum 
$\hat{\bf J}$, i.e., $(\hat{\bf J} \cdot \hat{\bf A}^{\prime}) = 0 = (\hat{\bf A}^{\prime} \cdot 
\hat{\bf J})$.  

Now, we are ready to deal with more general Poisson brackets which can be written in the `universal' 
form $[ \hat{\bf J}^{2}, \ldots [\hat{\bf J}^{2}, [ \hat{\bf J}^{2}, \hat{\bf A}]]\ldots]$. Analytical 
calculation of such Poisson brackets is more complicated process than our calculations performed above 
and results of such calculations explicitly depend upon the total number of $\hat{\bf J}^{2}$ operators 
included in these Poisson brackets. Below, this number is designated by the notation $N_{\hat{\bf J}^{2}}$. 
If the $N_{\hat{\bf J}^{2}}$ number is odd, i.e., $N_{\hat{\bf J}^{2}} = 2 k + 1$, then the multiple 
commutator, $[ \hat{\bf J}^{2}, \ldots [\hat{\bf J}^{2}, [ \hat{\bf J}^{2}, \hat{\bf A}]]\ldots]$ always 
contains the factor $2^{k-1} \imath$, where $\imath$ is the imaginary unit. Each of these terms in the 
final expression has the same structure: it includes the common factor-operator $\Bigl((\hat{\bf A} \times 
\hat{\bf J}) - (\hat{\bf J} \times \hat{\bf A})\Bigr)$ which is multiplied from the both sides (left and 
right) by the even powers of momenta $\hat{\bf J}$. In addition to this, there is an additional term which 
equals to the product of the scalar factor $(\hat{\bf J} \cdot \hat{\bf A})$ and different powers of 
vector-operator $\hat{\bf J}$.  

In the opposite case, when $N_{\hat{\bf J}^{2}}$ number is even (or $N_{\hat{\bf J}^{2}} = 2 k$), the multiple 
commutator, Eq.(\ref{eq8}), always contains the two following sums. The first sum includes only products of 
even powers of vector-operator $\hat{\bf J}$ and vector-operator $\hat{\bf A}$. Briefly, we can say that 
the first sum contains all products such as $C_{m,n} 2^{m+n} \hat{\bf J}^{2 m} \hat{\bf A} \hat{\bf J}^{2 n}$, 
where the both $m$ and $n$ are integer non-negative numbers and $C_{m,n}$ is some positive numerical 
coefficient. The second sum includes only terms each of which is the product of odd powers of $\hat{\bf J}$ 
and vector-operator $\hat{\bf A}$, e.g., $\hat{\bf J}^{2 m -1} \hat{\bf A} \hat{\bf J}^{2 n -1}$. The general 
expression for any term in this (second) sum takes the form $B_{m,n} 2^{m+n+2} \hat{\bf J}^{2 m -1} \hat{\bf A} 
\hat{\bf J}^{2 n -1}$, where now the both $m$ and $n$ are integer positive numbers and $B_{m,n}$ is some 
negative numerical coefficient.  

\section{Matrix elements of vector-operators in the basis of spherical states}

The equations derived in the previous Section allow one to obtain a number of relations between matrix elements 
of different vector-operators. Many of these relations were crucially important in early days of atomic 
spectroscopy (see, e.g., \cite{Wign1} - \cite{Bethe} and references therein). Currently, by using the well known 
method of tensor operators (see, e.g., \cite{LLQ}, \cite{Sob}, \cite{FroF}) one easily obtains the same formulas 
for all matrix elements required in actual atomic calculations. However, before this powerful method has finally 
been developed all scientists working in quantum mechanics used a different approach which was based on formulas 
derived below. If you are working with the matrix elements of some vector physical quantities only, then this old 
method is sufficient. Let us discuss this traditional method in detail which is appropriate from methodological 
point of view. First, let us introduce the basis set of quantum states $| \lambda J M \rangle$, where each of 
these states is an eigenfunction of the $\hat{\bf J}^{2}$ and $\hat{J}_z$ operators
\begin{eqnarray}
 \hat{\bf J}^{2} | \lambda J M \rangle = J (J + 1) |\lambda J M \rangle \; \; , \; \; {\rm and} \; \; \; \; 
 \hat{J}_z |\lambda J M \rangle = M | \lambda J M \rangle \; \; , \; \; \label{eq90} 
\end{eqnarray} 
where $\hat{J}_z$ is the $z-$component of the momentum $\hat{\bf J}$ which is a vector-operator. Here and below, 
the notation $\lambda$ designates the set of other quantum numbers which are either conserve in some problem, or 
play a significant role during construction of this basis set. In actual applications, it is convenient to 
represent these $|\lambda J M \rangle$ states in the following factorized form $|\lambda J M \rangle = | \lambda 
J \rangle \mid J M \rangle$, where the first state (i.e., $| \lambda J \rangle$ state) does not depend upon $M$. 
In reality, such factorized basis sets of `atomic' (or spherical) states $| \lambda J M \rangle = | \lambda J 
\rangle \mid J M \rangle$ states is often applied to solve many problems known in atomic and nuclear physics. 

In this Section we consider analytical and numerical calculations of matrix elements of an arbitrary vector-operator 
in the factorized basis of $|\lambda J \rangle \mid J M \rangle$ states. Briefly, we want to derive the explicit 
formulas for the matrix elements of an arbitrary self-conjugate vector-operator $\hat{\bf A}$ in the basis of 
factorized `atomic' states $|\lambda J M \rangle$. To achieve this goal let us obtain the explicit matrix formulas 
for the main both sides of main vector relation $[ \hat{J}_{i}, \hat{A}_{j}] = \imath \varepsilon_{ijk} \hat{A}_{k} 
= \imath \varepsilon_{kij} \hat{A}_{k}$ and Eq.(\ref{ParalPerpVO}) in the basis of factorized `atomic' states 
$|\lambda J M \rangle$. For the diagonal (upon $J$) matrix elements it is relatively easy to show (see, e.g., 
\cite{Bethe}) that an arbitrary $\alpha$-component of the vector-operator $\hat{\bf A}$ its matrix elements (with 
different $M$) are proportional to the corresponding matrix elements of the vector-operator $\hat{\bf J}$, i.e., we 
always have   
\begin{eqnarray}
  \langle \lambda J M_1 | \hat{A}_{\alpha} | \lambda J M \rangle = R(\lambda \lambda^{\prime} J) \; \langle 
  \lambda J M^{\prime} | \hat{J}_{\alpha} | \lambda J M \rangle =  R(\lambda \lambda^{\prime} J) \; \langle 
  J M^{\prime} | \hat{J}_{\alpha} | J M \rangle \; \; , \; \; \label{eq9000} 
\end{eqnarray} 
or in other words
\begin{eqnarray}
  \langle \lambda J M_1 | \hat{\bf A} | \lambda J M \rangle = R(\lambda \lambda^{\prime} J)  
  \; \langle J M^{\prime} | \hat{\bf J} | J M \rangle \; \; , \; \; \label{eq9001} 
\end{eqnarray} 
where $R(\lambda \lambda^{\prime} J)$ is some scalar constant which does not depend upon $M$. From Eq.(\ref{eq9001}) 
one obtains the following expression 
\begin{eqnarray}
  \langle \lambda J_1 M_1 | \hat{\bf J} \cdot \hat{\bf A} | \lambda J M \rangle = R(\lambda \lambda^{\prime} J) \; 
 \delta(J_1,J) \; J(J + 1) \ \; \; , \; \; \label{eq9002}  
\end{eqnarray} 
where $\delta(J_1,J)$ is the Kroneker delta-symbol \cite{KS}. This equation indicates clearly that the scalar product 
$\hat{\bf J} \cdot \hat{\bf A}$ of the two vector-operators is diagonal upon $M$ (as expected). Equation (\ref{eq9002}) 
can be applied to determine the scalar constant $R(\lambda \lambda^{\prime} J)$. As follows from Eqs.(\ref{eq9001}) - 
(\ref{eq9002}) and Eq.(\ref{ParalPerpVO}) above the longitudinal component of the vector-operator $\hat{\bf A}$ equals 
$R(\lambda \lambda^{\prime} J) \; \delta(J_1,J)$. It is crucially important here that all matrix elements of the 
vector-operator $\hat{\bf J}$ between $J M$ and $J_1 M_1$ functions, where $J_1 \ne J$ equal zero identically. Indeed, 
this operator $\hat{\bf J}$ cannot have any matrix elements which connect basis functions with different $J$.

Now, let us calculate the matrix elements of the transverse component of the vector-operator $\hat{\bf A}$, which is 
represented by the second term in our Eq.(\ref{eq7}). After a few simple transformations and rearrangement of terms 
in the arising equation one finds the following equation: 
\begin{eqnarray} 
 \Bigl[(J - J_1 + 1)^2 - 1\Bigr] \; \Bigl[(J - J_1)^2 - 1\Bigr] \; \langle \lambda_1 J_1 M_1 | \hat{\bf A} 
 |\lambda J M \rangle = 0 \; \; . \; \label{eq95} 
\end{eqnarray} 
As one can see from this equation the matrix elements of the vector-operator $\hat{\bf A}$ differ from zero 
only in those cases when $J_1 = J, J \pm 1$, i.e., when at least one coefficient in front of the $\langle 
\lambda_1 J_1 M_1 | \hat{\bf A} |\lambda J M \rangle$ matrix element equals zero. Otherwise, the matrix 
element $\langle \lambda_1 J_1 M_1 | \hat{\bf A} |\lambda J M \rangle$ must be equal zero by itself. This 
means that for any vector-operator $\hat{\bf A}$ all transitions such as $J_1 = J \pm 2; J_1 = J \pm 3$; etc, 
are strictly prohibited. Thus, Eq.(\ref{eq95}) leads to the well known sets of `selection rules' $J_1 
\rightarrow J, J \pm 1$ (or $\Delta J = J_1 - J = 0, \pm 1$) for transition probabilities, where such 
transitions are generated by the vector-operator $\hat{\bf A}$. Similar rules can be formulated for other 
similar tensor operators. Note also that all $0 \rightarrow 0$ transitions are strictly prohibited \cite{LLQ}, 
since they cannot be generated by the vector-operator $\hat{\bf A}$. These `selection rules' are crucially 
important for many actual problems known in atomic, quasi-atomic and nuclear physics.      

Finally, let us present the explicit formulas for the matrix elements of the $\hat{A}_{z}$ component of an 
arbitrary vector-operator $\hat{\bf A}$ in the factorized basis sets of `atomic' (or spherical) states. For 
$z-$component of such a vector one finds 
\begin{eqnarray} 
 \langle n_1 \; J \; M \; | \hat{A}_z | \; n \; J \; M \rangle &=& \frac{M}{\sqrt{J (J + 1) (2 J + 1)}} 
 \; \langle n_1 \; J \; || \hat{A} || \; n \; J \rangle \; \; , \; \label{eq951} \\
 \langle n_1 \; J \; M \; | \hat{A}_z | \; n \; J-1 \; M \rangle &=& \sqrt{\frac{J^{2} - M^{2}}{J (2 J 
 - 1) (2 J + 1)}} \; \langle n_1 \; J \; || \hat{A} || \; n \; J-1 \rangle \; \; , \; \label{eq952} \\
 \langle n_1 \; J-1 \; M \; | \hat{A}_z | \; n \; J \; M \rangle &=& \sqrt{\frac{J^{2} - M^{2}}{J (2 J 
 - 1) (2 J + 1)}} \; \langle n_1 \; J-1 \; || \hat{A} || \; n \; J \rangle \; \; , \; \label{eq953} 
\end{eqnarray} 
where the notation $\langle n_1 \; J-1 \; || \hat{A} || \; n \; J \rangle$ stands for the reduced matrix 
elements which do not depend upon the angular quantum number $M$ in this case), or component index of the 
vector-operator $\hat{\bf A}$ ($z$ in this case). In these formulas and everywhere below we replaced the 
`quantum numbers' $\lambda$ introduced above by the quantum number $n$ which is more appropriate for our 
atomic analysis which is performed in the last sub-section of Section VI. Analytical and numerical 
calculations of these reduced matrix elements are relatively simple, and it is well discussed in Chapter 
5 of book by Edmonds \cite{Edm} (see also references therein). Other matrix elements of $(\hat{\bf A})_z 
= A_z$ operator equal zero identically.

Analogously, for the $\hat{A}_{-} = \frac{1}{\sqrt{2}} (A_x - \imath A_y)$ component of the vector-operator 
$\hat{\bf A}$ one finds three following formulas 
\begin{eqnarray} 
 \langle n_1 \; J \; M-1 \; | \hat{A}_{-} | \; n \; J \; M \rangle &=& \sqrt{\frac{(J + M) (J - M + 1)}{J 
 (J + 1) (2 J + 1)}} \; \langle n_1 \; J \; || \hat{A} || \; n \; J \rangle \; \; , \; \label{eq951A} \\
 \langle n_1 \; J \; M-1 \; | \hat{A}_{-} | \; n \; J-1 \; M \rangle &=& \sqrt{\frac{(J - M) (J - M + 1)}{J 
 (2 J - 1) (2 J + 1)}} \; \langle n_1 \; J \; || \hat{A} || \; n \; J-1 \rangle \; \; , \; \label{eq952A} \\
 \langle n_1 \; J-1 \; M-1 \; | \hat{A}_{-} | \; n \; J \; M \rangle &=& - \sqrt{\frac{(J + M) (J + M - 1)}{J 
 (2 J - 1) (2 J + 1)}} \; \langle n_1 \; J-1 \; || \hat{A} || \; n \; J \rangle \; \; . \; \label{eq953A} 
\end{eqnarray} 
Three similar formulas for the  $\hat{A}_{+} = \frac{1}{\sqrt{2}} (A_x + \imath A_y)$ component of the 
vector-operator $\hat{\bf A}$ take the form 
\begin{eqnarray} 
 \langle n_1 \; J \; M+1 \; | \hat{A}_{+} | \; n \; J \; M \rangle &=& \sqrt{\frac{(J - M) (J + M + 1)}{J 
 (J + 1) (2 J + 1)}} \; \langle n_1 \; J \; || \hat{A} || \; n \; J \rangle \; \; , \; \label{eq951B} \\
 \langle n_1 \; J \; M+1 \; | \hat{A}_{+} | \; n \; J-1 \; M \rangle &=& \sqrt{\frac{(J + M) (J + M + 1)}{J 
 (2 J - 1) (2 J + 1)}} \; \langle n_1 \; J \; || \hat{A} || \; n \; J-1 \rangle \; \; , \; \label{eq952B} \\
 \langle n_1 \; J-1 \; M+1 \; | \hat{A}_{+} | \; n \; J \; M \rangle &=& - \sqrt{\frac{(J - M) (J - M - 1)}{J 
 (2 J - 1) (2 J + 1)}} \; \langle n_1 \; J-1 \; || \hat{A} || \; n \; J \rangle \; \; . \; \label{eq953B} 
\end{eqnarray} 
Other matrix elements of the $\hat{\bf A}_{-}$ and  $\hat{\bf A}_{+}$ operators equal zero identically. Note 
that all formulas from Eqs.(\ref{eq951}) - (\ref{eq953B}) contain only three different kinds of the reduced 
matrix elements: $\langle n_1 \; J-1 \; || \hat{A} || \; n \; J \rangle, \langle n_1 \; J \; || \hat{A} || \; 
n \; J \rangle$ and $ \langle n_1 \; J\; || \hat{A} || \; n \; J-1 \rangle$. There is a simple relation 
between the corresponding $A_{-}$ and $A_{+}$ components of any self-conjugate vector-operator $\hat{\bf A}$: 
\begin{eqnarray} 
 \langle n_1 \; J_1 \; M_1 | \hat{A}_{+} | n \; J \; M \rangle = \langle n \; J \; M | \hat{A}_{-} | n_1 \; 
  J_1 \; M_1 \rangle^{\ast} \; \; , \; \label{eq954} 
\end{eqnarray}
where the notation $\ast$ means the complex conjugate expression. From here one finds the following symmetry 
for the vector $\hat{A}_{-}$ and $\hat{A}_{+}$ operators. An arbitrary non-zero matrix element of the 
$\hat{A}_{+}$ operator is obtained from the corresponding matrix element of the $\hat{A}_{-}$ operator by the 
substitution $M \rightarrow - M$ and vice versa. This can also be found from direct comparison of 
Eqs.(\ref{eq953A}) and (\ref{eq953B})). 

In the method of tensor operators we can produce the following general and universal expression for the matrix 
elements of an arbitrary tensor operator $T$ in the $| \; n \; L \; M \rangle$ basis:
\begin{eqnarray} 
   \langle n_1 \; L_1 \; M_1 \; | \hat{T}^{(k)}_{q} | \; n \; L \; M \rangle = (-1)^{L_1 - M_1} 
 \left( \begin{array}{ccc}  L_1  &   k  &  L \\
                            - M_1 &  q  &  M  \\
  \end{array} \right)
   \; \; \langle n_1 \; L_1 \; || \hat{T}^{(k)} || \; n \; L \rangle \; \; . \; \label{eq954} 
\end{eqnarray}
where $\hat{T}^{(k)}_{q}$ is the $q-$th component of the spherical tensor $\hat{T}^{(k)}$ of rank $k$. The 
notation in the $()-$brackets on the right side of this formula is the $3 j-$symbol of Wigner which are defined 
exactly as in \cite{Edm}. This equation is the well known Wigner-Eckart theorem  (see, e.g., \cite{LLQ} and 
\cite{BarR}). In our current case of vector operators we have to choose $k = 1$ and $q = -1, 0, +1$. The same 
result can be derived from the well known formula for the scalar product of three spherical harmonics:  
\begin{eqnarray}
  \langle Y_{\ell_1, m_1}(\theta, \phi) | Y_{\ell, m}(\theta, \phi) |  Y_{\ell_2, m_2}(\theta, \phi) \rangle 
  &=& (-1)^{m_{1}} \;  \imath^{\ell_2 + \ell - \ell_1} \; \sqrt{\frac{(2 \ell + 1) (2 \ell_1 + 1) (2 \ell_2 + 
   1)}{4 \; \pi}} \; \times \; \nonumber \\
   & & \left( \begin{array}{ccc}  \ell_1  & \ell &  \ell_2 \\
                            - m_{1}  &   m  &   m_{2}  \\
         \end{array} \right) \;                     
   \left( \begin{array}{ccc}  \ell_1  & \ell &  \ell_2 \\
                             0  &  0  &  0  \\    
       \end{array} \right)  \; \; . \; \label{eq9555}                  
\end{eqnarray}

An obvious simplification can be found for the matrix elements of any scalar product of two vector quantities. 
In this case by using the following formula for the scalar product of the two arbitrary vector-operators 
$\hat{\bf A}$ and $\hat{\bf B}$
\begin{eqnarray} 
  \hat{\bf A} \cdot \hat{\bf B} = \frac12 ( \hat{A}_{+} \hat{B}_{-} + \hat{A}_{-} \hat{B}_{+} ) + \hat{A}_z 
  \hat{B}_z \; \; , \; \label{eq955} 
\end{eqnarray}
where $\hat{A}_{+} = \frac{1}{\sqrt{2}} ( \hat{A}_{x} + \imath \hat{A}_{y} )$ and $\hat{A}_{-} = 
\frac{1}{\sqrt{2}}( \hat{A}_{x} - \imath \hat{A}_{y} )$ (definition of the $\hat{B}_{+}$ and $\hat{B}_{-}$ 
components is absolutely similar). After a few simple transformations one obtains the following formula for 
the matrix elements of scalar product of any two vector quantities 
\begin{eqnarray} 
 \langle n_1 \; J_1 \; M_1 \; | \hat{\bf A} \cdot \hat{\bf B} | \; n \; J \; M \rangle = \frac{\delta(J_1,J) 
 \delta(M_1,M)}{2 J + 1} \; \sum_{n_2 J_2} \langle n_1 \; J \; || \hat{A} || \; n_2 \; J_2 \rangle\langle 
 n_2 \; J_2 \; || \hat{B} || \; n \; J \rangle \; \; , \; \label{eq955a} 
\end{eqnarray} 
or in other words 
\begin{eqnarray} 
 \langle n_{1} \; J \; M \; | \hat{\bf A} \cdot \hat{\bf B} | \; n \; J \; M \rangle = \frac{1}{2 J + 1} \; 
 \sum_{n_2 J_2} \langle n_{1} \; J \; || \hat{A} || \; n_2 \; J_2 \rangle\langle n_2 \; J_2 \; || \hat{B} || 
 \; n \; J \rangle \; \; , \; \label{eq955b} 
\end{eqnarray} 
which contains only pair products of the reduced matrix elements. Other matrix elements of the $\hat{\bf A} 
\cdot \hat{\bf B}$ scalar product equal zero identically. The both formulas, Eqs.(\ref{eq955a}) - 
(\ref{eq955b}), are important in applications, e.g., they are extensively used to determine the matrix 
elements of many scalar products.  

\section{Tensor quantities averaged over magnitudes of the orbital (angular) momenta}

There are many other interesting and important consequences of the main commutation relation $[ \hat{J}_{i}, 
\hat{A}_{j}] = \imath \varepsilon_{ijk} \hat{A}_{k}$ for the vector-operators in Quantum Mechanics. One of 
such consequences was used in early years of Quantum Mechanics to develop an effective approach for analytical 
evaluations of various expectation values of for the rotationally excited (bound) states. Suppose we have some 
closed quantum system, which contains electrons and does not interact noticeably with other systems. In the 
general case, in any similar system a set of conserving operators includes the orbital angular momentum 
$\hat{\bf \ell}$ which is vector. As follows from the results of two previous Sections any vector quantity in 
such system which varies slowly must have the same direction as $\hat{\bf \ell}$, or it is proportional to 
the vector $\hat{\bf \ell}$. 

Now, let us consider some internal (quantum) motion in the such a system. During this motion the radius-vector 
${\bf r}_{i}$ of the particle $i$ and vector of momenta ${\bf p}_{j}$ of the particle $j$ of all particles will 
change somehow. In general, it is extremely difficult, or even impossible, to describe in detail the complete 
time-evolution of these basic dynamical variables, i.e., the ${\bf r}_{i}, {\bf p}_{j}$ variables in our case. 
However, in many cases we just need to know only actual changes of a few slow varying physical quantities which 
are represented by some irreducible tensors (in respect to rotations) of different ranks. If all arising tensors 
have even ranks and each of them is either a polynomial function of the basic dynamical variables, then the    
original problem is drastically simplified and often we can obtain even its exact solution(s). 

To illustrate how this works in reality let us consider the irreducible tensor of second rank $t_{ij} = 
{\bf n}_{i} \; {\bf n}_{j}$ which is formed from the two unit vectors ${\bf n}_{i} = \frac{{\bf r}_{i}}{r_{i}}$ 
and ${\bf n}_{j} = \frac{{\bf r}_{j}}{r_{j}}$, which are, in fact, the directions of the two radius-vectors of  
particles $i$ and $j$, respectively. We want to describe the time-evolution of this tensor, which is designated 
below as $\overline{t}_{ij} = \overline{{\bf n}_{i} \; {\bf n}_{j}}$. The explicit formula for this tensor 
averaged over a state, for which the magnitude of angular momentum $\ell = \mid \vec{\ell} \mid$ is given, takes 
the form (see, e.g., \cite{LLQ})  
\begin{eqnarray}
 \overline{{\bf n}_{i} \; {\bf n}_{j}} = \frac13 \; \delta_{ij} - \frac{1}{(2 \ell - 1) (2 \ell + 3)} 
 \Bigl[ \hat{\ell}_{i} \hat{\ell}_{j} + \hat{\ell}_{j} \hat{\ell}_{i} - \frac23\; \delta_{ij} \; \ell 
 (\ell + 1) \Bigr] \; , \; \label{eq9} 
\end{eqnarray} 
where $\hat{\ell}_{i}$ are the Cartesian components of the momentum vector $\vec{\ell}$, while ${\bf n}_{i} 
= \frac{{\bf r}_i}{r_i}$ and ${\bf n}_{j} = \frac{{\bf r}_j}{r_j}$ are the unit vectors oriented along the 
original radius-vectors of particles ${\bf r}_{i}$ and ${\bf r}_{j}$, respectively. In early studies this 
procedure was called averaging of the ${\bf n}_{i} \; {\bf n}_{j}$ product over the magnitude of the partial 
angular momentum ${\bf \ell}$ (or ${\bf \ell}_1$). Note again that direction of the $\vec{\ell}$ vector is 
not known during this procedure.   

The equation, Eq.(\ref{eq9}), can be written in a slightly different form  
\begin{eqnarray}
 (\hat{\bf a} \cdot \overline{{\bf n}) ({\bf n}} \cdot \hat{\bf b}) &=& \frac13 (\hat{\bf a} \cdot 
 \hat{\bf b}) - \frac{1}{(2 \ell - 1) (2 \ell + 3)} \Bigl[ \hat{a}_{i} \hat{\ell}_{i} \hat{\ell}_{j} 
 \hat{b}_{j} + \hat{a}_{i} \hat{\ell}_{j} \hat{\ell}_{i} \hat{b}_{j} - \frac23\; \; \ell (\ell + 1) \; 
 (\hat{\bf a} \cdot \hat{\bf b}) \Bigr] \; \nonumber \\
 &=& \frac13 (\hat{\bf a} \cdot \hat{\bf b}) - \frac{2}{(2 \ell - 1) (2 \ell + 3)} \Bigl[ (\hat{\bf a} 
 \cdot \hat{\bf \ell}) (\hat{\bf \ell} \cdot \hat{\bf b}) - \frac13\; \; \ell (\ell + 1) \; (\hat{\bf a} 
 \cdot \hat{\bf b}) \Bigr] \; , \; \label{eq91} 
\end{eqnarray} 
where $\hat{\bf a}$ and $\hat{\bf b}$ are the two vector-operators which are conserved during 
time-evolution of our sub-system. The notation $\hat{\bf \ell} = (\hat{\ell}_x, \hat{\ell}_y, 
\hat{\ell}_z) = (\hat{\ell}_a, \hat{\ell}_b, \hat{\ell}_c) = (\hat{\ell}_1, \hat{\ell}_2, \hat{\ell}_3)$ 
mean the vector-operator of partial momenta. The `normalization' factor in this formula $\; N = - 
\frac{1}{(2 \ell - 1) (2 \ell + 3)}$ is derived explicitly in the Appendix B. During transformations of 
Eq.(\ref{eq91}) we have used the following equation 
\begin{eqnarray}
 \hat{a}_{i} \hat{\ell}_{j} \hat{\ell}_{i} \hat{b}_{j} = \hat{a}_{i} \hat{\ell}_{i} \hat{\ell}_{j} 
 \hat{b}_{j} + \hat{a}_{i} [\hat{\ell}_{j}, \hat{\ell}_{i}] \hat{b}_{j} = \hat{a}_{i} \hat{\ell}_{i} 
 \hat{\ell}_{j} \hat{b}_{j} + \imath \; \varepsilon_{jik} \; \hat{a}_{i} \hat{\ell}_{k} \hat{b}_{j} = 
 (\hat{\bf a} \cdot \hat{\bf \ell}) (\hat{\bf \ell} \cdot \hat{\bf b}) + \imath \; \hat{\bf a} \cdot 
 (\hat{\bf \ell} \times \hat{\bf b}) \; , \; \label{eq92}  
\end{eqnarray} 
and the fact that any term which is linear upon the vector-operator $\hat{\bf \ell} = (\hat{\ell}_1, 
\hat{\ell}_2, \hat{\ell}_3)$ gives zero when averaged over all possible directions of $\hat{\bf \ell}$.  

As an example of applications of this method, let us determine the hyperfine structure splitting for an 
atom which contain one atomic electron (outside the closed electron shells) with orbital angular moment 
$\ell \ge 1$. As is well known (see, e.g., \cite{VHein}, \cite{LLQ}) the hyperfine structure splitting 
arises in the result of magnetic interaction between the nuclear magnetic momentum ${\bf m}_{N}$ and 
total angular momentum $\hat{\bf j}$ of a single electron which has the orbital angular momentum $\ell 
\ge 1$. The total angular momentum $\hat{\bf j}$ is the vector sum of the electron spin $\hat{\bf s}$ 
and its orbital angular momentum $\hat{\bf \ell}$, i.e., $\hat{\bf j} = {\bf s} + \hat{\bf \ell} = 
\hat{\bf \ell} + {\bf s}$. The actual magnetic interaction in atoms is represented as a sum of the 
vector potential ${\bf A}$ and magnetic field strength ${\bf H}$. The explicit formula for the 
corresponding Hamiltonian of hyperfine structure $\hat{H}_{hfs}$ takes the form (see, e.g., 
\cite{Jacks}) 
\begin{eqnarray}
  \hat{H}_{hfs} &=& \frac{| e |}{m c} \; ({\bf A} \cdot \hat{\bf p}) + \frac{| e |}{m c} \; ({\bf H} 
  \cdot \hat{\bf s}) = \frac{| e |}{m c} \; \frac{({\bf m}_{N} \times {\bf n}) \cdot {\bf p}}{r^{2}} 
  + \frac{| e |}{m c} \; \frac{[3 ({\bf m}_{N} \cdot {\bf n}) {\bf n} - 
  {\bf m}_{N}] \cdot \hat{\bf s}}{r^{3}} \; , \nonumber \\ 
 &=& \frac{| e |}{m c} \frac{({\bf m}_{N} \cdot \hat{\ell})}{r^{3}} + \frac{| e |}{m c} \; \frac{[ 3  
 ({\bf m}_{N} \cdot {\bf n}) {\bf n} - {\bf m}_{N} ] \cdot \hat{\bf s}}{r^{3}}  \; , \; \label{Hhfs}
\end{eqnarray}
where ${\bf p}$ is the `mechanical' momentum of the electron, $\hat{\ell} = {\bf r} \times {\bf p}$ is 
the vector-operator of electron orbital (angular) momentum and ${\bf n} = \frac{{\bf r}}{r}$ is the 
unit-norm vector of this electron in the field of central nucleus. Below, we shall assume that this 
vector is always directed from the central nucleus towards the electron. 

After a few simple transformations one obtains from Eq.(\ref{Hhfs}) the following expression 
\begin{eqnarray}
  \hat{H}_{hfs} = \frac{2 \alpha \mu_B}{r^{3}} ({\bf m}_{N} \cdot [ \hat{\bf \ell} + 3 (\hat{\bf s} 
  \cdot {\bf n}) {\bf n} - \hat{\bf s}]) \label{Hhfs1}
\end{eqnarray}
where $\mu_B = \frac{| e |}{2 m c}$ is the Bohr magneton. This Hamiltonian contains some scalar products, 
which are not entirely clear how to interpret, since in our case the both magnetic quantum numbers (for 
electron) $m_{\ell}$ and $m_s$ are not good quantum numbers, i.e., they are not conserved. These produces 
a number of troubles in actual computations. To avoid these problems we have to introduce the set of 
proper $\mid j m_j \rangle$ basis functions for which the two following equations hold: ${\bf j}^{2} \mid 
j m_j \rangle = j (j + 1) \mid j m_j \rangle$ and ${\bf j}_{z} \mid j m_j \rangle = m_{j} \mid j m_j 
\rangle$. By using these `pure'  states $\mid j m_j \rangle$ states as the basis functions, we can 
construct the projector-operator to the pure `rotational' $\mid j m_j \rangle$-states. This projector is 
written in the form $\rho_{{\bf j}} = \frac{1}{j(j + 1)} \; \mid {\bf j} \rangle\langle {\bf j} \mid$ and
it is used to define the following expectation value
\begin{eqnarray}
& &V_{hfs} = \langle \rho_{{\bf j}} \hat{H}_{hfs} \rangle = \frac{1}{j(j + 1)} \; \langle \psi \mid 
{\bf j} \rangle \langle {\bf j} \mid \hat{H}_{hfs} \mid \psi \rangle \nonumber \\
&=& \frac{2 \alpha \mu_B}{j (j + 1)} \; ({\bf m}_{N} \cdot \hat{\bf j}) \; [ (\hat{\bf \ell} \cdot 
  \hat{\bf j}) + 3 (\hat{\bf s} \cdot {\bf n}) ({\bf n} \cdot \hat{\bf j}) - (\hat{\bf s} \cdot 
  \hat{\bf j}) \Bigr] \; \langle \frac{1}{r^{3}} \rangle \; , \label{Hhfs2}
\end{eqnarray}
This expectation value is, in fact, the Hamiltonian of the hyperfine structure. The differences between 
eigenvalues of this Hamiltonian equals to the hyperfine structure splittings. Note that all scalar products 
in Eq.(\ref{Hhfs2}) are properly defined, but this expression still cannot be used for direct calculations, 
since it contains the internal vector ${\bf n}$ of the electron system, which is included in the two 
different scalar products. Formally, we can introduce the spatial (electron) tensor $|{\bf n} \rangle\langle 
{\bf n}|$ of the second rank. In tensor calculus similar tensors are called diadas \cite{Kochin}. Now, by 
applying the formula, Eq.(\ref{eq91}), to this tensor one can finish calculations of the $V_{hfs}$ 
expectation value. The expression in square brackets from Eq.(\ref{Hhfs2}) is reduced to the form  
\begin{eqnarray}
  & &(\hat{\bf \ell} \cdot \hat{\bf j}) + (\hat{\bf s} \cdot \hat{\bf j}) 
  - \frac{1}{(2 \ell - 1) (2 \ell + 3)} \Bigl[ 6 (\hat{\bf s} \cdot 
  \hat{\bf \ell}) (\hat{\bf \ell} \cdot \hat{\bf j}) - 2 \ell (\ell + 1) \; (\hat{\bf s} \cdot 
  \hat{\bf j}) \Bigr] - (\hat{\bf s} \cdot \hat{\bf j}) \nonumber \\
  &=& \ell (\ell + 1) + (\hat{\bf s} \cdot \hat{\bf \ell}) + \frac{1}{(2 \ell - 1) (2 \ell + 3)} \Bigl[
    \frac32 \ell (\ell + 1) - 4 \ell (\ell + 1) \; (\hat{\bf s} \cdot \hat{\bf \ell}) - 
    6 \; (\hat{\bf s} \cdot \hat{\bf \ell})^{2} \Bigr] \; . \; \label{Hhfs3}
\end{eqnarray} 
where $\hat{\bf j} = \hat{\bf \ell} + {\bf s}$ (see above). It is clear from the formula, Eq.(\ref{Hhfs2}), 
that this expression is a quadratic polynomial upon the scalar $(\hat{\bf s} \cdot \hat{\bf \ell})$ product.  
 
As mentioned above we want to analyze the two following cases: (a) the case when $j = \ell + \frac12$, and 
(b) the case when $j = \ell - \frac12$. In the first case one easily finds that $(\hat{\bf s} \cdot \hat{\bf 
\ell}) = \frac12 \ell$, while in the second case we have $(\hat{\bf s} \cdot \hat{\bf \ell}) = - \frac{\ell 
+ 1}{2}$. The equation, Eq.(\ref{Hhfs2}), takes the form  
\begin{eqnarray}
  \ell (\ell + 1) + \frac12 \ell + \frac{\ell}{2 (2 \ell - 1) (2 \ell + 3)} [ 3 - 4 \ell (\ell + 1)] = 
  \ell (\ell + 1) + \frac12 \ell - \frac12 \ell = \ell (\ell + 1) \nonumber 
\end{eqnarray} 
in the first case, and the from 
\begin{eqnarray}
 \ell (\ell + 1) - \frac{\ell + 1}{2} + \frac{\ell + 1}{2 (2 \ell - 1) (2 \ell + 3)} [ 4 \ell^{2} + 
 4 \ell - 3)] = \ell (\ell + 1) - \frac{\ell + 1}{2} + \frac{\ell + 1}{2} = \ell (\ell + 1) \nonumber 
\end{eqnarray} 
in the second case. This means that in the both cases we obtained the same result which equals $\ell (\ell 
+ 1)$. Therefore, the expectation value $V_{hfs}$, Eq.(\ref{Hhfs2}), can be written in the `united' (or 
identical) form  
\begin{eqnarray}
  V_{hfs} = \frac{2 \alpha^{2} (\mu_B)^{2}}{M_p \; I_N} \; g_{N}\; \frac{\ell (\ell + 1)}{j (j + 1)} \; 
  \langle \frac{a^{3}_{0}}{r^{3}_{eN}} \rangle \; ({\bf I}_{N} \cdot \hat{\bf j}) \; \; , \; \label{Hhfs4}
\end{eqnarray}
where the nuclear magnetic moment is represented in the form ${\bf m}_{N} = \Bigl(\frac{\mu_B}{M_p}\Bigr) \; 
\Bigl(\frac{{\bf I}_N}{I_N}\Bigr) \; g_{N}$ and $g_{N}$ is the nuclear $g-$factor, or gyromagnetic factor 
of the nucleus. In Eq.(\ref{Hhfs4}) the notation $a_0 = \frac{\hbar^{2}}{m_e e^{2}} \approx$ 5.291772109031 
$\cdot 10^{-9}$ $cm$ stands for the Bohr radius. The index $e$ denotes the electron, while the index $N$ 
means the nucleus and $\langle \frac{a^{3}_{0}}{r^{3}_{eN}} \rangle$ is the electron-nucleus expectation 
value which is determined from atomic calculations. Also in this equation the notation $\mu_B = 
\frac{e \hbar}{2 m_e}$ stands for the Bohr magneton. In atomic units, where $\hbar = 1, m_e = 1$ and $e = 
1$ ($- e$ is the electric charge of the electron), the numerical value of Bohr magneton equals $\frac{1}{2}$ 
(exactly). The notation $\alpha$ is the dimensionless fine-structure constant which equals $\alpha = 
\frac{e^{2}}{\hbar c} = 7.2973525693 \cdot 10^{-3} (\approx \frac{1}{137.04}$). Also, in Eq.(\ref{Hhfs4}) 
the notation $M_p$ stands for the proton mass, while ${\bf I}_N$ is the nuclear spin. The $I_N$ number is 
the angular momentum quantum number of the nucleus which is also called the nuclear spin for short. However, 
in contrast with the vector ${\bf I}_N$ the number $I_N$ is either integer and/or semi-integer. In reality, 
the $I_N$ value is defined as the maximal numerical value of $\mid {\bf I}_N \mid$. The formula, 
Eq.(\ref{Hhfs4}), coincides with the similar results obtained earlier in \cite{Sob} and \cite{FroF} (see
also \cite{LLQ}). Some effective applications of this method for solution of actual problems from atomic 
and nuclear physics can be found, e.g., in the following papers \cite{Fermi1} - \cite{Blin}. 

The formula, Eq.(\ref{Hhfs4}), can be applied to the rotationally excited, one-electron states in atoms and 
ions, i.e. to the states with $\ell \ge 1$. Analogous formula for the bound states with $\ell = 0$ in 
one-electron atomic systems takes a different form
\begin{eqnarray}
  V_{hfs} = \Bigl(\frac{8 \pi \alpha^{2}}{3}\Bigr) \; \frac{(\mu_B)^{2} g_{N} \; g_{e}}{M_p \; I_N} \; 
  a^{3}_{0} \; \langle \delta({\bf r}_{eN}) \rangle \; ({\bf I}_{N} \cdot \hat{\bf s}) = 
  \Bigl(\frac{2 \pi \alpha^{2}}{3}\Bigr) \; \frac{ g_{N} \; g_{e}}{M_p \; I_N} \; 
  \langle \delta({\bf r}_{eN}) \rangle \; ({\bf I}_{N} \cdot \hat{\bf s}) , \label{Hhfs45}
\end{eqnarray}
where $\langle \delta({\bf r}_{eN}) \rangle$ is the expectation value of the electron-positron delta-function, 
while $\hat{\bf s}$ is the vector-operator of electron spin and expression on the right side of the last 
equation is written in atomic units. This formulas explicitly contains the electron $g-$factor ($g \approx$ 
-2.00231930436092(36)) and this is a fundamental difference between the formulas Eq.(\ref{Hhfs4}) and 
Eq.(\ref{Hhfs45}). 

\section{Systems with additional relations between some vector-operators and angular momentum}

In atomic, molecular and nuclear physics there are a number of problems where one finds some additional 
relations between the vector-operator $\hat{\bf A}$ and vector-operator of angular momentum $\hat{\bf L}$, 
or some of their components. In general, such relations between operators take a wide variety of forms.
However, in Quantum Mechanics the most important cases are those where the Hamiltonian of the original 
problem explicitly depends upon the both vector-operators $\hat{\bf A}$ and $\hat{\bf L}$, or at least it 
contains some components of these two vectors. In other words, if the Hamiltonian of some quantum system 
is expresses as a scalar function of the two vector-operators $\hat{\bf A}$ and $\hat{\bf L}$, i.e., 
$\hat{H} = f(\hat{\bf A}, \hat{\bf L})$, then we have an additional relation between the vector-operator 
$\hat{\bf A}$ and vector-operator of angular momentum $\hat{\bf L}$. The scalar function $f$ can be 
arbitrary, but it depends upon the two vector-operators $\hat{\bf A}$ and $\hat{\bf L}$ only. It also 
contains some physical and numerical parameters. Formally, in these case we have a few additional 
constraints between the vector-operator $\hat{\bf A}$ and vector-operator of angular momentum 
$\hat{\bf L}$. Such additional relations allow one to investigate similar systems to much deeper levels, 
while in some cases we can even obtain closed analytical expressions for their energy spectra and predict 
other properties. At first glance, it may seem that for such systems all possible bound state problems 
have already been solved.

In reality, among similar systems one finds a large number of unsolved, very interesting problems, which have 
never considered in earlier studies due to extremely difficult processes of their analysis and solutions. In 
addition to this, quite a few of such problems are either unknown to modern physics community, or they are 
are currently considered as `completely' solved. The general situation in similar systems can be understood  
from the consideration of the one-electron hydrogen-like atom/ion with the nuclear charge $Q e$. The actual 
Coulomb field is central. This means that the angular momentum of electron $\hat{\bf L}$ is conserved.    
However, as was well known to Laplace already in 1799 \cite{Lapl} (probably, he knew it even before 1791) 
during a single particle motion in the field of a central attractive potential $W(r) = - \frac{B}{r}$ there 
is one additional vector-integral of motion ${\bf A}$ which takes the form
\begin{eqnarray}
 {\bf A} = \sqrt{m} \; \Bigl( \frac{B \; {\bf r}}{r} - \frac{1}{m} [{\bf p}, {\bf L}] \Bigr) \; \; , 
 \; \; \label{Lapl} 
\end{eqnarray}
The vector form of this integral of motion means that we have three additional integrals of motion. As is 
well known the trajectory of any particle, which moves in the central field created by the potential $W(r) 
= \frac{B}{r}$, is a closed ellipse which has a constant orientation in space. The conserving vector ${\bf 
A}$ uniformly determines spatial orientation of this ellipse in outer space. In fact, the conserving vector 
${\bf A}$ is always directed along the major axis of the resulting ellipse, which is often called the 
Coulomb, or gravitational ellipse, since Laplace in \cite{Lapl} considered a weak gravitational (Newtonian) 
potential. In this study we investigate quantum two-body hydrogen-like systems with the Coulomb attractive 
potential which is written in the form $W_{C}(r) = - \frac{Q e^{2}}{\hbar r}$. 
     
In quantum mechanics the vector-integral ${\bf A}$ must be replaced by the following vector-operator 
\begin{eqnarray}
 \hat{\bf A} = \sqrt{m} \; \Bigl( \frac{\beta \; {\bf r}}{r} - \frac{1}{2 m} \; ( [\hat{\bf p}, \hat{\bf L}] - 
 [\hat{\bf L}, \hat{\bf p}] )\Bigr) = \frac{\sqrt{m} Q e^{2}}{\hbar} \; \Bigl( \frac{{\bf r}}{r} - 
 \frac{\hbar}{2 m \sqrt{m} Q e^{2}} \; ( [\hat{\bf p}, \hat{\bf L}] - [\hat{\bf L}, \hat{\bf p}] )\Bigr) 
 \; \; , \; \; \label{LRL} 
\end{eqnarray}
where $\beta = \frac{Q e^{2}}{\hbar}$ and $m$ is the mass of light, negatively charged particle (its electric 
charge is $- e$) which moves in the central field of the infinitely heavy `central' particle which has the 
electric charge $Q e$. In atomic units, where $\hbar = 1, m = m_{e} = 1, e = 1$ and also $4 \pi \epsilon_{0} 
= 1$, the last equation takes the form 
\begin{eqnarray}
 \hat{\bf A} = \frac{Q \; {\bf r}}{r} + \frac12 \; ( [ \hat{\bf p}, \hat{\bf L}] - [ \hat{\bf L}, \hat{\bf p}] 
 )\Bigr) \; \; . \; \; \label{RgLeLau} 
\end{eqnarray}
This vector-operator is called the Runge-Lenz vector \cite{Runge}, \cite{Lenz}. In the lowest-order approximation 
upon the fine-structure constant $\alpha$ the Runge-Lenz vector-operator is strictly conserved. Furthermore, there  
is the following constraint between the Hamiltonian $\hat{H}$ of the system and two vector-operators ${\bf L}$ and 
${\bf A}$
\begin{eqnarray}
  {\bf L}^{2} + \Bigl(- \frac{1}{Q^{2} \; Ry}\Bigr) \; \hat{H} \; {\bf A}^{2} = - 1 + Q^{2} \; Ry \; 
  {\hat{H}}^{-1} \; \; \; , \; \; \label{LplusA} 
\end{eqnarray}
where $Ry = \frac{e^{4} m_{e}}{2 \hbar^{2}}$ is the so-called Rydberg constant (see below). In spectral form the 
same equation takes the form 
\begin{eqnarray}
  {\bf L}^{2} + \frac{(- 2 E)}{Q^{2} \; (2 Ry)} \; {\bf A}^{2} = - 1 - \frac{Q^{2} \; (2 Ry)}{(- 2 E)} \; \; \; , 
  \; \; \label{LplusA1} 
\end{eqnarray}
where $2 Ry = \frac{e^{4} m_{e}}{\hbar^{2}}$ is the doubled Rydberg constant. Another constraint between the 
vector-operators ${\bf L}$ and ${\bf A}$ is `almost obvious': ${\bf L} \cdot {\bf A} = 0 = {\bf A} \cdot {\bf L}$.  
These facts and constraints were used by W. Pauli \cite{Pauli} to derive his famous, pure algebraic approach and
obtain the well known Bohr's formula for the energy spectrum of an arbitrary hydrogen-like two-body system. Note 
that the quantum mechanics in its final (or Schr\"{o}dinger) form has not been created yet at the time when the 
article \cite{Pauli} was completed by 25 year-old Pauli. For the Runge-Lenz vector-operator we note that there is 
an interesting development in two recent papers \cite{Efim1} and \cite{Efim2}, where their author based on the 
well-known work of Fock \cite{Fock} and using the explicit form of the Runge-Lenz vector-operator in the momentum 
representation, discovered a number of new properties of this important operator. 

It was shown in \cite{Pauli} that the six operators $\hat{L}_{x}, \hat{L}_{y}, \hat{L}_{z}$ and $\hat{A}_{x}, 
\hat{A}_{y}, \hat{A}_{z}$ obey the following commutation relations 
\begin{eqnarray}
 [ \hat{L}_{i}, \hat{L}_{j}] = \imath \; \varepsilon_{ijk} \; \hat{L}_{k} \; \; , \; \; [ \hat{L}_{i}, 
 \hat{A}_{j}] = \imath \; \varepsilon_{ijk} \; \hat{A}_{k} \; \; , \; \; [ \hat{A}_{i}, \hat{A}_{j}] = 
 - 2 \imath \hat{H} \; \varepsilon_{ijk} \; \hat{A}_{k} \; \; , \; \; \label{Commute1}
\end{eqnarray}
where the notation $\hat{H}$ means the Hamiltonian of the two-body atom/ion which equals (in atomic units) 
$\hat{H} = - \frac{1}{2} \Delta - \frac{Q}{r}$, where $\Delta$ is the Laplace operator. Below we restrict 
ourselves to the consideration of the bound states only. Such bound states have a fixed negative energy $E$, 
where $- E > 0$. For such states we can consider the Hamiltonian $\hat{H} = E$ as a constant negative number. 
This fact allows us to introduce the three new operators $\hat{N}_{x}, \hat{N}_{y}, \hat{N}_{z}$, where 
$\hat{N}_{i} = (- 2 \hat{H})^{\frac12} \hat{A}_{i} = (- 2 E)^{\frac12} \hat{A}_{i}$ and $i = x, y, z$ (or $i$ 
= 1, 2, 3). These three operators $\hat{N}_{x}, \hat{N}_{y}, \hat{N}_{z}$ will be used here instead of the 
three components of the Runge-Lenz vector-operator $\hat{A}_{x}, \hat{A}_{y}, \hat{A}_{z}$. At this point we 
can define the two following three-dimensional vector-operators $\hat{\bf L} = ( \hat{L}_{x}, \hat{L}_{y}, 
\hat{L}_{z})$ and $\hat{\bf N} = ( \hat{N}_{x}, \hat{N}_{y}, \hat{N}_{z})$. The commutators between the 
components of these vector-operators are 
\begin{eqnarray}
 [ \hat{L}_{i}, \hat{L}_{j}] = \imath \; \varepsilon_{ijk} \; \hat{L}_{k} \; \; , \; \; [ \hat{L}_{i}, 
 \hat{N}_{j}] = \imath \; \varepsilon_{ijk} \; \hat{N}_{k} \; \; , \; \; [ \hat{N}_{i}, \hat{N}_{j}] = 
 \imath\; \varepsilon_{ijk} \; \hat{L}_{k} \; \; . \; \; \label{Commute2}
\end{eqnarray} 
The advantage of the new operators $\hat{\bf N}$ is obvious, since the Poisson brackets, Eq.(\ref{Commute2}), 
do not contain any physical constant from the original problem.  

From the definition of the $\hat{\bf N}$ operators, Eqs.(\ref{LplusA}) - (\ref{LplusA1}) and commutation 
relations, Eq.(\ref{Commute2}), between the vector-operators $\hat{\bf L}$ and $\hat{\bf N}$ one finds that 
for the $\hat{\bf L}$ and $\hat{\bf N}$ vector-operators the two following equations are obeyed:
\begin{eqnarray}
  \hat{\bf L} \cdot \hat{\bf N} = \hat{\bf N} \cdot \hat{\bf L} = 0 \; \; \; {\rm and} \; \; \; 
  \hat{\bf L}^{2} + \hat{\bf N}^{2} = - 1 + \frac{E_{0}}{2 \; H} = - 1 - \frac{E_{0}}{2 \; E} \; \; , 
  \; \; \label{Commute3}
\end{eqnarray} 
where $E_{0} = \frac{Q^{2}}{2} \; \frac{e^{4} m_{e}}{\hbar^{2}} = \frac{Q^{2}}{2} \; R_y$ and the factor 
$R_y = \frac{e^{4} m_{e}}{\hbar^{2}}$ defines the atomic unit of energy and exactly coincides with the
doubled Rydberg constant. As follows from these definitions in atomic units, where $e = 1, m_e = 1$ and
$\hbar = 1$, we have $R_y = 1$ and $E_{0} = \frac{Q^{2}}{2}$ and $Q e = Q$ is the electric charge of the 
central atomic nucleus. From now on, all equations in this study are written in atomic units only. 

At the final step of the Pauli procedure we introduce the two following vector-operators $\hat{\bf J}^{(1)} 
= \frac12 (\hat{\bf L} + \hat{\bf N})$ and $\hat{\bf J}^{(2)} = \frac12 (\hat{\bf L} - \hat{\bf N})$. For 
Cartesian components of these two vectors the following commutation relations are obeyed 
\begin{eqnarray}
  [ \hat{J}^{(1)}_{i}, \hat{J}^{(1)}_{j} ] = \imath \; \varepsilon_{ijk} \; \hat{J}^{(1)}_{k} \; \; \; , \; \; 
  [ \hat{J}^{(2)}_{i}, \hat{J}^{(2)}_{j} ] = \imath \; \varepsilon_{ijk} \; \hat{J}^{(2)}_{k} \; \; \; , \; \;   
  [ \hat{J}^{(1)}_{i}, \hat{J}^{(2)}_{j} ] = 0 \; \; \; . \; \; \label{Commute4}
\end{eqnarray} 
Note that the commutation relations from the first and second group of these equations exactly coincide with the 
well known commutation relations of the rotation group in our three-dimensional space \cite{GelfMS} which is 
designated as $SO(3)$-group. Thus, as follows from Eq.(\ref{Commute4}) we have six generators of the two $SO(3)$ 
groups (three generators for each $SO(3)$-group) which are independent of each other. Their independence directly 
follows the third group of equations in Eq.(\ref{Commute4}). Therefore, the actual symmetry group of the 
one-center Coulomb two-body atomic system is $SO(3) \otimes SO(3)$. The generators of the first $SO(3)$ group are 
$\hat{\bf J}^{(1)}_{i}$, where $i$ = 1, 2, 3, while analogous generators of the second $SO(3)$ group are 
$\hat{\bf J}^{(2)}_{j}$, where $j$ = 1, 2, 3. As is well known from the general group theory the direct product 
of the two $SO(3)$ groups coincides with the $SO(4)$ group, i.e., $SO(3) \otimes SO(3) = SO(4)$, where $SO(4)$ 
group is the compact group of rotations in the four-dimensional Euclidean space. In applications to the hydrogen 
atom and hydrogen-like ions, any actual representations of the compact $SO(4)$ group is defined by a pair of 
integer (or semi-integer) numbers $(j_1, j_2)$. This fact is often designated by the $D(j_1, j_2)$ notation. All 
representations of the $SO(4)$ group are finite-dimensional.  
 
Now, by taking into account the relations mentioned in Eq.(\ref{Commute3}) one finds 
\begin{eqnarray}
  \Bigl( \hat{\bf J}^{(1)} \Bigr)^{2} = C^{(1)}_{2} = \frac{{\bf L}^{2} + {\bf N}^{2}}{4} + \frac14 = 
  - \frac{Q^{2}}{8 \; H} = - \frac{Q^{2}}{8 \; E} = \Bigl( \hat{\bf J}^{(2)} \Bigr)^{2} 
  = C^{(2)}_{2} \; \; \; , \; \; \label{Commute5a}
\end{eqnarray} 
where $C^{(1)}_{2}$ and $C^{(2)}_{2}$ are the two Casimir operators $C_2$ of the two independent compact 
$SO(3)-$groups which are the scalars and each of them equals $j_{k} ( j_{k} + 1)$, where $k$ = 1, 2, while 
$j_{k}$ are semi-integer numbers. In general, for any closed operator algebra the Casimir operator is an 
invariant operator of the second order, i.e., it commutes with all operators from this algebra. In respect 
to Schur's lemma, for any irreducible representation of this algebra such an operator is a multiple of the 
identity operator $\hat{I}$ (or $\hat{E}$). From here and Eq.(\ref{Commute5a}) we obtain the two following 
equations 
\begin{eqnarray}
  \Bigl( j_{1} + \frac12 \Bigr)^{2} = - \frac{Q^{2}}{8 \; H} = - \frac{Q^{2}}{8 \; E} \; \; \; , 
  \; \; \Bigl(j_{2} + \frac12 \Bigr)^{2} = - \frac{Q^{2}}{8 \; H} = - \frac{Q^{2}}{8 \; E} \; \; \; 
  . \; \; \label{Commute5b}
\end{eqnarray} 
These equations are correct if (and only if) $j_{1} = j_{2} = j$. Then after a few simple transformations one 
finds in atomic units:
\begin{eqnarray}
  E = H = - \frac{E_{0}}{(2 j + 1)^{2}} = - \frac{Q^{2}}{2 (2 j + 1)^{2}} \; \; \; . \; \; \label{Commute5c}
\end{eqnarray} 
In this equation from the definition \cite{GelfMS} we have to assume that $j = 0, \frac12, 1, \frac32, 2, 
\ldots$. Now, by introducing the integer numbers $n = 2 j + 1$, where $n = 1, 2, 3, \ldots$, we obtain the 
well known formula for the energy spectrum of bound states in one-electron atoms and/or ions: $E_{n} = - 
\frac{Q^{2}}{2 n^{2}}$, where $n$ is called the principal quantum number. The multiplicity of each bound 
state equals $(2 j_1 + 1) (2 j_2 + 1) = (2 j + 1)^{2} = n^{2}$. Since $j_1 = j_2 = \frac{n - 1}{2}$, then 
the representations of the compact $SO(4)$ group, which are of interest in applications to hydrogen-like 
atoms and ions, are $D\Bigl( \frac{n - 1}{2}, \frac{n - 1}{2} \Bigr)$. Other representations of this group 
are not considered in our study. Note that the formula, Eq.(\ref{Commute5c}), correctly describes the energy 
spectrum of bound states in an arbitrary two-body Coulomb system. Similar systems are often called the 
hydrogenic (or one-electron) atoms and ions. 

\subsection{One-electron Coulomb problem in other systems of orthogonal coordinates}

The method applied by Pauli in \cite{Pauli} is essentially based on the use of spherical coordinates in the 
non-relativistic Sch\"{o}dinger equation $H \psi = E \psi$ for one-electron hydrogen atom. In spherical 
coordinates three spatial variables of this problem are truly separated. Later, it was found that similar 
separation of spatial variables is also possible in some other systems of coordinates, e.g., in parabolic, 
spheroidal and spheroconical systems of coordinates. In general, it is very interesting to repeat the 
approach developed by Pauli in these coordinate systems. Let us briefly describe the generalized version of 
the Pauli method which works in different systems of orthogonal coordinates. First of all, we note that in 
spherical coordinates the bound state wave function $\psi$ of hydrogenic atom/ion with the nuclear charge 
$Q e = Q$ must obey the four following equations  
\begin{eqnarray}
 & &\Bigl(\hat{\bf L}^{2} + \hat{\bf N}^{2}\Bigr) \; \psi = - f\Bigl(\frac{H}{Q^{2}}\Bigr) \; \psi = 
  - f\Bigl(\frac{E}{Q^{2}}\Bigr) \; \psi \; \; \; (A) \; \; \; , \; \; \; (\hat{\bf L} \cdot 
  \hat{\bf N}) \; \psi = 0 \; \; \; \; (B) \; \; \; , \; \nonumber \\ 
 & &\; \; \hat{\bf L}^{2} \; \psi = \ell (\ell + 1) \; \psi \; \; \; \; (C) \; \; , \; \; \; \; 
 \hat{L}^{2}_{z} \; \psi = m^{2} \; \psi \; \; \; \; (D) \; \; , \; \; \label{fourequats} 
\end{eqnarray}
where $f(x) = 1 + \frac{1}{2 x}$ is the scalar function, while $H$ is the Hamiltonian operator. The equation 
$(A)$ plays the role of the Sch\"{o}dinger equation. In general, this equation is uniformly reduced to the 
actual  Sch\"{o}dinger equation and can be considered as its real equivalent. The equation $(B)$ is the 
operator identity. The third equation $(C)$ expresses the fact that operator of separation of angular 
variables $\hat{\bf L}^{2}$ (in spherical coordinates) must also be diagonal in hydrogenic wave functions. 
It is clear that the operators $H, \hat{\bf L}^{2}$ and $\hat{L}^{2}_{z}$ are (or can be considered) as the 
partial Hamiltonians of the two- and one-dimensional motions, respectively. The last two operators 
($\hat{\bf L}^{2}$ and $\hat{L}^{2}_{z}$) linearly included in the total Hamiltonian of the three-body 
problem. In reality this means that we represent the unknown motion in the original three-dimensional space 
as a linear combination of some well known motions in the one-dimensional radial space (variable $r$) and 
two-dimensional space in spherical coordinates $\theta, \phi$. The bound state wave functions of hydrogen 
atom and hydrogen-like ions in spherical coordinates are often designates as the three-component $\mid n, 
\ell, m \rangle$ vector.  

Now, let us consider the parabolic system of coordinates $\xi, \eta, \phi$, which are defined by the 
following equations (see, e.g., Chapter 21 in \cite{AS} and \$ 48 in \cite{LLM})  
\begin{eqnarray}
  \xi = r \; + \; z \; , \eta = r \; - \; z \; \; , \; \; \phi = \arctan\Bigl(\frac{x}{y}\Bigr) \; 
  \; , \; \; {\rm where} \; \; r = \sqrt{x^{2} + y^{2} + z^{2}} \; \; \label{parabol}
\end{eqnarray}
and vise versa
\begin{eqnarray}
  x = \sqrt{\xi\; \eta} \; \cos\phi \; \; , \; \; y = \sqrt{\xi \; \eta} \; \sin\phi \; \; , \; \; 
  z = \frac12 (\xi - \eta) \; \; . \; \; \label{parabol1}
\end{eqnarray}
These coordinates are used in various problems known for the hydrogen-like (or one-electron) atoms and 
ions, e.g., to analyze the Stark effect and total ionization of one-electron atomic systems in a strong 
electric field. In parabolic coordinates the bound state wave function $\psi$ of any hydrogenic atom/ion 
is determined from the same four equations, Eq.(\ref{fourequats}), where equations $(A), (B)$ and $(D)$ 
have exactly the same form as above, but they must explicitly be re-written in the parabolic coordinates. 
However, equation (C) takes a different form: $\hat{A}_{z} \; \psi = a \; \psi$, where the operator 
$\hat{A}_{z}$ is $z-$component of the Runge-Lenz vector-operator (see above) and $a$ is its eigenvalue. 
It is clear that the operator $\hat{N}_{z}$ can be used here instead of the traditional $\hat{A}_{z}$ 
operator. 

The operator $\hat{A}_{z}$ (or $\hat{N}_{z}$) is responsible in this case for the complete separation 
of parabolic variables. Note that all these equations, including the Sch\"{o}dinger equation $(A)$, 
are explicitly symmetric in respect to the $\xi \Leftrightarrow \eta$ substitution. Another interesting 
moment which we want to note here is the fact that the eigenfunctions of the hydrogen atoms and 
hydrogen-like ions, which appear in the parabolic system of coordinates (when we separate variables), 
are the wave functions of the following operators $\hat{H}, (\hat{\bf J}^{(1)})_{z}$ and 
$(\hat{\bf J}^{(2)})_{z}$, where the vector-operators $\hat{\bf J}^{(1)}$ and $\hat{\bf J}^{(2)}$ are 
defined above. The bound state wave functions of hydrogen atom and hydrogen-like ions in spherical 
coordinates are often designates as the three-component $\mid k, k_1, k_2 \rangle$ vector, where $k = 
\frac{n - 1}{2}, n = 2 k + 1, k_1 = \frac12( m + n_1 - n_2 )$ and $k_2 = \frac12( m + n_2 - n_1 )$. 
Here $n_1$ and $n_2$ are the two parabolic quantum numbers which arise, if the problem of one-electron 
atom/ion is considered in parabolic coordinates. For these $n_1$ and $n_2$ numbers the two following 
equations hold: $n_1 + n_2 = n - \mid m \mid - 1$ and $n_1 - n_2 = k_1 - k_2$. Note also that the both 
$k_1$ and $k_2$ numbers in the $\mid k, k_1, k_2 \rangle$ vectors always vary between - $k$ and $k$ 
values, or between the $- \frac{n-1}{2}$ and $\frac{n-1}{2}$ values. 

It is interesting and crucially important for our purposes in this study to derive the explicit 
relations between the spherical $\mid n, \ell, m \rangle$ vectors in terms of the parabolic $\mid k, 
k_1, k_2 \rangle$ vectors and vice versa. Such relations can be found, if we consider the equation 
$\hat{\bf L} = \hat{\bf J}^{(1)} + \hat{\bf J}^{(2)}$ mentioned above as the vector-coupling relation 
between three vector-operators of momenta $\hat{\bf L}, \hat{\bf J}^{(1)}$ and $\hat{\bf J}^{(2)}$.  
Also as follows from this equation we obtain that $\ell_{min} = 0$ (since $n_{min} = 1$) and 
$\ell_{max} = n - 1$ (since $j_1 = j_2 = \frac{n - 1}{2}$). Now, based on the vector-coupling relation 
one finds the two required expressions 
\cite{Park}: 
\begin{eqnarray}
  \mid n, \ell, m \rangle = \sum_{k_1 + k_2 = m} \langle k, k_1, k_2 \mid n, \ell, m \rangle 
  \mid k, k_1, k_2 \rangle = \sum_{k_1 + k_2 = m} {\cal C}^{\ell m}_{k k_1; k k_2} \mid k, k_1, k_2 
  \rangle \; \; , \; \; \label{CGC1}
\end{eqnarray}
where $k = \frac{n-1}{1}$, and 
\begin{eqnarray}
  \mid k, k_1, k_2 \rangle = \sum^{2 k}_{\ell = \mid m \mid} \langle n, \ell, m \mid k, k_1, k_2 
  \rangle \mid  n, \ell, m \rangle = \sum^{2 k}_{\ell = \mid m \mid} {\cal C}^{\ell m}_{k k_1; k k_2} 
  \mid n, \ell, m \rangle \; \; \; , \; \label{CGC2}
\end{eqnarray}
where $n = 2 k + 1$, while the numerical coefficients in these formulas are the Clebsh-Gordan 
coefficients (see, e.g., \cite{Edm}). Here we assume that all bound state wave functions $\mid n, \ell, 
m \rangle$ and $\mid k, k_1, k_2 \rangle$ have unit norm. The formulas, Eqs.(\ref{CGC1}) - (\ref{CGC2}), 
allow one to derive the explicit expression for the matrix elements of the $\hat{A}_{z}$ vector-operator 
in the spherical $\mid n, \ell, m \rangle$ basis, or in the $(n, \ell, m)$-representation: 
\begin{eqnarray}
 & & \langle n, \ell_{1}, m \mid \hat{A}_{z} \mid n, \ell, m \rangle = \sum_{k_1 + k_2 = m} 
 \langle n, \ell_{1}, m \mid k, k_1, k_2 \rangle \langle k, k_1, k_2 \mid \hat{A}_{z} 
 \mid k, k_1, k_2 \rangle \times \nonumber \\
 & & \langle k, k_1, k_2 \mid n, \ell, m \rangle =  
 \Bigl[ \sum_{k_1 + k_2 = m} {\cal C}^{\ell_{1} m}_{k k_1; k k_2} \Bigl( \frac{k_1 - k_2}{2 k + 1} 
 \Bigr) {\cal C}^{\ell m}_{k k_1; k k_2} \Bigr]\nonumber \\
 & & = \Bigl\{ \frac{1}{n} \; \Bigl[ \sum_{k_1 + k_2 = m} {\cal C}^{\ell_{1} m}_{k k_1; k k_2} \; 
 (k_1 - k_2) \; {\cal C}^{\ell m}_{k k_1; k k_2} \Bigr] \Bigr\} \; \; \; . \; \label{CGCT0}  
\end{eqnarray}
After a few additional transformations we obtain the final formula for the matrix elements of the 
$\hat{A}_{z}$ operator in the basis of spherical $\mid n, \ell, m \rangle$ states:  
\begin{eqnarray}
 & &\langle n, \ell_{1}, m \mid \hat{A}_{z} \mid n, \ell, m \rangle = \delta_{\ell_{1}, \ell + 1} 
 \; \; \Biggl\{ \frac{[n^{2} - (\ell + 1)^{2}] [(\ell + 1)^{2} - m^{2}]}{ n \; 
 [2 (\ell + 1)^{2} - 1]} \Biggr\} \nonumber \\ 
&+& \delta_{\ell_{1}, \ell - 1} \; \; \Biggl\{ \frac{[n^{2} - (\ell - 1)^{2}] [(\ell - 1)^{2} 
- m^{2}]}{n \; [2 (\ell - 1)^{2} - 1]} \Biggr\} \; \; \; . \; \label{CGCT1} 
\end{eqnarray} 
As we predicted above all diagonal matrix elements ($\sim \delta_{\ell, \ell}$) in this matrix 
equal zero. The formula, Eq.(\ref{CGCT1}), will be used below. 

The last case which we consider here includes the ellipsoidal (or spheroidal) coordinates. These 
coordinates are often used to solve various two-center problems for one-electron systems. 
Definition of the ellipsoidal (or spheroidal) coordinates is significantly more complicated. For 
instance, the azimuth angle $\phi$ is defined differently in the prolate and oblate ellipsoidal 
coordinates (see, e.g., \cite{AS}). To introduce the ellipsoidal coordinates we have to assume 
that a very heavy atomic nucleus with the positive electric charge $Q e$ is always located in one 
of the two focuses of the ellipse (or hyperbola). In fact, these two focuses correspond to two 
families of confocal ellipses and hyperbolas, and the distance between the two focuses is denoted 
below as $2 R$. For the two-body atoms and ions, which contain only one electron, the second focus 
is empty, or in other words, it contains a model (or fictional) atomic nucleus with zero electric 
charge. Formal definition of the ellipsoidal coordinates $\xi$ and $\eta$ is: 
 \begin{eqnarray}
  \xi = \frac{r_{1} + r_{2}}{2 R} \; \; , \; \; \eta = \frac{r_{1} - r_{2}}{2 R} \; \; , \; \; 
  \phi = \arctan\Bigl(\frac{y}{x}\Bigr) \; \; , \; \; \label{spheroid1}
\end{eqnarray}
where $r_{1}$ and $r_{2}$ are the scalar distances from the electron to the two focuses, while the exact 
definition of the $\phi$ coordinate corresponds to the system of prolate ellipsoidal coordinates. For the 
system of oblate ellipsoidal coordinates $\phi = \arctan\Bigl(\frac{x}{y}\Bigr)$. Below in this study we 
apply the system of prolate ellipsoidal coordinates only. The inverse relations between ellipsoidal and 
Cartesian (electron) coordinates take the form      
\begin{eqnarray}
  x = R \; \sqrt{(\xi^{2} - 1) \; (1 - \eta^{2})} \; \cos\phi \; \; , \; \; y = R \; \sqrt{(\xi^{2} 
  - 1) \; (1 - \eta^{2})} \; \sin\phi \; \; , \; \; z = R \; \xi \; \eta \; \; , \; \; \label{spheroid2}
\end{eqnarray}
where $1 \le \xi < + \infty, -1 \le \eta \le 1$ and $0 \le \phi \le 2 \pi$. In ellipsoidal coordinates 
the bound state wave function $\psi$ of hydrogenic atom/ion is determined from the same four equations, 
Eq.(\ref{fourequats}). Again, the three of these equations (equations $(A), (B)$ and $(D)$) have exactly 
the same form as above (they must explicitly be re-written in the ellipsoidal coordinates). However, the 
equation (C) takes a different form $\hat{\Lambda}(R) \psi = (\hat{\bf L}^{2} + \sqrt{- 2 E} \; R \; 
\hat{A}_{z}) \; \psi = (\hat{\bf L}^{2} + R \; \hat{N}_{z}) \; \psi = \lambda(R) \; \psi$, where the 
operator $\hat{A}_{z}$ is $z-$component of the Runge-Lenz vector-operator, the operator $\hat{N}_{z} = 
\frac{1}{\sqrt{- 2 E}} A_z$ is defined above and $\lambda(R)$ is the eigenvalue of the $\hat{\Lambda}(R)$ 
operator. The operator $\hat{\Lambda}(R)$ is responsible for separation of the ellipsoidal variables $\xi$ 
and $\eta$. This operator depends upon the distance $R$ between two focuses and its explicit form has been 
found in \cite{Eric} and \cite{Couls}. For $R \Rightarrow 0$ one easily finds that $\hat{\Lambda}(R) 
\Rightarrow \hat{\bf L}^{2}$, while for $R \Rightarrow \infty$ we have $\hat{\Lambda}(R) \Rightarrow 
\sqrt{- 2 E} \; R \; \hat{A}_{z}$, or $\hat{\Lambda}(R) \Rightarrow R \; \hat{N}_{z}$. These two limiting 
cases correspond a separation of variables in the spherical and parabolic coordinates, respectively (see 
above). Note also that for ellipsoidal coordinates it is useless to discuss the permutation $\xi 
\Leftrightarrow \eta$ symmetry, since these two variables are defined on different intervals.     

\subsection{One-electron Coulomb two-center problem}

An ultimate and obvious success achieved by Pauli in \cite{Pauli} for the two-body Coulomb systems 
(or hydrogen-like systems) stimulated many other scientists at that time to generalize his method 
and study the bound state spectra in other one-, few- and many-electron atomic and molecular 
systems. The aim of such studies was to produce some closed analytical expressions for the bound 
state energies in similar systems. Such an expression may include all conserving quantum numbers 
(other than energy) which are needed to label this bound state. In some sense this problem has 
finally been solved in the so-called multi-dimensional hyperspherical coordinates \cite{Fro2018}, 
\cite{Fro2018A}. In these papers by using the methods of matrix mechanics we have shown that for 
any $N_e$-electron atomic system its total energy spectrum is represented by the Bohr-like formula 
(see, Eq.(55) from \cite{Fro2018} and Eqs.(55) - (58) from  \cite{Fro2018A}) which explicitly 
contains the excitation quantum excitation number $n_{r}$ which is an integer non-negative number. 
Other quantum numbers are the total and partial hyperangular momenta \cite{Knirk}, which are not 
conserved in few-electron (atomic) systems, where $N_e \ge 1$. 

Another class of promising quantum systems consists of one-electron Coulomb (three-body) ions which 
contain two very heavy, positively charged nuclei (or centers) and one negatively charged electron 
which is bound to these nuclei. Since 1928 it was known that the wave functions of such ions are 
represented in the form of product of two hydrogenic wave functions. This fact was clear indication  
that analytical solutions of the original problem does exist and it is quite simple. In fact, the 
one-electron (Coulomb) problem with two heavy centers can also be solved exactly (or analytically) 
by using some advanced, pure algebraic methods developed for the rotation groups. However, in order 
to obtain and describe all bound states in similar three-body ions one has to deal with hydrogen 
atoms in the three-dimensional pseudo-Euclidean space. In actual applications this pseudo-Euclidean 
space is replaced by a complex space where the two Cartesian coordinates $x$ and $y$ become pure 
imaginary. In addition to this, operations with the non-compact $SO(2,2)-$ and $SO(3,1)-$groups and 
their infinite-dimensional representations becomes mandatory in such an analysis. 

The original Schr\"{o}dinger equation for the one-electron (Coulomb) two-center problem $(Q_{1}, 
Q_{2})$ in Cartesian coordinates takes the form 
\begin{eqnarray}
  & &\hat{H} \; \psi(x, y, z) = \Bigl[ -\frac12 \Delta - \frac{{Q}_1}{r_{1}} - \frac{{Q}_2}{r_{2}} 
  \Bigr] \psi(x, y, z) =  \Bigl[ -\frac12 \Bigl( \frac{\partial^{2}}{\partial x^{2}} + 
  \frac{\partial^{2}}{\partial y^{2}} + \frac{\partial^{2}}{\partial z^{2}} \Bigr) \nonumber \\
 & &- \frac{{Q}_1}{\sqrt{x^{2} + y^{2} + (z - R)^{2}}} - \frac{{Q}_2}{\sqrt{x^{2} + y^{2} + 
  (z + R)^{2}}} \Bigr] \; \psi(x, y, z) = E \; \psi(x, y, z) \; \; , \; \label{SchrodCart}
\end{eqnarray}
where we have assumed that the two atomic (bare) nuclei with electrical charges $Q_1 e^{2} = Q_{1}$ 
and $Q_2 e^{2} = Q_{2}$ are located at the spatial $(0, 0, -R)$ and $(0, 0, R)$ points, respectively. 
Note that there is a certain arbitrariness in the positions of these two heavy, Coulomb centers. 
Indeed, we can chose spatial locations of the two electrical charges $Q_1$ and $Q_{2}$ differently, 
e.g., these nuclei are located at the spatial $(0, 0, R)$ and $(0, 0, -R)$ points, respectively. It 
is clear that no actual (or observable) physical property of the original Coulomb two-center system 
will be changed, if we replace one nucleus by another. In other words, if we chose spatial positions 
of these two Coulomb centers differently, i.e., if we interchange the nuclear charges $Q_{1} 
\leftrightarrow Q_{2}$ in the original Schr\"{o}dinger equation, then nothing will change for the 
Coulomb two-center problem. Invariance of the original Schr\"{o}dinger equation in respect to this 
action means some new symmetry, which is typical for all similar systems which contain two Coulombs 
centers. This new symmetry is the reversal of the molecular axis, i.e., the $z-$axis on which the 
both nuclei are located. 

Conservation of the `observable' properties in the Coulomb two-center problem during reversal of 
the molecular axis is a well known fact (see, e.g., \cite{LLQ}) which is extensively used in 
molecular spectroscopy since its early days. This `principle' can be formulated in the form: if 
some vector-operator $\hat{\bf B}$ is truly conserved in one-electron atom (one-center system), 
then in the one-electron (Coulomb) two-center problem $(Q_{1}, Q_{2})$ the absolute eigenvalues 
of the scalar $({\bf n} \cdot \hat{\bf B})$ operator are also conserved. In actual applications 
it is better to discuss the scalar $({\bf n} \cdot \hat{\bf B})^{2}$ which is truly conserved. 
As is well known in any one-electron hydrogen-like ion (or hydrogen atom) the vector-operators 
of orbital angular momentum $\hat{\bf L}$ and Runge-Lenz `vector' $\hat{\bf A}$ are conserved. 
For one-electron (Coulomb) two-center problem $(Q_{1}, Q_{2})$ our principle predicts conservation 
of the $({\bf n} \cdot \hat{\bf L})^{2}$ and $({\bf n} \cdot \hat{\bf A})^{2}$ operators. The two
corresponding quantum numbers e $\; m^{2} \;$ and e $\; a^{2} \;$ are truly conserved in this 
problem. In other words, the bound state wave function(s) and any actual expectation value 
determined for one-electron two-center ions can depend upon the $m^{2}$ and $a^{2}$ values only. 
These simple fact is important to understand all steps of our analysis below. 

\subsection{Spheroidal coordinates in the one-electron (Coulomb) two-center problem}

Now, let us consider the one-electron quasi-molecular ion (or system), which contains two infinitely 
heavy atomic nuclei. Below, such systems are called the two-center Coulomb ions (or systems). The 
electric charges of these nuclei $Q_{1}$ and $Q_{2}$. Below we shall assume that $Q_{1} > Q_{2}$, 
where $Q_{1}$ and $Q_{2}$ are positive and coprime integers. The condition $Q_{1} > Q_{2}$ is 
introduced here only for the sake of simplicity. Analysis of the case when $Q_{1} = Q_{2}$ is very 
similar, but this requires a large number of special reservations, which only complicate our 
description of the general method. For one-electron problem we can apply formulas which are known 
from the non-relativistic Quantum Mechanics (see, e.g., \cite{LLQ}). In particular, the electron's 
Laplacian operator in spheroidal coordinates Eq.(\ref{spheroid1}) takes the form 
\begin{eqnarray}
 \Delta = \frac{1}{R^{2} \; (\xi^{2} - \eta^{2})} \; \Bigl\{\Bigl[ \frac{\partial}{\partial \xi} 
 (\xi^{2} - 1) \frac{\partial}{\partial \xi} \; + \; \frac{\partial}{\partial \eta} (1 - \eta^{2}) 
 \frac{\partial}{\partial \eta} \Bigr]  \; + \frac14 \; \Bigl[ \frac{1}{\xi^{2} - 1} + \frac{1}{1 
 - \eta^{2}} \Bigr] \; \frac{\partial^{2}}{\partial \phi^{2}} \Bigr\} \; . \; \label{LaplSpherdal}
\end{eqnarray}
The operator of potential energy $V$ in spheroidal coordinates is  
\begin{eqnarray}
 V = - \frac{Q_1 \; e^{2}}{R ( \xi - \eta )} - \frac{Q_2 \; e^{2}}{R ( \xi + \eta )} \; \; . \; \; 
 \label{Pot1}
\end{eqnarray} 
From here one finds that the product $R^{2} \; (\xi^{2} - \eta^{2}) \; V$ takes the form 
\begin{eqnarray}
 R^{2} \; (\xi^{2} - \eta^{2}) \; V = - e^{2} \; R \; [ (Q_1 + Q_{2}) \; \xi  + (Q_1 - Q_{2}) \; 
 \eta ] \; \; . \; \; \label{Pot2}
\end{eqnarray} 
Now, it is easy to write the non-relativistic Schr\"{o}dinger equation for the one-electron, two-center 
$(Q_1, Q_2)-$ion. The explicit form of this equation is 
\begin{eqnarray}
 \Bigl\{ \frac12 \; R^{2} \; (\xi^{2} - \eta^{2}) \; \Delta &-& e^{2} \; R \; [ (Q_1 + Q_{2}) \; \xi 
 + (Q_1 - Q_{2}) \eta ] \nonumber \\ 
  &-& E \; R^{2} \; (\xi^{2} - \eta^{2}) \Bigr\} \Psi(\xi, \eta, \phi; R; Q_1; Q_2) = 0 \; \; , \; 
  \; \label{Schrod3}
\end{eqnarray} 
where $E(R)$ is the corresponding total energy, i.e. eigenvalue of the original Schr\"{o}dinger 
equation.   

As directly follows from Eqs.(\ref{LaplSpherdal}) and (\ref{Schrod3}) the spheroidal variables $\eta, 
\eta$ and $\phi$ are truly separated from each other. The wave function $\Psi(\xi, \eta, \phi; R; 
Q_1; Q_2)$ is always written in the factorized form $\Psi(\xi, \eta, \phi; R; Q_1; Q_2) = \; 
X(\xi; R; Q_{1} + Q_{2}) \; Y(\eta; R; Q_{1} - Q_{2}) \; \exp(\pm \imath \; m \; \phi)$, where the 
`partial' wave functions $X(\xi)$ and $Y(\eta)$ obey the following one-dimensional equations 
\begin{eqnarray}
 \frac{\partial}{\partial \xi} \Bigl[ (\xi^{2} - 1) \frac{\partial X(\xi)}{\partial \xi} \Bigr] 
 + \Bigl[ \frac{E \; R^{2}}{2} \; \xi^{2} + \Bigl( Q_1 + Q_2 \Bigr) \; \xi + A(R) - 
 \frac{m^{2}}{\xi^{2} - 1} \Bigr] X(\xi) = 0 \; \; , \; \; \label{Eq1sphdal}
\end{eqnarray}
and 
\begin{eqnarray}
 \frac{\partial}{\partial \eta} \Bigl[ (1 - \eta^{2}) \frac{\partial Y(\eta)}{\partial \eta} \Bigr] 
 + \Bigl[ - \frac{E \; R^{2}}{2} \; \eta^{2} - \Bigl( Q_1 - Q_2 \Bigr) \; \eta - A(R) - 
 \frac{m^{2}}{1 - \eta^{2}} \Bigr] Y(\eta) = 0 \; \; , \; \; \label{Eq2sphdal}
\end{eqnarray}
where $A(R)$ is the so-called separation parameter which also depends upon the both $Q_1$ and $Q_2$ 
electric charges. The both one-dimensional functions $X(\xi)$ and $Y(\eta)$ depend upon the 
inter-nuclear distance $R$, but they also are the functions of some linear combinations of the two 
electric charges $Q_1$ and $Q_2$. In fact, as follows from Eq.(\ref{Eq1sphdal}) the function 
$X(\xi)$ of spheroidal $\xi$ variable is a function of the electric $Q_1 + Q_2$ charge. In other 
words, the $X(\xi; R, Q_1 + Q_2)$ function is an eigenfunction of some hydrogen-like (or one-center) 
atomic system where the electric charge of its central nucleus equals $Q_1 + Q_2$ (or $Q_1 e^{2} + 
Q_2 e^{2}$). Analogously, from Eq.(\ref{Eq2sphdal}) one finds that the function $Y(\eta; R, Q_1 - 
Q_2)$ of spheroidal $\eta$ variable can be a function of the electric $Q_1 - Q_2$ charge only, i.e., 
it is also an eigenfunction of the Schr\"{o}dinger one-dimensional equation which is written for 
the one-center Coulomb ion (or atom) which contains one electron. In this case the electric charge 
of the central, infinitely heavy nucleus equals $Q_1 - Q_2$. 

In the case when $Q_{2} = 0$ the both equations Eqs.(\ref{Eq1sphdal}) and (\ref{Eq2sphdal}), become 
identical to each other and the corresponding factorized wave function of the hydrogen ion can be 
written in the form: $\Psi(\xi, \eta, \phi; R; Q_1; Q_2) = \; X(\xi; R; Q_{1}) \; Y(\eta; R; Q_{1}) 
\; \exp(\pm \imath \; m \; \phi)$, where the function $X(\xi; R; Q_{1})$ takes the form 
\begin{eqnarray} 
  X(\xi; R; Q_{1}) = \Bigl(\xi^{2} - 1\Bigr)^{\frac{m}{2}} \; P(\xi) \; \exp\Bigl(- 
 \frac{R \; Q \; \xi}{2 \; n_{1}} \Bigr) \; \; , \; \; \label{Xfunct1}
\end{eqnarray} 
where $m = \mid m \mid (\ge 0)$ is integer and $P(\xi)$ is a polynomial function, i.e., the finite 
sum of non-negative powers of $\xi$. The function $Y(\eta; R; Q_{1})$ is written in a similar form 
\begin{eqnarray} 
  Y(\eta; R; Q_{1}) = \Bigl(1 - \eta^{2}\Bigr)^{\frac{m}{2}} \; Q(\eta) \; \exp\Bigl(- 
 \frac{R \; Q \; \eta}{2 \; n_{2}} \Bigr) \; \; , \; \; \label{Xfunct2}
\end{eqnarray} 
where $Q(\eta)$ is a polynomial function. These two functions are regular in the three following 
points: $\xi = -1, 1, +\infty$ and $\eta = -1, 1, \infty$, but for the hydrogen atom and 
hydrogen-like ions the two spheroidal variables vary between 1 and $+\infty$ and -1 and 1, 
respectively. Furthermore, any true hydrogen-like (or one-electron) wave function $X(\xi; R; Q_{1})$ 
has no more than $n_{1} - \mid m \mid - 1$ nodes in the $1 \le \xi < +\infty$ interval. In each node  
the polynomial function $P(\xi)$ equals zero. Analogously, any true hydrogen-like wave function 
$Y(\eta; R; Q_{1})$ has no more than $n_{2} - \mid m \mid - 1$ such nodes in the $-1 \le \eta \le 1$
interval. Again, in each node the polynomial function $Q(\eta)$ equals zero. More details of this 
method and its applications can be found in \cite{Demkov} and \cite{Fro83}. 

As follows directly from the equations, Eqs.(\ref{Eq1sphdal}) and (\ref{Eq2sphdal}), there is an 
obvious symmetry between these two equations. Moreover, for the hydrogen-like $X(\xi; R; Q_{1})$ and 
$Y(\eta; R; Q_{1})$ the corresponding equations in spheroidal coordinates are identical, and since 
these two functions are regular at $\xi = 1 = \eta$, then these two functions coincide with each 
other, i.e., in Eqs.(\ref{Xfunct1}) and (\ref{Xfunct2}) we can use only one notation, e.g., $X(\xi; 
R; Q_{1})$ and $X(\eta; R; Q_{1})$ instead of two. However, this can be confusing, since these two 
functions are formally defined at different intervals and they have different number of nodes.  
In the general case, the replacement of variables $\xi \rightarrow \imath \; \eta$ transforms 
Eq.(\ref{Eq1sphdal}) into Eq.(\ref{Eq2sphdal}), while similar replacement of variables $\eta 
\rightarrow \imath \; \xi$ transforms Eq.(\ref{Eq2sphdal}) into Eq.(\ref{Eq1sphdal}). Such a 
replacement leads to the new symmetry which does exist for any one-electron, two-center Coulomb 
systems. 

In order to discuss this symmetry and its consequences we have to re-define the both spheroidal $\xi$ 
and $\eta$ variables, since in Eq.(\ref{spheroid2}) these two variables are $a$ $priori$ non-symmetric. 
Indeed, they are varied at different intervals and variable $\xi$ always exceeds unity, while $\eta$ is 
less than unity. In reality, if we want to introduce and discuss a true symmetry between spheroidal 
variables $\xi$ and $\eta$ in the two-center problem, we have to assume in Eq.(\ref{spheroid2}) that 
these two variables vary in some identical intervals, e.g., $-1 \le \xi < + \infty, -1 \le \eta < + 
\infty$. It is looks good, but as one can see directly from Eq.(\ref{spheroid2}) now the both 
electron's Cartesian coordinates $x$ and $y$ can instantaneously be either real, or imaginary. This 
means that if we want to solve the one-electron, two-center (Coulomb) problem, then we have to include 
the hydrogenic (or one-center) wave functions obtained in the both Euclidean and pseudo-Euclidean 
spaces. In Cartesian coordinates the corresponding Schr\"{o}dinger equation for the same two-center 
system in pseudo-Euclidean space takes the form 
\begin{eqnarray}
  & &\hat{H} \; \psi(x, y, z) = \Bigl[ -\frac12 \Delta - \frac{{Q}_1}{r_{1}} - \frac{{Q}_2}{r_{2}} 
  \Bigr] \psi(x, y, z) =  \Bigl[ \; \frac12 \; \Bigl( \; \frac{\partial^{2}}{\partial x^{2}} \; 
  + \; \frac{\partial^{2}}{\partial y^{2}} \; - \; \frac{\partial^{2}}{\partial z^{2}} \Bigr) 
  \nonumber \\
 & &- \frac{{Q}_1}{\sqrt{ \mid - x^{2} - y^{2} + (z - R)^{2} \mid }} - \frac{{Q}_2}{\sqrt{ \mid 
  - x^{2} - y^{2} + (z + R)^{2} \mid }} \Bigr] \; \psi(x, y, z) = E \; \psi(x, y, z) \; \; . \; 
  \label{SchrodCartC}
\end{eqnarray}
The differences and similarities between this equation and Eq.(\ref{SchrodCart}) are obvious. In 
particular, the Laplace operator from Eq.(\ref{SchrodCart}) now becomes the Laplace-Beltrami 
operator (see, Eq.(\ref{SchrodCartC})). In general, the wave functions of the Schr\"{o}dinger 
equation, Eq.(\ref{SchrodCartC}), must be added to analogous wave functions of the Schr\"{o}dinger 
equation, Eq.(\ref{SchrodCart}). Otherwise, it is impossible to solve the original one-electron 
(Coulomb), two-center problem completely and accurately. The complete solution of the one-electron 
atomic problem in pseudo-Euclidean space can be found, e.g., in \cite{Trus1} and \cite{Trus2} (see, 
also \cite{Fro90}). Furthermore, those papers also contain applications of the results to actual 
solution of the one-electron, two-center (Coulomb) problem. Currently, such an approach is known 
s the algebraic method for the one-electron, two-center (Coulomb) problem. 

Note that applications of pure analytical (or algebraic) methods based on hydrogenic states in order 
to obtain the closed formulas for the energies of bound states in the one-electron, two-center 
(Coulomb) problem began in 1928 when A.H. Wilson published his paper \cite{Wilson}. In that paper he 
noticed the fact that the wave functions of any one-electron, two-center system $(Q_{1}, Q_{2})$ are  
always represented as the product of the two one-electron (or hydrogenic) functions written in 
spheroidal coordinates $\xi$ and $\eta$, respectively. His remarkable observation followed from the 
fact that the equation for the function of $\xi$ includes only the charge $Q_{1} + Q_{2}$, while a 
similar formula for the function of $\eta$ includes only the charge $Q_{1} - Q_{2}$. This Wilson's 
observation was absolutely correct (see our discussion after Eq.(\ref{Eq2sphdal}). In reality, this 
conclusion also applies to all wave functions which represent the unbound electron's states. Based on 
this fact Wilson tried to obtain all bound state solution for an arbitrary one-electron, two-center 
(Coulomb) system $(Q_{1}, Q_{2})$ in the form of products of hydrogenic functions only. In addition 
to this, he was trying to obtain analytical expressions for the corresponding total energies. 

However, his paper \cite{Wilson} contains two fundamental mistakes which consequently led its author 
to meaningless results. First, the whole Wilson's paper is essentially based on an obvious `symmetry' 
between the two spheroidal variables $\xi$ and $\eta$ in the two-center problem, but such a symmetry 
was not defined strictly and properly in \cite{Wilson}. Another Wilson's mistake was even more serious, 
since in \cite{Wilson} he accepted only a tiny sub-class of the spheroidal wave functions. Formally, 
among all functions of spheroidal variables, which have finite norms, Wilson chose only those that 
were represented in the following form: $\psi(\xi, \eta, \phi) = F_{n_{1}}(\xi; Q_1 + Q_2) \; 
G_{n_{2}}(\eta; Q_1 - Q_2) \; \exp(\pm \imath\; m \; \phi)$, where the functions $F_{n_{1}}(\xi; Q_1 
+ Q_2)$ and $G_{n_{2}}(\eta; Q_1 - Q_2)$ functions include finite polynomials. The first function 
($F$) has the $n_{1} - \mid m \mid - 1$ nodes, while the second function ($G$) has the $n_{2} - \mid 
m \mid - 1$ similar nodes. All other functions, which were derived in the form of infinite series of 
spheroidal coordinates, were simply rejected by Wilson as `non-regular', or `non-hydrogenic' despite 
the fact that each of those functions had a finite norm and correct number of nodes. 

Solutions of the two-center problem, i.e., the corresponding wave functions, have been 
constructed as the products of `regular' spheroidal functions of the $\xi$ or $\eta$ variables only. 
The original two-center problem has instantly been reduced to the problem of two hydrogen-like ions 
with different nuclear charges. Then it became clear that the wave functions of the `new' problem 
can be constructed only for some selected values of the interparticle distances $R$ in the two-center 
one-electron system. For instance, in one-electron two-center Coulomb system with $Q_1 = 5$ and $Q_2 
= 1$, the hydrogen-like solution does exist at $R \frac{\sqrt{10}}{3}$ and for $n_1 = 3, n_2 = 2$ 
(the principal quantum numbers). It is written in the form: 
\begin{eqnarray}
  \Psi = N \; ( 5 \; \xi^{2} - 2 \; \sqrt{10} \; \xi - 7 ) ( 5 \; \eta + \sqrt{10} ) \exp\Bigl[
  - \frac{\sqrt{10}}{3} (\xi + \eta) \Bigr]  \; \; , \; \; \label{AnalitSol} 
\end{eqnarray}
where $N$ is a normalization constant. For other interparticle distances $R$ we do not have similar 
hydrogenic (or finite) solutions. In general, for each one-electron two-center system there are a 
very few `special' values of $R$ where hydrogenic solutions can be found. For large number of 
one-electron two-center Coulomb problems it is impossible to find even one special value of $R$ 
\cite{Demkov}. To respect this fact Wilson suggested \cite{Wilson} that the bound states in the 
two-center, one-electron problem can exist only at these `special' interparticle distances $R$, 
i.e., he discovered the law of `quantization' of inter-nuclear distance $R$. In other words, 
one-electron two-center Coulomb ions could be stable only at certain values of inter-particle 
distances $R$. Such a conclusion and other mistakes made by Wilson in his paper were strongly 
criticized by E. Teller in \cite{Teller}. Later Demkov defined such a criticism as a `dance on the 
bones'. However, Teller could not explain why many `finite-term' hydrogenic wave functions obtained 
by Wilson (see, e.g., Eq.(\ref{AnalitSol})), were correct. This fact can easily be checked by direct 
substitutions of these wave functions in the Schr\"{o}dinger equation. Nevertheless, in \cite{Teller} 
Teller also developed an effective numerical method for solution of the Schr\"{o}dinger equation for 
arbitrary two-center, pure adiabatic ions. This method survived numerous modifications,  and till 
now it is successfully applied for similar systems. 

After Wilson's paper \cite{Wilson}, there were no other attempts to develop algebraic methods to 
solve the problem of one-electron motion in the field of two very heavy Coulomb centers. Only fifty 
years later Truskova in \cite{Trus1} reconsidered this problem and could re-write it in the Pauli-like 
form. By modifying these equation we have found that in our usual three-dimensional Euclidean space 
the corresponding equations are written in the form 
\cite{Trus1}, \cite{Fro90}
\begin{eqnarray}
  & &\frac12 \Bigl[ \hat{\bf L}^{2}_{1} + \hat{\bf N}^{2}_{1} \Bigr] \Psi =  \frac12 \Bigl[ -1 + 
  \frac{(Q_{1} + Q_{2})^{2}}{\varepsilon^{2}_{1}} \Bigr] \Psi = [ \phi_{1} (\phi_{1} + 1) + 
  \tilde{\phi}_{1} (\tilde{\phi}_{1} + 1) ] \Psi \; \; , \; \nonumber \\
  & &( \hat{\bf L}_{1} \cdot \hat{\bf N}_{1}) \Psi = [ \phi_{1} (\phi_{1} + 1) -  \tilde{\phi}_{1} 
  ( \tilde{\phi}_{1} + 1) ] \Psi = 0 \; \; , \; \nonumber \\
  & &\frac12 \Bigl[ \hat{\bf L}^{2}_{2} + \hat{\bf N}^{2}_{2} \Bigr] \Psi =  \frac12 \Bigl[ -1 + 
  \frac{(Q_{1} - Q_{2})^{2}}{\varepsilon^{2}_{2}} \Bigr] \Psi = [ \phi_{2} (\phi_{2} + 1) + 
  \tilde{\phi}_{2} (\tilde{\phi}_{2} + 1) ] \Psi \; \; , \; \nonumber \\
  & &( \hat{\bf L}_{2} \cdot \hat{\bf N}_{2} ) \Psi = [ \phi_{2} (\phi_{2} + 1) - \tilde{\phi}_{2} 
  ( \tilde{\phi}_{2} + 1) ] \Psi = 0 \; \; , \; \label{MainEq} \\
 & &\hat{\Lambda}_{1} \Psi =  \Bigl[ \hat{\bf L}^{2}_{1} + \varepsilon_{1} \; R \; 
 (\hat{\bf A}_{1})_{z} \Bigr] \Psi = \lambda_{1} \Psi = \Bigl[ \hat{\bf L}^{2}_{1} + R \; 
 (\hat{\bf N}_{1})_{z} \Bigr] \Psi = \lambda_{1} \Psi \; \; , \; \nonumber \\
  & &\hat{\Lambda}_{2} \Psi = \Bigl[ \hat{\bf L}^{2}_{2} + \varepsilon_{2} \; R \; 
  (\hat{\bf A}_{2})_{z} \Bigr] \Psi = \lambda_{2} \Psi = \Bigl[ \hat{\bf L}^{2}_{2} + R \; 
  (\hat{\bf N}_{2})_{z} \Bigr] \Psi = \lambda_{2} \Psi \; \; , \; \nonumber \\
  & &\Bigl[ (L^{2}_{1})_{z} \Bigr] \Psi = m^{2}_{1} \Psi \; \; \; , \; \; \; \Bigl[ (L^{2}_{2})_{z} 
  \Bigr] \Psi = m^{2}_{2} \Psi  \; \; , \; \nonumber 
\end{eqnarray}
where $R$ is the distance between two nuclei, while $\Psi$ is a product of the two wave functions,  
i.e., $\Psi = \psi_{1}(x_1, y_1, z_1) \psi_{2}(x_2, y_2, z_2)$. The first wave function 
$\psi_{1}(x_1, y_1, z_1)$ corresponds to the one-electron hydrogen-like ion with the nuclear charge 
$Q_{1} + Q_{2} = Q_A$, while the second wave function $\psi_{2}(x_2, y_2, z_2)$ is the one-electron 
wave function of the hydrogen-like ion with the nuclear charge $Q_{1} - Q_{2} = Q_B (\ge 0)$. It is 
crucially important to remember that our system of hydrogenic equations, Eq.(\ref{MainEq}), will be 
truly equivalent to the original Schr\"{o}dinger equation for the two Coulomb two-center problem(s), 
if (and only if) the three following constraints are obeyed: $\varepsilon_{1} = \varepsilon_{2} = 
\sqrt{- 2 E}, m^{2}_{1} = m^{2}_{2}$ and $\lambda_{1} = \lambda_{2}$. 

An obvious advantage of this approach is a possibility to deal with the wave functions and operators 
written in Cartesian coordinates. On the other hand, one can also use spherical, parabolic and/or 
spheroidal coordinates. The explicit forms of the vector-operators $\hat{\bf L} = (\hat{L}_{x}, 
\hat{L}_{y}, \hat{L}_{z})$ and $\hat{\bf A} =  (\hat{A}_{x}, \hat{A}_{y}, \hat{A}_{z})$, which appear 
in Eq.(\ref{MainEq}), in Cartesian coordinates are (in atomic units):
\begin{eqnarray}
 & &\hat{L}_{x} = - \imath \Bigl( y \frac{\partial}{\partial z} - z \frac{\partial}{\partial y} \Bigr) 
  \; \; , \hat{L}_{y} = - \imath \Bigl( z \frac{\partial}{\partial x} - x \frac{\partial}{\partial z} 
  \Bigr) \; \; ,  \hat{L}_{z} = - \imath \Bigl( y \frac{\partial}{\partial x} - x 
  \frac{\partial}{\partial x} \Bigr) \; \; , \; \; \; \label{LLL} \\
 & &\hat{A}_{x} = \frac{Q \; x}{r} + \frac{Q}{2} \Bigl[ (\hat{\bf L} \times \hat{\bf p})_{x} 
  - (\hat{\bf p} \times \hat{\bf L})_{x} \Bigr] = \frac{Q \; x}{r} + \frac{Q}{2} \Bigl[ 
  \hat{\bf L}_{y} \hat{\bf p}_{z} - \hat{\bf L}_{z} \hat{\bf p}_{y} + \hat{\bf p}_{z} 
  \hat{\bf L}_{y} -  \hat{\bf p}_{y} \hat{\bf L}_{z} \Bigr] \nonumber \\
 & &= \frac{Q \; x}{r} - Q \Bigl[ z \frac{\partial^{2}}{\partial z \partial x} + 
 y \frac{\partial^{2}}{\partial y \partial x} + \frac{\partial}{\partial x} - x \Bigl( 
 \frac{\partial^{2}}{\partial y^{2}} + \frac{\partial^{2}}{\partial z^{2}} \Bigr)\Bigr] \; \; , \; 
 \label{AX} \\
 & &\hat{A}_{y} = \frac{Q \; y}{r} + \frac{Q}{2} \Bigl[ (\hat{\bf L} \times \hat{\bf p})_{y} 
  - (\hat{\bf p} \times \hat{\bf L})_{y} \Bigr] = \frac{Q \; y}{r} + \frac{Q}{2} \Bigl[ 
  \hat{\bf L}_{z} \hat{\bf p}_{x} - \hat{\bf L}_{x} \hat{\bf p}_{z} + \hat{\bf p}_{x} 
  \hat{\bf L}_{z} -  \hat{\bf p}_{z} \hat{\bf L}_{x} \Bigr] \nonumber \\
 & &= \frac{Q \; y}{r} - Q \Bigl[ x \frac{\partial^{2}}{\partial x \partial y} + 
 z \frac{\partial^{2}}{\partial z \partial y} + \frac{\partial}{\partial y} - y \Bigl( 
 \frac{\partial^{2}}{\partial x^{2}} + \frac{\partial^{2}}{\partial z^{2}} \Bigr)\Bigr] \; \; , \; 
 \label{AY} \\
 & &\hat{A}_{z} = \frac{Q \; z}{r} + \frac{Q}{2} \Bigl[ (\hat{\bf L} \times \hat{\bf p})_{z} 
  - (\hat{\bf p} \times \hat{\bf L})_{z} \Bigr] = \frac{Q {\bf z}}{r} + \frac{Q}{2} \Bigl[ 
  \hat{\bf L}_{x} \hat{\bf p}_{y} - \hat{\bf L}_{y} \hat{\bf p}_{x} + \hat{\bf p}_{y} 
  \hat{\bf L}_{x} -  \hat{\bf p}_{x} \hat{\bf L}_{y} \Bigr] \nonumber \\
 & &= \frac{Q \; z}{r} - Q \Bigl[ x \frac{\partial^{2}}{\partial x \partial z} + 
 y \frac{\partial^{2}}{\partial y \partial z} + \frac{\partial}{\partial z} - z \Bigl( 
 \frac{\partial^{2}}{\partial x^{2}} + \frac{\partial^{2}}{\partial y^{2}} \Bigr)\Bigr] \; \; , \;  
 \label{AZ} 
\end{eqnarray}
where ${\bf r} = (x, y, z)$ and $r = \sqrt{x^{2} + y^{2} + z^{2}}$. The commutation relations between 
these operators are 
\begin{eqnarray}
  & &[ \hat{L}_{i}, \hat{L}_{j} ] = \imath \varepsilon_{ijk} \hat{L}_{k} \; \; , \;  [ \hat{L}_{i}, 
  \hat{A}_{j} ] = \imath \varepsilon_{ijk} \hat{A}_{k} \; \; \label{LLLA} \\
  & &[ \hat{A}_{i}, \hat{A}_{j} ] = \imath \varepsilon_{ijk} \Bigl( - {\bf p}^{2} + \frac{2 Q}{r} 
  \Bigr) \hat{L}_{k} = (- 2 \hat{H}) \; \; \imath \varepsilon_{ijk} \hat{L}_{k} = (- 2 E) \; \; 
  \imath \varepsilon_{ijk} \hat{L}_{k} \; \; , \; \label{AAA} 
\end{eqnarray}
where $(i, j, k) = (x, y, z)$, or (1, 2, 3), while ${\bf p}^{2} - \frac{2 Q}{r} = - 
\frac{\partial^{2}}{\partial x^{2}} - \frac{\partial^{2}}{\partial y^{2}} - 
\frac{\partial^{2}}{\partial z^{2}} - \frac{2 Q}{r}$  is the doubled Hamiltonian of this problem in 
our three-dimensional Euclidean space. This scalar operator commute with all components of the both 
vector-operators $\hat{\bf L}$ and  $\hat{\bf A}$. For stationary bound states one can replace in 
the last formula $\hat{H} \rightarrow E$, where $E$ is the total energy of this bound state. Since 
for bound states $E < 0$, then instead of $\hat{A}_{i}$ operators we introduce the three new 
operators $\hat{N}_{i} = \sqrt{- 2 E} \; \hat{A}_{i}$ ($i$ = 1, 2, 3) which are also used in 
Eq.(\ref{MainEq}). The commutation relations for the six $\hat{L}_{i}, \hat{N}_{i}$, where $i =$ 1, 
2, 3, operators, take the form 
\begin{eqnarray}
  & &[ \hat{L}_{i}, \hat{L}_{j} ] = \imath \varepsilon_{ijk} \hat{L}_{k} \; \; \; , \; 
  \; \; [ \hat{L}_{i}, \hat{N}_{j} ] = \imath \varepsilon_{ijk} \hat{N}_{k} \; \; \; ,
  \; \; \;  [ \hat{N}_{i}, \hat{N}_{j} ] = \imath \varepsilon_{ijk} \hat{L}_{k} \; \; 
  \; . \; \; \; \label{ComAAA} 
\end{eqnarray}
The first and last equations from these commutation relations, Eq.(\ref{ComAAA}), indicate clearly 
that we have two related $SO(3)-$algebras, which are sub-algebras of some general algebraic 
construction. The Casimir operators of the second order $C_2$ for these sub-algebras equal $C_{2} 
= \hat{\bf L}^{2} = \hat{L}^{2}_{x} + \hat{L}^{2}_{y} + \hat{L}^{2}_{z}$ and $C_{2} = 
\hat{\bf N}^{2} = \hat{N}^{2}_{x} + \hat{N}^{2}_{y} + \hat{N}^{2}_{z}$, respectively. In order to 
reveal this general algebraic construction “in the flesh” it is necessary to introduce the two new 
vector-operators ${\bf J}^{(1)} = \frac12 ( \hat{\bf L} + \hat{\bf N} )$ and ${\bf J}^{(2)} = 
\frac12 ( \hat{\bf L} - \hat{\bf N} )$. The commutation relations for the six components of these
two vector-operators ${\bf J}^{(1)}$ and ${\bf J}^{(1)}$ exactly coincide with commutation relations 
of the 3 + 3 generators of two independent $SO(3)-$algebra discussed above (see, Eq.(\ref{Commute4})). 
This means that we are dealing with the $SO(4)-$algebra, which correspond to the $SO(4)$ group of 
rotations in the four-dimensional Euclidean space. Note also, as follows from Eq.(\ref{MainEq}) in 
order to solve analytically the one-electron (Coulomb) two-center problem $(Q_1, Q_2)$ one has to 
consider the two cases when in the equations above $Q_A = Q_1 + Q_2$ and $Q_B = Q_1 - Q_2$.  

The system of equations, Eq.(\ref{MainEq}), looks very complex, but in reality almost all equations in 
this system are obeyed automatically. For instance, the second and fourth equations in Eq.(\ref{MainEq}) 
mean that we always have $\phi_{1} = \tilde{\phi}_{1} = j_{1}$ and $\phi_{2} = \tilde{\phi}_{2} = 
j_{2}$, where $j_{1}$ and $j_{2}$ are semi-integer numbers. The first and third equations determine 
the bound state wave functions of the two hydrogen-like ions with the nuclear charges $Q_1 + Q_2 = Q_A$ 
and $Q_1 - Q_2 = Q_B$, respectively. In general, the bound state wave function of the one-electron, 
hydrogen-like ion with the nuclear charge $Q_{1} + Q_{2} = Q_A$ is written in the form (in spherical 
coordinates and in atomic units):  
\begin{eqnarray}
  \mid n_1 \ell_1 m_1 \rangle = R_{n_{1} \ell_{1}}(r_1; Q_A) \; Y_{\ell_{1} m_1}(\theta_{1}, \phi_{1}) 
  \; \; \; , \; \; \label{psi-1}
\end{eqnarray}
where $Q_{A} = Q_{1} + Q_{2} , r_1 = \sqrt{x^{2}_1 + y^{2}_1 + z^{2}_1}$ ($0 \le r < + \infty$) is the 
spatial radius of the electron's orbit and $Y_{\ell_{1} m_1}(\theta_{1}, \phi_{1})$ are the regular 
spherical harmonics which are the well known functions of the two spherical angles $\theta_{1}$ and 
$\phi_{1}$ (see, e.g., \cite{Rose}):
\begin{eqnarray}
  & &Y_{\ell_{1} m_1}(\theta_{1}, \phi_{1}) \; = \hat{P}^{m}_{\ell_{1}}(\theta_{1}) \; \Bigl[ 
  \frac{1}{\sqrt{2 \pi}} \; \exp(\imath \; m_{1} \; \phi_{1})) \Bigr] \nonumber \\
  & &= (-1)^{m_1} \sqrt{\frac{(2 \ell_{1} + 1)! (\ell_{1} - m_{1})!}{2 (\ell_{1} + m_{1})!}} \; 
  P^{m}_{\ell_{1}}(\cos \theta_{1}) \; \Bigl[ \frac{1}{\sqrt{2 \pi}} \; \exp(\imath \; m_{1} \; 
  \phi_{1} ) \Bigr] \; \; , \; \; \label{Spherharm}
\end{eqnarray}
where $\hat{P}^{m}_{\ell}(\theta)$ are the unit-norm associated Legendre polynomials, $- \frac{\pi}{2} 
\le \theta_1 \le \frac{\pi}{2}$ and $0 \le \phi \le 2 \pi$. Also in Eq.(\ref{psi-1}) the functions 
$R_{n_{1} \ell_{1}}(r_1; Q_A)$ are the unit norm, radial functions which here we choose in the form 
 \begin{eqnarray}
  R_{n_{1} \ell_{1}}(r_1; Q_A) = - \sqrt{\frac{(n_1 - \ell_1 - 1)!}{2 \; n_1 \; [(n_1 + 
  \ell_1)!]^{3}}} \; \Bigl(\frac{2 Q_A}{n_1}\Bigr)^{\frac32} \exp\Bigl(\frac{-Q_A r_1}{n_1}\Bigr)  
  \Bigl(\frac{2 Q_A r_1}{n_1}\Bigr)^{\ell_1} \; L^{2 \ell_1 + 1}_{n_1 + \ell_1}
  \Bigl(\frac{2 Q_A r_1}{n_1}\Bigr) \; . \; \; \label{psi-1rad}
\end{eqnarray}
The bound state wave function of the one-electron, hydrogen-like ions with the nuclear charge $Q_{1} - 
Q_{2} = Q_B$ ion takes a very similar form 
\begin{eqnarray}
  \mid n_2 \ell_2 m_2 \rangle = R_{n_{1} \ell_{1}}(r_2; Q_B) \; Y_{\ell_{2} m_{2}}(\theta_{2}, 
  \phi_{2})\; \; , \; \; \label{psi-2}
\end{eqnarray}
where $Q_{B} = Q_{1} - Q_{2} , r_2 = \sqrt{x^{2}_2 + y^{2}_2 + z^{2}_2}$ is the same spatial radius of 
electron defined for the second hydrogen-like ion, while $Y_{\ell_{2} m_{2}}(\theta_{2}, \phi_{2})$ are 
the spherical harmonics which depend upon the two spherical angles $\theta_{2}$ and $\phi_{2}$. In this
case the radial part of the total wave function is 
 \begin{eqnarray}
  R_{n_{2} \ell_{2}}(r_2; Q_B) = - \sqrt{\frac{(n_2 - \ell_2 - 1)!}{2 \; n_2 \; [(n_2 + 
  \ell_2)!]^{3}}} \; \Bigl(\frac{2 Q_B}{n_2}\Bigr)^{\frac32} \exp\Bigl(\frac{-Q_B r_2}{n_2}\Bigr)  
  \Bigl(\frac{2 Q_B r_2}{n_2}\Bigr)^{\ell_2} \; L^{2 \ell_2 + 2}_{n_2 + \ell_2}
  \Bigl(\frac{2 Q_B r_2}{n_2}\Bigr) \; , \; \; \label{psi-2rad}
\end{eqnarray}
where all notations have the same sense as in Eq.(\ref{psi-1rad}). 

Let us choose the wave functions of the two hydrogenic-like atoms in the form of Eq.(\ref{psi-1}) 
and Eq.(\ref{psi-2}). In this case the product of these two wave functions is written in the form 
$\Psi = \mid n_1 \ell_1 m_1 \rangle \mid n_2 \ell_2 m_2 \rangle$ and the first four equations in 
Eq.(\ref{MainEq}) become identities, i.e., they hold automatically. Now, if we also choose in these 
wave functions $m^{2}_{1} = m^{2}_{2}$, or $\mid m_{1} \mid = \mid m_{2} \mid$, then the last two 
equations in the system Eq.(\ref{MainEq}) also obey automatically. In reality, this constraint is 
trivial, since the dependence of each hydrogenic state $\mid n \ell m \rangle$ upon the azimuthal 
angle $\phi$ is written in the form of complex exponents which explicitly contains the azimuthal 
quantum number $m$.  

Finally, we have to deal only with the fifth and sixth equations in Eq.(\ref{MainEq}). In order 
to solve these two remaining equations we have to deal with the two operators $\hat{\Lambda}_k = 
\hat{\bf L}^{2}_{k} + \varepsilon_{k} \; R \; (\hat{A}_{k})_{z} = \hat{\bf L}^{2}_{k} + R \; 
(\hat{N}_{k})_{z} $, where $k = 1, 2$, which must be represented in the basis of our 
$\Psi-$functions. Thus, the original complex problem of solving eight equations in this approach 
has been reduced to the form: determine the eigenvalues of two  matrices, which represent the 
$\hat{\Lambda}_1 = \hat{\bf L}^{2}_{1} + \varepsilon_{1} \; R \; (\hat{A}_{1})_{z}$ and 
$\hat{\Lambda}_2 = \hat{\bf L}^{2}_{2} + \varepsilon_{2} \; R \; (\hat{A}_{2})_{z}$ operators in 
the basis of hydrogenic $\mid n_1 \ell_1 m_1 \rangle$ and $\mid n_2 \ell_2 m_2 \rangle$ states, 
respectively. The actual solutions of the two-center $(Q_1, Q_2)$ problem corresponds to the case 
when the eigenvalues of these two problems will be equal to each other.  

In other words, to the fifth and sixth equations in Eq.(\ref{MainEq}) we have to find the 
simultaneous solutions of the two following eigenvalue problems with equal eigenvalue(s):
\begin{eqnarray}
 \hat{\Lambda}_{1}\sum_{\ell_1} C_{\ell_1} \mid n_1 \ell_1 m_1 \rangle = 
 \Bigl[ \hat{\bf L}^{2}_{1} + \varepsilon_{1} \; R \; (\hat{A}_{1})_{z} \Bigr] 
 \sum_{\ell_1} C_{\ell_{1}} \mid n_1 \ell_1 m_1 \rangle = \lambda \sum_{\ell_1} 
 C_{\ell_1} \mid n_1 \ell_1 m_1 \rangle \; \; \; , \; \; \label{L2Az1}
\end{eqnarray}
and 
\begin{eqnarray}
  \hat{\Lambda}_{2} \sum_{\ell_2} C_{\ell_{2}} \mid n_2 \ell_2 m_2 \rangle = 
  \Bigl[ \hat{\bf L}^{2}_{2} + \varepsilon_{2} \; R \; (\hat{A}_{2})_{z} \Bigr] 
  \sum_{\ell_2} C_{\ell_{2}} \mid n_2 \ell_2 m_2 \rangle = \lambda \sum_{\ell_2} 
  C_{\ell_{2}} \mid n_2 \ell_2 m_2 \rangle \; \; . \; \; \label{L2Az2}
\end{eqnarray} 
In matrix form these two operators are: 
\begin{eqnarray}
  & &\Bigl[ \hat{\Lambda}_{1} \Bigr]_{\ell_{1}, \ell} = \delta_{\ell_{1}, \ell} 
  \; \ell (\ell + 1) + \delta_{\ell_{1}, \ell + 1} \; 
  \Bigl(\frac{Q_A}{n^{2}_{1}}\Bigr) \; R \; \Biggl\{ \frac{[n^{2}_{1} - 
  (\ell + 1)^{2}] [(\ell + 1)^{2} - m^{2}]}{[2 (\ell + 1)^{2} - 1]} \Biggr\} 
  \nonumber \\ 
  & & + \delta_{\ell_{1}, \ell - 1} \; \Bigl(\frac{Q_A}{n^{2}_{1}}\Bigr) \; R \; 
  \Biggl\{ \frac{[n^{2}_{1} - (\ell - 1)^{2}] [(\ell - 1)^{2} - m^{2}]}{[2 (\ell 
  - 1)^{2} - 1]} \Biggr\} \; \; \; , \; \label{MatForm1} 
\end{eqnarray} 
where  $Q_A = Q_1 + Q_2$ and the matrix indexes $\ell$ and $\ell_1$ vary between 0 and 
$n_1 - 1$, i.e., the dimension of this $\hat{\Lambda}_{1}$ matrix equals $n_1 \times 
n_1$. Analogous formula for the second matrix is 
\begin{eqnarray}
  & &\Bigl[ \hat{\Lambda}_{2} \Bigr]_{\ell_{2}, \ell} = \delta_{\ell_{2}, \ell} 
  \; \ell (\ell + 1) + \delta_{\ell_{2}, \ell + 1} \; 
  \Bigl(\frac{Q_B}{n^{2}_{2}}\Bigr) \; R \; \Biggl\{ \frac{[n^{2}_{2} - 
  (\ell + 1)^{2}] [(\ell + 1)^{2} - m^{2}]}{[2 (\ell + 1)^{2} - 1]} \Biggr\} 
  \nonumber \\ 
  & & + \delta_{\ell_{2}, \ell - 1} \; \Bigl(\frac{Q_B}{n^{2}_{2}}\Bigr) \; R \; 
  \Biggl\{ \frac{[n^{2}_{2} - (\ell - 1)^{2}] [(\ell - 1)^{2} - m^{2}]}{[2 (\ell 
  - 1)^{2} - 1]} \Biggr\} \; \; \; , \; \label{MatForm2} 
\end{eqnarray} 
where $Q_B = Q_1 - Q_2$ and the matrix indexes $\ell$ and $\ell_2$ vary between 0 and 
$n_2 - 1$, i.e., the dimension of this $\hat{\Lambda}_{2}$ matrix equals $n_2 \times 
n_2$. Note that the two $\hat{\Lambda}_{1}$ and $\hat{\Lambda}_{2}$ matrices, which 
play a crucial role in this method, have different dimensions. 

Thus, despite the apparent complexity of our method, it is reduced to construction of 
the two systems of one-electron, hydrogen-like functions, or spherical states $\mid 
n_1 \ell_1 m_1 \rangle$ and $\mid n_2 \ell_2 m_2 \rangle$, where $\mid m_1 \mid = 
\mid m_2 \mid$. In the case of three-dimensional Euclidean space this part of the 
problem is trivial, since such states are well known in Quantum Mechanics for years. 
At the second stage of this method we have to solve (simultaneously) the two eigenvalue 
problems for the $\hat{\Lambda}_1$ and $\hat{\Lambda}_2$ operators, Eqs.(\ref{MatForm1}) 
and (\ref{MatForm2}). This step is used to construct the `correct' linear combinations 
which contain some finite number(s) of products of the spherical $\mid n_1 \ell_1 m 
\rangle$ and $\mid n_2 \ell_2 m \rangle$ states. These correct linear combinations of 
the $\mid n_1 \ell_1 m \rangle \mid n_2 \ell_2 m \rangle$ states represent the actual 
wave functions of the one-electron, two-center problem $(Q_1, Q_2)$. When this procedure 
is completed, then we also obtain (automatically and for free) the bound states of 
another one-electron, two-center (Coulomb) problem $(Q_1, -Q_2)$. 

Now, let us consider the same problem in the three-dimensional pseudo-Euclidean space. As 
mentioned above in contrast with the Pauli case \cite{Pauli}, in order to obtain all bound 
state wave functions for the one-electron, two-center (Coulomb) problem we need to operate 
in the orthogonal three-dimensional, pseudo-Euclidean space with the metric (-1, -1, 1). 
To achieve our goal we can introduce the imaginary $x$ and $y$ coordinates. Then, the  
vector-operator of position vector is written in the form ${\bf r} = ( \imath x, \imath y, 
z)$, while the three components of momentum vector ${\bf p}$ take the form: $p_{x} = - 
\frac{\partial}{\partial x}, p_{y} = - \frac{\partial}{\partial y}$ and $p_{z} = - \imath 
\frac{\partial}{\partial z}$. Note that the basic Poisson bracket between coordinates ${\bf r} 
= ( \imath x, \imath y, z)$ and `new' momenta ${\bf p} = ( p_x, p_y, p_z)$ do not change. In 
other words, they are exactly the same as in the orthogonal Euclidean $3 D-$space i.e., $[p_x, 
{\bf r}_{x} ] = -\imath, [p_y, {\bf r}_{y} ] = -\imath, [p_z, {\bf r}_{z} ] = -\imath$. All 
other Poisson brackets between any pair of the $p_{i}$ and ${\bf r}_{j}$ variables, where $(i, 
j) = ( x, y, z)$ equal zero identically. 

In three-dimensional, orthogonal, pseudo-Euclidean space we have six components of the two  
vector-operators $\hat{\bf L} = (\hat{L}_{x}, \hat{L}_{y}, \hat{L}_{z})$ and $\hat{\bf A} =  
(\hat{A}_{x}, \hat{A}_{y}, \hat{A}_{z})$ which are written in the following forms (in atomic 
units):
\begin{eqnarray}
 & &\hat{L}_{x} = \Bigl( y \frac{\partial}{\partial z} + z \frac{\partial}{\partial y} \Bigr) \; \; , 
 \; \; \hat{L}_{y} = - \Bigl( z \frac{\partial}{\partial x} - x \frac{\partial}{\partial z} 
  \Bigr) \; \; , \; \; \hat{L}_{z} = - \imath \Bigl( y \frac{\partial}{\partial x} - x 
  \frac{\partial}{\partial x} \Bigr) \; \; , \; \; \label{LLLPES} \\
 & &\hat{A}_{x} = \imath \frac{Q \; x}{r} + \imath \frac{Q}{2} \Bigl[ (\hat{\bf L} \times 
 \hat{\bf p})_{x} - (\hat{\bf p} \times \hat{\bf L})_{x} \Bigr] \nonumber \\
 & &= \imath \frac{Q \; x}{r} + \imath Q \Bigl[ z \frac{\partial^{2}}{\partial z 
 \partial x} + y \frac{\partial^{2}}{\partial y \partial x} + \frac{\partial}{\partial x} 
 + x \Bigl( \frac{\partial^{2}}{\partial z^{2}} - \frac{\partial^{2}}{\partial y^{2}} \Bigr)\Bigr] 
 \; \; , \; \label{AXPES} \\
 & &\hat{A}_{y} = \imath \frac{Q \; y}{r} + \frac{Q}{2} \Bigl[ (\hat{\bf L} \times 
 \hat{\bf p})_{y} - (\hat{\bf p} \times \hat{\bf L})_{y} \Bigr] \nonumber \\
 & &= \imath \frac{Q \; y}{r} + \imath Q \Bigl[ x \frac{\partial^{2}}{\partial x \partial y} 
 + z \frac{\partial^{2}}{\partial z \partial y} + \frac{\partial}{\partial y} + y \Bigl( 
  \frac{\partial^{2}}{\partial z^{2}} - \frac{\partial^{2}}{\partial x^{2}} \Bigr)\Bigr] \; 
  \; , \; \label{AY} \\
 & &\hat{A}_{z} = \frac{Q \; z}{r} + \frac{Q}{2} \Bigl[ (\hat{\bf L} \times \hat{\bf p})_{z} 
  - (\hat{\bf p} \times \hat{\bf L})_{z} \Bigr] = \nonumber \\
 & &= \frac{Q \; z}{r} - Q \Bigl[ x \frac{\partial^{2}}{\partial x \partial z} + 
 y \frac{\partial^{2}}{\partial y \partial z} + \frac{\partial}{\partial z} - z \Bigl( 
 \frac{\partial^{2}}{\partial x^{2}} + \frac{\partial^{2}}{\partial y^{2}} \Bigr)\Bigr] 
 \; \; , \;  \label{AZ} 
\end{eqnarray}
where ${\bf r} = (\imath x, \imath y, z)$ and $r = \sqrt{\mid - x^{2} - y^{2} + z^{2} \mid}$. As 
follows from these equations the commutation relations between components of the $\hat{\bf L}$ 
and $\hat{\bf A}$ vector-operators take the form:
\begin{eqnarray}
  & &[ \hat{L}_{x}, \hat{L}_{y} ] = - \imath \hat{L}_{z} \; \; \; , \; \; \;  [ \hat{L}_{y}, 
  \hat{L}_{z} ] = \imath \hat{L}_{x} \; \; , \; \; \;  [ \hat{L}_{z}, \hat{L}_{x} ] = \imath 
  \hat{L}_{y} \; \; , \; \; \label{PBracket1} \\ 
 & &[ \hat{L}_{x}, \hat{A}_{y} ] = - \imath \hat{A}_{z} \; \; \; , \; \; \;  [ \hat{L}_{y}, 
  \hat{A}_{z} ] = \imath \hat{A}_{x} \; \; , \; \; \;  [ \hat{L}_{z}, \hat{A}_{x} ] = \imath 
  \hat{A}_{y} \; \; , \; \; \label{PBracket2} \\ 
  & &[ \hat{A}_{x}, \hat{A}_{y} ] = - \imath (- 2 \hat{H}) \hat{L}_{z} \; \; \; , \; \; \;  
  [ \hat{A}_{y}, \hat{A}_{z} ] = \imath  (- 2 \hat{H}) \hat{L}_{x} \; \; , \; \; \; [ 
  \hat{A}_{z}, \hat{A}_{x} ] = \imath (- 2 \hat{H}) \hat{L}_{y} \; \; , \; \; 
  \label{PBracket3} 
\end{eqnarray}
where $2 \hat{H} = {\bf p}^{2} - \frac{2 Q}{r} = \frac{\partial^{2}}{\partial x^{2}} + 
\frac{\partial^{2}}{\partial y^{2}} - \frac{\partial^{2}}{\partial z^{2}} - \frac{2 Q}{r}$  
is the doubled Hamiltonian of this problem in three-dimensional pseudo-Euclidean space. 
For bound (stationary) states when $E < 0$ we can replace $- \hat{H} = - E$ in 
Eq.(\ref{PBracket3}) and introduce a vector set of three new vector-operators $\hat{N}_{i} 
= (- 2 \hat{H})^{\frac12} \hat{A}_{i}$ ($i$ = 1, 2, 3). For the complete set of six 
operators $\hat{\bf L}_{i}, \hat{\bf N}_{i}$ ($i$ = 1, 2, 3) one finds the following 
commutation relations:
\begin{eqnarray}
  & &[ \hat{L}_{i}, \hat{L}_{j} ] = - \imath \kappa^{ab}_{ijk} \hat{L}_{k} \; \; \; , \; 
  \; \; [ \hat{L}_{i}, \hat{N}_{j} ] = - \imath \kappa^{ab}_{ijk} \hat{N}_{k} \; \; \; ,
  \; \; \;  [ \hat{N}_{i}, \hat{N}_{j} ] = - \imath \kappa^{ab}_{ijk} \hat{L}_{k} \; \; 
  \; , \; \; \; \label{PBracket4} 
\end{eqnarray}
where the new tensor symbol $\kappa^{ab}_{ijk}$ is defined as follows $\kappa^{ab}_{ijk} 
= \Bigl[ 1 - 2 \Bigl( \delta^{a}_{i} \delta^{b}_{j} + \delta^{a}_{j} \delta^{b}_{i} 
\Bigl)\Bigr] \; \varepsilon_{ijk}$. The indexes $a$ and $b$ stand for the indexes of two 
(of three) operators which included in the corresponding Casimir operator(s) with the same 
signs (see below). In our present case equal these indexes equal $a = $1 and $b = 2$, 
respectively. The commutation relations, Eq.(\ref{PBracket4}), correspond to the 
$SO(2,1)-$algebra (see, e.g., \cite{BarR} and \cite{Perelom}). The $SO(2,1)-$group, its 
properties and representations have been described in detail in \cite{VBarg} and 
\cite{GelfGra}. 
 
For these six vector-operators $\hat{L}_{i}, \hat{N}_{i}$, where $i$ = 1, 2, 3, one finds 
the following system of equations in the three-dimensional pseudo-Euclidean space: 
\begin{eqnarray}
  & &\frac12 \Bigl[ \hat{\bf L}^{2}_{1} + \hat{\bf N}^{2}_{1} \Bigr] \Psi =  \frac12 \Bigl[ -1 + 
  \frac{(Q_{1} + Q_{2})^{2}}{\varepsilon^{2}_{1}} \Bigr] \Psi = [ \phi_{1} (\phi_{1} + 1) + 
  \tilde{\phi}_{1} (\tilde{\phi}_{1} + 1) ] \Psi \; \; , \; \nonumber \\
  & &( \hat{\bf L}_{1} \cdot \hat{\bf N}_{1}) \Psi = [ \phi_{1} (\phi_{1} + 1) -  \tilde{\phi}_{1} 
  ( \tilde{\phi}_{1} + 1) ] \Psi = 0 \; \; , \; \nonumber \\
  & &\frac12 \Bigl[ \hat{\bf L}^{2}_{2} + \hat{\bf N}^{2}_{2} \Bigr] \Psi =  \frac12 \Bigl[ -1 + 
  \frac{(Q_{1} - Q_{2})^{2}}{\varepsilon^{2}_{2}} \Bigr] \Psi = [ \phi_{2} (\phi_{2} + 1) + 
  \tilde{\phi}_{2} (\tilde{\phi}_{2} + 1) ] \Psi \; \; , \; \nonumber \\
  & &( \hat{\bf L}_{2} \cdot \hat{\bf N}_{2} ) \Psi = [ \phi_{2} (\phi_{2} + 1) - \tilde{\phi}_{2} 
  ( \tilde{\phi}_{2} + 1) ] \Psi = 0 \; \; , \; \label{MainEqPSE} \\
 & &\hat{\Lambda}_{1} \Psi = \Bigl[ \hat{\bf L}^{2}_{1} + R \; 
 (\hat{\bf N}_{1})_{z} \Bigr] \Psi = \lambda_{1} \Psi \; \; , \; \nonumber \\
  & &\hat{\Lambda}_{2} \Psi = \Bigl[ \hat{\bf L}^{2}_{2} + R \; 
 (\hat{\bf N}_{2})_{z} \Bigr] \Psi = \lambda_{2} \Psi \; \; , \; \nonumber \\
  & &\Bigl[ (L^{2}_{1})_{z} \Bigr] \Psi = m^{2}_{1} \Psi \; \; \; , \; \; \; \Bigl[ 
  (L^{2}_{2})_{z} \Bigr] \Psi = m^{2}_{2} \Psi \; \; , \; \nonumber 
\end{eqnarray}
where $(\hat{\bf N}_{i})_{z} =  (\hat{N}_{i})_{z}$ ($i$ = 1, 2), while $R$ is the distance between 
two nuclei, while $\Psi$ is a product of the two wave functions,  i.e., $\Psi = \psi_{1}(x_1, y_1, 
z_1) \psi_{2}(x_2, y_2, z_2)$. The three additional constraints $\varepsilon^{2}_{1} = 
\varepsilon^{2}_{2}, \lambda_{1} = \lambda_{2}$ and $m^{2}_{1} = m^{2}_{2}$ must be obeyed for the 
scalar values in these eight equations. Otherwise, the system of equations,  Eqs.(\ref{MainEqPSE}), 
will not be equivalent to the original Schr\"{o}dinger equation for the one-electron problem of the 
two Coulomb centers. This system of equations formally coincides with similar system of 
Eqs.(\ref{MainEq}) for the three-dimensional, orthogonal, Euclidean space. However, some operators 
in these equations have different sense, e.g., $\hat{\bf L}^{2} = - \hat{L}^{2}_{x} - \hat{L}^{2}_{y} 
+ \hat{L}^{2}_{z} \; , \; \hat{\bf N}^{2} = - \hat{N}^{2}_{x} - \hat{N}^{2}_{y} + \hat{N}^{2}_{z}$, 
etc. 

Thus, we have reduced the original Schr\"{o}dinger equation, Eq.(\ref{SchrodCart}), for the 
one-electron (quantum) motion in the field of the two Coulomb centers to the Pauli-like form, 
Eq.(\ref{MainEqPSE}). Indeed, in our three-dimensional Euclidean space we obtain the system of 
equations, Eqs.(\ref{MainEq}), while in three-dimensional pseudo-Euclidean space we derive 
another system of equations, Eq.(\ref{MainEqPSE}). Solutions of this system of equations are 
determined in the same way as we did above for the dimensional Euclidean space. Only in the 
case of Eqs.(\ref{MainEqPSE}) we have to apply the one-electron (or hydrogenic), bound state 
wave functions $| n_{r}, \nu, m \rangle$ and/or $| \kappa, \nu, m \rangle$ in three-dimensional 
pseudo-Euclidean space, which have explicitly been derived (and described) in the Appendix C. 
If we apply these wave functions (or $| n_{r}, \nu, m \rangle$ states) with the `correctly' 
chosen physical parameters, then six (of eight) equations from Eq.(\ref{MainEqPSE}) essentially 
become identities. The actual problems arise with the fifth and sixth equations, where we have to 
diagonalize simultaneously the two following operators: $\hat{\Lambda}_{1} = \hat{\bf L}^{2}_{1} 
+ R \; (\hat{N}_{1})_{z}$ and $\hat{\Lambda}_{2} = \hat{\bf L}^{2}_{2} + R \; (\hat{N}_{2})_{z}$.     

For the two operators $\hat{\Lambda}_{1}$ and $\hat{\Lambda}_{2}$ we need to solve (simultaneously) 
their eigenvalue problems in the basis of one-electron (or hydrogenic), bound state functions 
$| \kappa_{1}, \nu_{1}, m_{1} \rangle$ and $| \kappa_{2}, \nu_{2}, m_{2} \rangle$, respectively. 
The explicit form of the $\hat{\Lambda}_{1}$ and $\hat{\Lambda}_{2}$ operators in these basis sets 
can be found relatively easy, if one takes into account the equations: $\hat{\bf L}^{2}_{i} \; 
| \kappa_{j}, \nu_{j}, m_{j} \rangle = \delta_{ij} \; \nu_{i} ( \nu_{i} + 1 ) \; | \kappa_{j}, 
\nu_{j}, m_{j} \rangle$ for $i \ne j = (1, 2)$. However, any explicit derivation of analytical 
formulas for $z-$components of the both $\hat{\bf N}_{1}$ and $\hat{\bf N}_{2}$ vector-operators 
contains some non-trivial steps. Here we have to use our formula, Eq.(\ref{AZ}), where all Cartesian 
coordinates are replaced according to the formula, Eq.(\ref{hypspher}). Finally, we obtain the two 
differential equations written in variables $r, \theta$ and $\phi$. In the basis of products of the 
one-electron (hydrogenic), bound state functions $| \kappa_{1}, \nu_{1}, m_{1} \rangle | \kappa_{2}, 
\nu_{2}, m_{2} \rangle$ one finds two infinite-dimensional matrices which represent the differential 
$\hat{\Lambda}_{1}$ and$\hat{\Lambda}_{2}$ operators, respectively. To simplify the arising equations
one has to apply the two following conditions $m^{2}_{1} = m^{2}_{2}$ and $\frac{(Q_1 + Q_{2})^{2}}{2 
\; \kappa^{2}_{1}} = \frac{(Q_1 - Q_{2})^{2}}{2 \; \kappa^{2}_{2}}$, which have been derived above. 

Unfortunately, in this case we cannot apply the same direct (or algebraic) approach, which was used 
to solve this problem for the two hydrogen-like ions with discrete spectra. The reason is obvious, 
since our current knowledge of the properties of coefficients of vector coupling for the 
representations of non-compact $SO(2,1)-$group are still at a quite primitive level. Briefly, in this 
area of research the new `Giulio Racah' is desperately needed which will help us to solve a number of 
actual problems. In particular, we can derive some closed analytical expressions for the energy terms 
$E(R)$ in an arbitrary one-electron, two-center (Coulomb) problem $(Q_1, Q_2)$ which will be 
represented as a function of one (real) quantum number, e.g., $\kappa_{1}$, only. 

\section{Discussion and Conclusions}

We have considered analytical and numerical calculations of the matrix elements of vector physical 
quantities which are represented by some finite sets of the vector (Hermitian) operators. In this 
study we dealt only with systems that have an explicit spherical symmetry. This problem is of 
constant interest in Quantum mechanics, since its early days (see, e.g., \cite{BHJ}). In general, 
any new analysis of this `old' quantum problem brings us some new important, useful and often 
unexpected results.  

Our analysis begins from the approach which is widely known as the second quantization for angular 
momentum. Currently, this approach and its numerous derivations are extensively used for solving 
a large number of actual problems in theoretical physics. Then we discuss the method which has 
been developed to determine the multiple commutators of different even powers of momenta 
$\hat{\bf J}^{2}$ and vector-operator $\hat{\bf A}$ in the general form. Analytical calculations 
of some expectation values averaged over partial angular momenta is also considered in this study. 
In early years of Quantum Mechanics this method was an effective and useful tool, which has often 
been applied to various problems. 

We also discuss the current status of pure algebraic methods which were developed for analytical 
solutions of the Coulomb two-center problem for one-electron atoms and ions. Remarkably, but for 
similar three-body systems it is possible to obtain a complete set of hydrogen-like solutions (or 
one-electron wave functions known for hydrogen atom and hydrogen-like ions in our usual Euclidean 
and pseudo-Euclidean spaces. The corresponding approach is based on a few transparent and obvious 
ideas tested for one-electron, hydrogen-like, two-body systems.  

\appendix
 \section{On the mixed product of three vectors}
\label{A} 

In this Appendix we consider one of the problems which arises during generalization of the mixed 
product of three vectors to the case of three vector-operators. First, we note that as is well 
known from vector analysis for arbitrary six vectors ${\bf a}, {\bf b}, {\bf c}, {\bf d}, {\bf e}$ 
and ${\bf f}$ we can write the following formula (see, e.g., Eq.(41) from \$ 8, Chapter 1 in 
\cite{Kochin})
\begin{eqnarray}      
 [{\bf a} \cdot ({\bf b} \times {\bf c})] [{\bf d} \cdot ({\bf e} \times {\bf f})] =
 \left| \begin{array}{ccc} {\bf a} \cdot {\bf d} \; & \; {\bf b} \cdot {\bf d} \; & \; {\bf c} \cdot {\bf d} \\
                           {\bf a} \cdot {\bf e} \; & \; {\bf b} \cdot {\bf e} \; & \; {\bf c} \cdot {\bf e} \\
                           {\bf a} \cdot {\bf f} \; & \; {\bf b} \cdot {\bf f} \; & \; {\bf c} \cdot {\bf f} \\
  \end{array} \right| \; \; . \; \label{A1}
\end{eqnarray} 
If now we assume that ${\bf d} = {\bf a}, {\bf e} = {\bf b}$ and ${\bf f} = {\bf c}$, then from the 
last equation it is easy to find that 
\begin{eqnarray}      
 [{\bf a} \cdot ({\bf b} \times {\bf c})]^{2} =
 \left| \begin{array}{ccc} {\bf a}^{2} \; & \; {\bf a} \cdot {\bf b} \; & \; {\bf a} \cdot {\bf c} \\
                           {\bf a} \cdot {\bf b} \; & \; {\bf b}^{2} \; & \; {\bf b} \cdot {\bf c} \\
                           {\bf a} \cdot {\bf c} \; & \; {\bf b} \cdot {\bf c} \; & \; {\bf c}^{2} \\
  \end{array} \right| \; \; . \; \label{A2}
\end{eqnarray} 
Now, by applying this formula, Eq.(\ref{A2}), to the case when ${\bf a} = {\bf J}, {\bf b} = {\bf J}$ and 
${\bf c} = {\bf A}$ we obtain  
\begin{eqnarray}      
 [ {\bf J} \cdot ({\bf J} \times {\bf A})]^{2} =
 \left| \begin{array}{ccc} {\bf J}^{2} \; & \; {\bf J}^{2} \; & \; {\bf J} \cdot {\bf A} \\
                           {\bf J}^{2} \; & \; {\bf J}^{2} \; & \; {\bf J} \cdot {\bf A} \\
                 {\bf J} \cdot {\bf A} \; & \; {\bf J} \cdot {\bf A} \; & \; {\bf A}^{2} \\
  \end{array} \right| = 0 \; \; , \; \label{A2A}
\end{eqnarray} 
since this matrix has two identical lines (1st and 2nd lines) and its determinant equals zero identically. 
Analogously, if in Eq.(A2) we choose ${\bf a} = {\bf J}, {\bf b} = {\bf A}$ and ${\bf c} = {\bf J}$, then 
one finds 
\begin{eqnarray}      
 [ {\bf J} \cdot ({\bf A} \times {\bf J})]^{2} =
 \left| \begin{array}{ccc} {\bf J}^{2} \; & \; {\bf J} \cdot {\bf A} \; & \; {\bf J}^{2} \\
            {\bf J} \cdot {\bf A} \; & \; {\bf A}^{2} \; & \; {\bf J} \cdot {\bf A} \\
                          {\bf J}^{2} \; & \; {\bf J} \cdot {\bf A} \; & \; {\bf J}^{2} \\
  \end{array} \right| = 0 \; \; , \; \label{A2B}
\end{eqnarray} 
since this matrix has two identical lines (1st and 3rd lines) and its determinant equals zero. It seems 
obvious to assume that we just proved that $\hat{\bf J} \cdot (\hat{\bf J} \times \hat{\bf A}) = 0$ and 
$\hat{\bf J} \cdot (\hat{\bf A} \times \hat{\bf J}) = 0$ for the two vector-operators. However, this is 
not true, since by using the fundamental $[ \hat{J}_{i}, \hat{A}_{j}] = \imath \varepsilon_{ijk} 
\hat{A}_{k}$ relation for the vector-operators in Quantum Mechanics it is easy to show that $\hat{\bf J} 
\cdot (\hat{\bf J} \times \hat{\bf A}) = \imath (\hat{\bf J} \cdot \hat{\bf A})$ and $\hat{\bf J} \cdot 
(\hat{\bf A} \times \hat{\bf J}) = - \imath (\hat{\bf A} \cdot \hat{\bf J})$. Therefore, we also find 
that $\hat{\bf J} \cdot [\hat{\bf J}^{2}, \hat{\bf A}] = - 2 \; (\hat{\bf J} \cdot \hat{\bf A}) = - 2 
\; (\hat{\bf A} \cdot \hat{\bf J})$. In general, none of the two scalar products mentioned in the last 
equation equals zero. This result is a clear indication of troubles which one faces by trying to 
generalize the definition of mixed product of three vector-operators to Quantum Mechanics. 

\section{Explicit derivation of the normalization constant in Eq.(\ref{eq9})}
\label{B} 

Now, let us obtain the numerical value of the `normalization' constant $N$ in the formula, Eq.(\ref{eq9}) 
from the main text. For these purposes we multiply the equation 
\begin{eqnarray}
 0 = \frac13 \; \hat{\ell}_{i} \hat{\ell}_{i} + A \Bigl[ \hat{\ell}_{i} \hat{\ell}_{i} \hat{\ell}_{j} 
 \hat{\ell}_{j} + \hat{\ell}_{i} \hat{\ell}_{j} \hat{\ell}_{i} \hat{\ell}_{j} - \frac23 \; \ell 
 (\ell + 1) \; \hat{\ell}_{i} \hat{\ell}_{i} \Bigr] \; \; \label{AF1}  
\end{eqnarray}
on the left by $\hat{\bf \ell}_{i}$ and on the right by $\hat{\bf \ell}_{j}$. This equation is almost 
identical to Eq.(\ref{eq9}), but contains an unknown `normalization' factor $A$. Since the vector 
$\hat{\bf \ell} = {\bf r} \times {\bf p}$ is orthogonal to the vector ${\bf n} = \frac{{\bf r}}{r}$, we 
can write the two additional identities: $n_{i} \ell_{i} = 0$ and $\ell_{j} n_{j} = 0$. These identities 
allow us to reduce Eq.(\ref{AF1}) to the form 
\begin{eqnarray}
 - \frac13 \; \hat{\ell}^{2} = A \Bigl[\Bigl(\hat{\ell}^{2}\Bigr)^{2} + \hat{\ell}_{i} \hat{\ell}_{j} 
  \hat{\ell}_{i} \hat{\ell}_{j} - \frac23 \ell (\ell + 1) \; \hat{\ell}^{2} \Bigr] = 
  A \Bigl\{ 2 \ell^{2} (\ell + 1)^{2} + \hat{\ell}_{i} [\hat{\ell}_{j}, \hat{\ell}_{i}] \hat{\ell}_{j} 
 - \frac23 \ell^{2} (\ell + 1)^{2} \Bigr\} , \nonumber 
\end{eqnarray}
where we make an obvious substitution $\hat{\ell}^{2} = \ell (\ell + 1)$. Since $[\hat{\ell}_{j}, 
\hat{\ell}_{i}] = \imath \varepsilon_{jik} \hat{\ell}_{k}$, then we can write the last equation in the 
form 
\begin{eqnarray}
 - \; \ell (\ell + 1) = A \Bigl[ 4 \; \ell^{2} (\ell + 1)^{2} + 3 \imath \varepsilon_{jik} 
 \hat{\ell}_{i} \hat{\ell}_{k} \hat{\ell}_{j} \Bigr] \; , \; \label{AF3} 
\end{eqnarray}

In order to determine the second term ($\simeq \varepsilon_{jik} \hat{\ell}_{i} \hat{\ell}_{k} 
\hat{\ell}_{j}$) on the right side of this equation we apply the following trick. First, we re-write 
this expression twice  
\begin{eqnarray}
 \varepsilon_{ikj} \hat{\ell}_{i} \hat{\ell}_{k} \hat{\ell}_{j} = \frac12 \varepsilon_{ikj} 
 \hat{\ell}_{i} \hat{\ell}_{k} \hat{\ell}_{j} + \frac12 \varepsilon_{ikj} \hat{\ell}_{i} 
 \hat{\ell}_{k} \hat{\ell}_{j} = \frac12 \varepsilon_{ikj} \hat{\ell}_{i} \hat{\ell}_{k} 
 \hat{\ell}_{j} + \frac12 \varepsilon_{ijk} \hat{\ell}_{i} \hat{\ell}_{j} \hat{\ell}_{k} 
 \; , \; \label{AF4} 
\end{eqnarray}
where the first equation is a formal re-writing, while in the last term from the second equation we 
re-designate the two dummy indexes $k \leftrightarrow j$. At the second step of our transformations, 
in the last equation we simply interchange the two indexes in the second $\varepsilon_{ijk}$ symbol 
by using the rule $\varepsilon_{ijk} = - \varepsilon_{ikj}$ and obtain 
\begin{eqnarray}
 \imath \varepsilon_{ikj} \hat{\ell}_{i} \hat{\ell}_{k} \hat{\ell}_{j} = \frac12 \varepsilon_{ikj} 
 \hat{\ell}_{i} \hat{\ell}_{k} \hat{\ell}_{j} + \frac12 \varepsilon_{ijk} \hat{\ell}_{i} \hat{\ell}_{j} 
 \hat{\ell}_{k} = \imath \frac12 \varepsilon_{ikj} \hat{\ell}_{i} \Bigl(\hat{\ell}_{k} \hat{\ell}_{j} - 
 \hat{\ell}_{j} \hat{\ell}_{k}\Bigr) = \imath \frac12 \varepsilon_{ikj} \hat{\ell}_{i} [ \hat{\ell}_{k}, 
 \hat{\ell}_{j}] \; . \; \label{AF5} 
\end{eqnarray}
For the Poisson bracket we write $[ \hat{\ell}_{k}, \hat{\ell}_{j}] = \imath \varepsilon_{kjn} \hat{\ell}_{n}$. 
The final result equals 
\begin{eqnarray} 
 - \frac12 \varepsilon_{ikj} \varepsilon_{kjn} \hat{\ell}_{i} \hat{\ell}_{n} = - \frac12 \varepsilon_{kji} 
 \varepsilon_{kjn} \hat{\ell}_{i} \hat{\ell}_{n} = - \frac12 (2 \delta_{in}) \hat{\ell}_{i} \hat{\ell}_{n} 
 = - \hat{\ell}^{2} = - \ell (\ell + 1) \; , \; \label{AF6} 
\end{eqnarray}
where we used the relation $\varepsilon_{kji} \varepsilon_{kjn} = 2 \delta_{in}$ (see, e.g., \cite{Kochin}). 

Finally, the equation, Eq.(\ref{AF3}), takes the form 
\begin{eqnarray}
 - \; \ell (\ell + 1) = A \Bigl[ 4 \; \ell^{2} (\ell + 1)^{2} - 3 \ell (\ell + 1) \Bigr] \; . \; \label{AF7} 
\end{eqnarray}
By reducing both sides of this equation by the factor $\ell (\ell + 1)$, one finds  
\begin{eqnarray}
 - 1 = A \Bigl[ 4 \; \ell (\ell + 1) - 3 \Bigr] = A (2 \ell - 1) (2 \ell + 3) \; , \; \label{AF8} 
\end{eqnarray}
or in other words, for the `normalization' constant $A$ we obtain $A = - \frac{1}{(2 \ell - 1) (2 \ell + 3)}$, 
i.e., exactly the same numerical value which is used in the main text (see, e.g., Eqs.(\ref{eq9}) and 
(\ref{eq91})). 

\section{Hydrogen-like ions in pseudo-Euclidean space}
\label{C} 

In this Appendix we consider the hydrogen atom and hydrogen-like ions in three-dimensional pseudo-Euclidean 
space. This part seems absolutely necessary, since currently there are many physicists, who are ready to 
spend hours discussing some tiny details of solving the non-relativistic Sch\"{o}dinger equation for usual 
hydrogen-like atomic systems, but become absolutely helpless when similar problem is considered in 
three-dimensional pseudo-Euclidean space. In reality, the hydrogen-like atomic systems in three-dimensional 
pseudo-Euclidean space are already a necessary part of the modern physical curriculum. Here we briefly 
discuss solutions of the non-relativistic Sch\"{o}dinger equation for the hydrogen atom in three-dimensional 
pseudo-Euclidean space. Our goal is to show that this process is straightforward and does not contain any 
extremely complicated steps and/or does not produce any cumbersome results. These two facts is crucially 
important for success of our method which is developed in the end of Section VI from the main text. Also,
we want to make our analysis as brief as possible. Therefore, some important features of general solutions 
can be lost.  

First, consider the derivation of angular part $\Psi_{A}$ of the total wave function $\Psi$ of the one-electron 
hydrogen atom (or hydrogen-like ion) in three-dimensional pseudo-Euclidean space. In Eq.(\ref{LLLPES}) from the 
main text we replace Cartesian variables by using our spherical coordinates defined in Eq.(\ref{hypspher}) below. 
The new formulas for these three operators take the form: 
\begin{eqnarray}
  L_{x} = \sin \phi \; \frac{\partial}{\partial \theta} + \coth \theta \; \cos \phi \; 
  \frac{\partial}{\partial \phi} \; , \; 
   L_{y} = - \cos \phi \; \frac{\partial}{\partial \theta} + \coth \theta \; \sin \phi \; 
  \frac{\partial}{\partial \phi} \; , \; 
   L_{z} = - \imath \frac{\partial}{\partial \phi} \; . \; \label{AppC1} 
\end{eqnarray}
From here one finds 
\begin{eqnarray}
 L_{+} =  L_{x} + \imath L_{y} = - \imath \exp (\imath \phi) \; \frac{\partial}{\partial \theta} + 
 \coth \theta \; \exp (\imath \phi) \; \frac{\partial}{\partial \phi} \; \; \; 
 \label{AppC+} 
\end{eqnarray}
and  
\begin{eqnarray}
 L_{-} =  L_{x} - \imath L_{y} = \imath \exp (- \imath \phi) \; \frac{\partial}{\partial \theta} + 
 \coth \theta \; \exp (- \imath \phi) \; \frac{\partial}{\partial \phi} \; \; , \; 
 \label{AppC-} 
\end{eqnarray}
where we have applied (twice) the De Moivre's formula. By using these expressions for the $L_{x}, L_{y}$ and 
$L_{z}$ operators, one can reproduce the commutation relations, Eq.(\ref{PBracket1}), from the main text. As 
directly follows from these commutation relations the arising three-operator algebra is $SO(2,1)$ (see, e.g., 
\cite{Perelom}). The Casimir operator of this algebra $C_{2}$ equals 
\begin{eqnarray}
 C_{2} = L_{+} \; L_{-} + L^{2}_{z} + L_{z} = \frac{1}{\sinh^{2} \theta} \; \frac{\partial^{2}}{\partial 
  \phi^{2}} \; + \; \frac{1}{\sinh \theta} \; \frac{\partial}{\partial \theta} \Bigl( \sinh \theta \; 
  \frac{\partial}{\partial \theta}\Bigr) \; \; , \; \label{AppCas} 
\end{eqnarray}
The general theory of group representation (see, e.g., \cite{BarR} and references therein) predicts that this 
Casimir operator is a numerical constant which we shall designate below as $\nu (\nu + 1)$, where $\nu$ is a 
real number (in general, $\nu$ can be a complex number). Thus, for the angular part $\Psi_{A}$ of the total 
wave function $\Psi$ we have the following Sch\"{o}dinger equation $C_2 \Psi_{A} = {\bf L}^{2} \Psi_{A} = \nu 
(\nu + 1 ) \Psi_{A}$ and one additional condition $L^{2}_{z} \Psi_{A}= m^{2} \Psi_{A}$. Solutions of these 
two equations in Euclidean space are well known. In order to describe them in a compact form let us consider 
the differential equation which determines the associated Legendre polynomials $P^{m}_{\ell}(\theta)$ 
\begin{eqnarray}
  \frac{1}{\sin \theta} \; \frac{d}{d \theta} \; \Bigl[ \sin \theta \; \frac{d P^{m}_{\ell}(\theta)}{d \theta} 
  \Bigr] \; + \; \Bigl[ \ell (\ell + 1) - \frac{m^{2}}{\sin^{2} \theta} \Bigr] P^{m}_{\ell}(\theta) = 0 \; \; 
  . \; \; \; \label{APC0} 
\end{eqnarray}
where $\ell$ and $m$ are integer and $- \ell \le m \le \ell$. This differential equation plays a central role 
for all quantum systems with central interaction potentials in three-dimensional Euclidean space. 

Now, by replacing in Eq.(\ref{APC0}) the variable $\theta$ with new variable $\imath \theta$ we obtain the new 
equation, which is applicable in three-dimensional pseudo-Euclidean space:  
\begin{eqnarray}
  \Bigl\{ - \frac{d}{d (\cosh \theta)} \; \Bigl[ \sinh^{2} \theta \; \; \frac{d}{d (\cosh \theta)} \Bigr] \; 
  + \; \Bigl[ \nu (\nu + 1) - \frac{m^{2}}{\cosh^{2} \theta - 1} \Bigr] \Bigr\} Q^{| m |}_{\nu}(\cosh \theta) 
  = 0 \; , \; \label{APC} 
\end{eqnarray}
where $| m |$ and $m^{2}$ are the non-negative integer, while $\nu$ is an arbitrary, in principle, real number. 
This new equation has some similarities with Eq.(\ref{APC0}), but its internal structure and properties are 
quite different. In particular, its variable $\cosh \theta$ varies between $+1$ and $+\infty$, i.e., now it is 
non-compact variable. Therefore, we cannot assume that the parameter $\nu$ in Eq.(\ref{APC}) can still be 
integer (in contrast with $\ell$ in Eq.(\ref{APC0})). Moreover, the notation $Q^{| m |}_{\nu}(\cosh \theta)$ 
in Eq.(\ref{APC}) stands for some new angular functions which have no direct relation(s) with the associated 
Legendre polynomials $P^{m}_{\ell}(\theta)$ mentioned above.  

After a few additional transformations one reduces this equation to the following form 
\begin{eqnarray}
 \Bigl\{ ( 1 - \cosh^{2} \theta ) \; \frac{d^{2} }{d (\cosh \theta)^{2}} - 2 \; \cosh \theta \;  
 \frac{d }{d (\cosh \theta)} + \nu ( \nu + 1 ) - \frac{m^{2}}{1 -  \cosh^{2} \theta} \Bigr\} \; 
 Q^{| m |}_{\nu}(\cosh \theta) = 0 ,  \nonumber %\label{APC1} 
\end{eqnarray}
or 
\begin{eqnarray}
 \Bigl\{ ( 1 - w^{2} ) \; \frac{d^{2} }{d w^{2}} - 2 \; w \;  \frac{d }{d w} + \nu ( \nu + 1 ) - 
 \frac{m^{2}}{1 -  w^{2}} \Bigr\} \; Q^{| m |}_{\nu}(w) = 0 \; , \; \label{APC1} 
\end{eqnarray}
where $w = \cosh \theta$. These two equations exactly coincide with the differential equation for the 
associated Legendre functions (see, e.g., Eq.(8.700) on page 1013 in \cite{GR}). Solutions of this equation
are well known (see, e.g., \cite{GR} and \cite{AS}). Indeed, the corresponding unit-norm functions 
$Q^{| m |}_{\nu}(\cosh \theta)$ are written in the form:  
\begin{eqnarray}
  Q^{| m |}_{\nu}(\cosh \theta) &=& \frac{ \exp(\imath \; \pi \; | m |) \sqrt{\pi} \; 
  \Gamma\Bigl( \nu + | m | + 1 \Bigr)}{2^{\nu + 1} \; \Gamma\Bigl( \nu + \frac32 \Bigr)} \; \; 
  \frac{(\tanh \theta)^{\frac{| m |}{2}}}{(\cosh \theta)^{\nu + \frac{| m |}{2} + 1}} \; \times \nonumber \\
  & & {}_{2}F_{1}\Bigl( \frac{ \nu + | m | + 1}{2} , \frac{ \nu + | m | + 2}{2} ; \nu + \frac32 ; 
    \frac{1}{(\cosh \theta)^{2}} \Bigr) \; , \; \label{APC2}  
\end{eqnarray}
where $\tanh \theta = \frac{\sinh \theta}{\cosh \theta}, \Gamma(x)$ is the Euler's integral of the first kind 
defined exactly as in Eq.(8.310) from \cite{GR}. Also in this equation the notation $\; {}_{2}F_{1}(a, b ; c ; 
x)$ is the usual (2,1)-hypergeometric function (see, e.g., \cite{GR} and \cite{AS}). 

In general, the associated Legendre functions $Q^{| m |}_{\nu}(x)$, where $m$ is integer, can be derived from 
the `zero' Legendre function $Q^{0}_{\nu}(x) = Q_{\nu}(x)$ by using the relation:
\begin{eqnarray}
 Q^{| m |}_{\nu}(x) = (-1)^{| m |} \; (1 - x^{2})^{\frac{| m |}{2}} \; \frac{d^{| m |}}{d x^{| m |}} \; 
 Q_{\nu}(x) \; , \; \label{APC3}  
\end{eqnarray}
where $| m |$ is a positive integer. Another important relation, which is used often in various applications,  
is 
\begin{eqnarray}
 Q^{-m}_{\nu}(x) = (-1)^{| m |} \; \frac{\Gamma( \nu - | m | + 1 )}{\Gamma( \nu + | m | + 1 )} \; 
 Q^{| m |}_{\nu}(x) \; . \; \label{APC33}  
\end{eqnarray}
This equation is used to produce the associated Legendre functions $Q^{-m}_{\nu}(x)$ with negative upper indexes. 
Now, it is clear that the angular basis set $Z^{m}_{\nu}(\theta, \phi) = Q^{m}_{\nu}(\theta) \; 
\frac{1}{\sqrt{2 \pi}} \; \exp(\imath \; m \; \phi)$ is a complete system of angular functions. In other words, 
in three-dimensional pseudo-Euclidean space the basis functions $Z^{m}_{\nu}(\theta, \phi) = Q^{m}_{\nu}(\theta) 
\; \frac{1}{\sqrt{2 \pi}} \; \exp(\imath \; m \; \phi)$ play the same role which the spherical 
$Y^{m}_{\ell}(\theta, \phi)$ harmonics play in our three-dimensional Euclidean space.  

Now, we are ready to describe analytical solutions of the non-relativistic Sch\"{o}dinger equation for the 
hydrogen atom and hydrogen-like one-electron ions in three-dimensional pseudo-Euclidean space. In Cartesian 
coordinates this equation takes the form 
\begin{eqnarray}
  & &\hat{H} \; \psi(x, y, z) = \Bigl( -\frac{\hbar^{2}}{2 \mu} \Delta - \frac{{Q e^{2}}}{r} \Bigr) 
  \psi(x, y, z) = \Bigl[ -\frac{\hbar^{2}}{2 \mu} \Bigl( - \frac{\partial^{2}}{\partial x^{2}} - 
  \frac{\partial^{2}}{\partial y^{2}} + \frac{\partial^{2}}{\partial z^{2}} \Bigr) \nonumber \\
 & &- \frac{Q e^{2}}{\sqrt{|- x^{2} - y^{2} + z^{2}|}} \Bigr] \; \psi(x, y, z) = E \; \psi(x, y, z) \; \; 
 , \; \label{APCSchrod}
\end{eqnarray} 
where $\hat{H}$ is the Hamiltonian, $E$ is its eigenvalue which is called the energy (or total energy) of the 
hydrogen atom. Also in this equation $\mu = m_e \Bigl( 1 + \frac{1}{M} \Bigr)$ is the reduced mass of the 
electron, $M$ is the nuclear mass expressed in the electron's mass ($M \gg 1$), while $e$ is the elementary 
electric charge. The explicit form of Eq.(\ref{APCSchrod}) tells us about the preferable form of spherical 
coordinates in this case
\begin{eqnarray}
  x = r \; \sinh \theta \; \cos \phi \; \; \; , \; \; \; y = r \; \sinh \theta \; \sin \phi \; \; \; , 
  \; \; \; z = r \; \cosh \theta \; . \; \label{hypspher}
\end{eqnarray} 
where $r$ is the radius, $\theta$ and $\phi$ are the polar and azimuthal angles, respectively. In these 
coordinates one finds $- x^{2} - y^{2} + z^{2} = r^{2}$, while the radial part of the non-relativistic 
Sch\"{o}dinger equation is written in a familiar form 
\begin{eqnarray}
 - \frac{\hbar^{2}}{2 \mu} \; \frac{1}{r^{2}} \; \frac{d}{d r} \Bigl( r^{2} \frac{d R}{d r} \Bigr)
 - \frac{{Q \; e^{2}}}{r} \; R + \frac{\hbar^{2}}{2 \mu} \; \frac{\nu (\nu + 1)}{r^{2}} \; R = 
 E \; R \; \; , \; \; \label{APCSchrodR}
\end{eqnarray} 
where $R(r)$ is the radial part of the total wave function. In fact, this equation exactly coincides with 
the Sch\"{o}dinger equation known for the hydrogen atom in three-dimensional Euclidean space (see, e.g., 
\cite{Shiff}). Following \cite{Shiff} we introduce the two following parameters $\beta, \lambda$ and one 
dimensionless variable $\rho$
\begin{eqnarray}
  \beta = \frac{\sqrt{8 \; \mu \; | E |}}{\hbar} \; \; \; , \; \; \lambda = \frac{Q \; e^{2}}{\hbar} 
  \; \sqrt{\frac{\mu}{2 | E |}} \; \; , \; \; \rho = \beta \; r \; \; . \; \label{APCparam}
\end{eqnarray}

After substitution of these parameters and variable $\rho$ the differential equation for radial part of 
the total wave function takes a different form   
\begin{eqnarray}
 - \frac{1}{\rho^{2}} \; \frac{d}{d \rho} \Bigl( \rho^{2} \frac{d R}{d \rho} \Bigr) + 
 \Bigl[ \frac{\lambda}{\rho} - \frac14 - \frac{\nu (\nu + 1)}{\rho^{2}} \Bigr] \; R = 0 \; \; , \; \; 
 \label{APCradial}
\end{eqnarray} 
where $R = R(\rho)$ is the unknown radial function. It was shown in \cite{Shiff} that the correct radial
function $R(\rho)$ is represented in the form $R(\rho) = \rho^{s} \; \Phi(\rho) \; \exp(- \frac12 \rho)$. 
In this case after a few additional transformations we obtain the formula  
\begin{eqnarray}
 \rho^{2} \; \frac{d^{2} \Phi(\rho)}{d \rho^{2}} \; + \; \rho \; [ 2 (s + 1) - \rho ] \; 
 \frac{d \Phi(\rho)}{d \rho} + [ \rho (\lambda - s - 1) + s (s + 1) - \nu (\nu + 1)] \Phi(\rho) = 0 
 \; \; . \; \; \label{APCradial2}
\end{eqnarray} 
If $\rho$ is set equal to zero in this equation, then as follows from Eq.(\ref{APCradial2}) the following 
equation must be obeyed: $s (s + 1) - \nu (\nu + 1) = 0$. From here one finds that $s = \nu$ and/or $s = 
- \nu - 1$. First, let assume that $\nu \ge 0$. In this case we have to chose $s = \nu$ and 
Eq.(\ref{APCradial2}) is reduced to the form
\begin{eqnarray}
 \rho \; \frac{d^{2} \Phi(\rho)}{d \rho^{2}} \; + [ 2 (\nu + 1) - \rho ] \; \frac{d  \Phi(\rho)}{d \rho} 
 - ( \nu + 1 - \lambda ) \; \Phi(\rho) = 0 \; \; . \; \; \label{APCradial3}
\end{eqnarray}  
The actual radial function $\Phi(\rho)$ must be square integrable when the radial variable $\rho$ varies 
between zero and infinity, or for $0 \le \rho < + \infty$. The appropriate solution of this equation does 
exist if (anf only if) the following condition is obeyed: $- \lambda + \nu + 1 = - n_{r}$, or $\lambda = 
n_{r} + \nu + 1$. Here and below $n_{r} = 0, 1, 2, \ldots$ is the radial quantum number (also excitation 
number) which is non-negative integer. This equation is also called the main quantum constraint (or 
condition) between all essential quantum numbers in hydrogen atom. This condition determines the following 
explicit formula for the energy spectrum: 
\begin{eqnarray}
  E = - \frac{1}{1 + \frac{m_e}{M}} \; \Bigl(\frac{e^{4}}{m_e \; \hbar^{2}}\Bigr) \; 
  \frac{Q^{2}}{2 \; ( n_{r} + \nu + 1 )^{2}} = - \Bigl(\frac{Ry}{1 + \frac{m_e}{M}}\Bigr) \; 
  \frac{Q^{2}}{2 \; ( n_{r} + \nu + 1 )^{2}} \; \; , \; \; \label{APCradial4}
\end{eqnarray}
where $Ry = \frac{e^{4}}{m_e \; \hbar^{2}}$ is the Rydberg constant and it is assumed that all energies of 
bound states are negative, i.e., here and  everywhere below we have $E < 0$. This formula coincides with 
the well known Bohr's formula for the energy spectrum of the usual hydrogen atom in three-dimensional 
Euclidean space. However, in Eq.(\ref{APCradial3}) the quantum number $\nu$ is a non-negative real number. 
For these bound states the radial part of the total wave function is written in the form 
\begin{eqnarray}
  R(\rho) = \rho^{\; \nu} \; {}_{1}F_{1}(- n_{r}, 2 \nu + 2; \rho) \; \exp(- \frac12 \rho) = 
  \rho^{\; \nu} \; \exp(- \frac12 \rho) \; D^{n_r}_{n_r + 2 \nu + 1} \; L^{2 \nu + 1}_{n_r}(\rho) 
  \; \; , \; \; \label{APCradial5}
\end{eqnarray} 
where the notation ${}_{1}F_{1}( a, b; z)$ designates the confluent hypergeometric function (see, e.g., 
\cite{GR}, \cite{AS}), another notation $L^{a}_{n}(x)$ stands for the Laguerre polynomials \cite{GR}, 
while the numerical coefficients $D^{k}_{x}$ are defined by the formula $D^{k}_{x} = \frac{k!}{x (x - 1) 
\ldots (x - k + 1)}$. To obtain the final answer we have to replace in this equation $\rho = \beta \; r$. 

For the hydrogen atom in three-dimensional Euclidean space the two formulas, Eqs.(\ref{APCradial4}) and 
(\ref{APCradial5}) where $\nu = \ell$, provide the final answer. However, in three-dimensional 
pseudo-Euclidean space we have to consider the second case, when the quantum number $\nu$ is negative, but 
the sum $- \nu - 1$ is non-negative. In this case instead of our Eq.(\ref{APCradial3}) one obtains the 
following differential equation 
\begin{eqnarray}
 \rho \; \frac{d^{2} \Phi(\rho)}{d \rho^{2}} \; + [ - 2 \nu - \rho ] \; \frac{d \Phi(\rho)}{d \rho} 
 - ( - \nu - \lambda ) \; \Phi(\rho) = 0 \; \; . \; \; \label{APCradial3}
\end{eqnarray}  
This leads us to the following spectral equation: $- \lambda - \nu = - n_{r}$, or $\lambda = n_{r} - \nu$, 
where $\nu$ is a real negative number. Here and below $n_{r} = 0, 1, 2, \ldots$ is the radial quantum number 
which is non-negative integer. The energy spectrum is described by the formula:
\begin{eqnarray}
  E = - \frac{1}{1 + \frac{m_e}{M}} \; \Bigl(\frac{e^{4}}{m_e \; \hbar^{2}}\Bigr) \; 
  \frac{Q^{2}}{2 \; ( n_{r} - \nu )^{2}} = - \Bigl(\frac{Ry}{1 + \frac{m_e}{M}}\Bigr) \; 
  \frac{Q^{2}}{2 \; ( n_{r} - \nu )^{2}} \; \; . \; \; \label{APCradial45}
\end{eqnarray}
For similar bound states the radial part of the total wave function takes the form 
\begin{eqnarray}
  R(\rho) = \rho^{ - \nu - 1} \; {}_{1}F_{1}(- n_{r}, - 2 \nu ; \rho) \; \exp(- \frac12 \rho) = 
  \rho^{- \nu - 1} \; \exp(- \frac12 \rho) \; D^{n_r}_{n_r - 2 \nu - 1} \; L^{- 2 \nu - 1}_{n_r}(\rho) 
  \; \; , \; \; \label{APCradial51}
\end{eqnarray} 
where the numerical coefficients $D^{k}_{x}$ have been defined above. Further investigation of formulas  
Eq.(\ref{APCradial5}) and Eq.(\ref{APCradial51}) indicate clear that the first formula can be applied in 
those case, when $\nu \ge -\frac12$. In those cases when $- \nu - 1 \ge -\frac12$, or $\nu \le -\frac12$ 
(i.e., for negative $\nu$) we have to use the formula, Eq.(\ref{APCradial51}). This also follows from the 
`boundary condition' $\nu = - \nu - 1$ which predicts the boundary value of $\nu$ for this problem (the 
point where we have to change the radial part of the total wave function. From quantum mechanical point 
of view this is another indication that the square integrability of the radial part of the total wave 
function is more important to make the correct choice of $\nu$ in the formulas above, then its regularity 
at $r = 0$. 

As follows from our formulas derived in this Appendix the spectrum of bound state states of hydrogen atom 
and hydrogen-like ions in three-dimensional pseudo-Euclidean space is a continuous spectrum (continuum of
bound states). Indeed, the both spectral formulas Eqs.(\ref{APCradial4}) and (\ref{APCradial45}) contain 
the parameter $\nu$ which is an arbitrary real number. From these formulas one finds that if $E(\nu, 
n_{r})$ is the energy of some bound state, then all $E(\overline{\nu}, n_{r})$ energies located in the 
interval $\nu - \Delta \nu, \nu + \Delta \nu$ are the actual energies of some other bound states. Here 
the notation $\Delta \nu$ is the difference in the quantum number $\nu$, which can be arbitrary small. 
In other words, any finite energy interval $\nu - \Delta \nu, \nu + \Delta \nu$ contains an infinite 
number of bound states, which cannot be numbered. This means that any finite vicinity of a given bound 
state always contains an infinite number (continuum) of similar bound states. Such an energy spectrum of 
bound states is a real continuum and it does not include any discrete series of bound states. Discrete 
series of bound states are observed for hydrogen-like one-electron ion/atom in three-dimensional 
Euclidean space. This is the first and most important difference between hydrogen-like ions/atoms in 
three-dimensional Euclidean and pseudo-Euclidean spaces. The fact that we can see the discrete atomic 
spectra means that our space is truly Euclidean. 

There are some other differences between these two cases. On the other hand, one easily finds a number 
of similarities between energy spectra and properties of hydrogen-like atomic systems in the 
three-dimensional Euclidean and pseudo-Euclidean spaces. First of all, we note an obvious similarity in 
radial part of the total wave functions. Indeed, our formulas for these `radial' wave functions 
Eqs.(\ref{APCradial5}) and (\ref{APCradial51}) almost coincide with the corresponding radial functions 
of the usual hydrogen atom in three-dimensional Euclidean space. An obvious replacement $\ell 
\rightarrow \nu$ of the integer angular quantum number $\ell$ by the real angular quantum number $\nu$ 
can easily be adopted in our solution scheme. For instance, the `radial' wave functions 
Eqs.(\ref{APCradial5}) and (\ref{APCradial51}) are reduced to the Laguerre polynomials \cite{GR}, which 
means that for the radial part of total wave functions the actual differences between Euclidean and 
pseudo-Euclidean spaces are relatively small and can easy be handled. 

Another similarity is, probably, more interesting for development of the general atomic theory in 
three-dimensional pseudo-Euclidean space. Let us define three following (radial) operators $S, U$ 
and $T$: 
\begin{eqnarray}
  \hat{S} &=& \frac12 \; r \; \Bigl[ p_{r}^{2} + \frac{\nu (\nu + 1)}{r^{2}} + 1 \Bigr] \; \; \; ,
  \; \; \; \hat{T} = p_{r} \; r  \nonumber \\
  \hat{U} &=& \frac12 \; r \; \Bigl[ p_{r}^{2} + \frac{\nu (\nu + 1)}{r^{2}} - 1 \Bigr] \; \; \; , 
  \; \; \; \label{SO21} 
\end{eqnarray}
where $p_{r} = - \imath \Bigl( \frac{d}{d r} + \frac{1}{r} \Bigr)$ is the radial momentum. For this 
operator one fins the two following equations which are useful in applications: $[ p_{r} , r ] = - 
\imath$ and $p^{2}_{r} = - \frac{d^{2}}{d r^{2}} - \frac{2}{r} \; \frac{d}{d r}$. The commutation 
relations between these three operators are: 
\begin{eqnarray}
  [ \hat{S}, \hat{U} ] = \imath \hat{T} \; \; \; , \; \; \; [ \hat{U}, \hat{T} ] = - \imath \hat{S} 
  \; \; \; , \; \; \; [ \hat{T}, \hat{S} ] = \imath \hat{U} \; \; , \; \label{SO21A} 
\end{eqnarray}
which exactly coincide with the commutation relations of the $SO(2,1)$ algebra. Finally, we conclude 
that the three radial operators $\hat{S}, \hat{U}$ and $\hat{T}$ form the $SO(2,1)$ algebra. The 
Casimir operator $\hat{C}_2$ of this algebra equals $\hat{C}_2 = \hat{S}^{2} - \hat{T}^{2} - 
\hat{U}^{2} = \nu ( \nu + 1) \; \hat{I}$, where $\hat{I}$ is the unit operator. The proof of this 
theorem is simple and it is based on direct calculations of three commutators, Eq.(\ref{SO21A}), and 
Casimir operator $\hat{C}_2$. This Casimir operator equals to the corresponding Casimir operator 
$\hat{C}_2$ of the rotation $SO(2,1)-$group which describes rotations in three-dimensional 
pseudo-Euclidean space. Such an exact coincidence of the Casimir operators constructed for two 
different groups was well explained by Moshinsky and Quesne \cite{MoshQues} for the hydrogen-like 
one-electron ions in three-dimensional Euclidean space. Remarkably, but the same result is also 
correct for the hydrogen-like atomic systems in three-dimensional pseudo-Euclidean space. Finally, 
the algebra of `hidden' symmetry of the hydrogen atom in three-dimensional pseudo-Euclidean space 
is $SO(2,1) \oplus SO(2,1)$, which contains six generators, while the group of `hidden' symmetry 
is $SO(2,1) \otimes SO(2,1) = SO(2,2)$. For bound states in hydrogen-like atomic systems, which 
are considered in three-dimensional pseudo-Euclidean space, this non-compact $SO(2,2)$ group plays 
a very important, central role. The same role the compact $SO(4)-$group plays for bound states in 
one-electron hydrogen-like atoms and ions in three-dimensional Euclidean space (see, e.g., 
\cite{Pauli}, \cite{Fock}). 

There is also an obvious similarity in notations of the bound states in hydrogen atoms which are 
considered in the both Euclidean and pseudo-Euclidean (three-dimensional) spaces, and this fact is 
extensively used in Section IV above. First, we note that our notations of bound hydrogenic states 
by the three-components (integer) `vectors' $| n_{r}, \ell, m \rangle$ in the main text were very 
useful. Here we want to show that a similar system of notations can be introduced for one-electron 
atomic systems in three-dimensional pseudo-Euclidean space. In this vector the notation $n_{r}$ is 
a non-negative integer, which is the eigenvalue of $\hat{S}$ operator Eq.(\ref{SO21}). It is also 
called the radial quantum number, or number of radial excitations. The second index $\ell$ is the 
angular quantum number which is also non-negative integer. The last (third) `coordinate' (or index) 
in the vector $| n_{r}, \ell, m \rangle$ notation is the magnetic quantum number. The total energy 
of any hydrogen-like one-electron system equals $E_{n} = - \frac{Q^{2}}{2 \; (n_{r} + \ell + 1)^{2}} 
= - \frac{Q^{2}}{2 \; n^{2}}$ (in atomic units), where $n = n_{r} + \ell + 1$ is the principal 
quantum number. 

This system of notations can be generalized to the three-dimensional pseudo-Euclidean space. Briefly, 
the bound state wave functions are designated as the triplet of quantum numbers $| n_{r}, \nu, m 
\rangle$ and/or $| \kappa, \nu, m \rangle$, where the indexes (or quantum numbers) $n_{r} \ge 0$ and 
$m$ are integer and have the same meaning as above. Two other indexes used in these notations $\nu$ 
and $\kappa$ are real. Here $\nu$ an arbitrary real number, while $\kappa = n_{r} + \nu + 1$ (for 
$\nu \ge -\frac12$) and $\kappa = n_{r} - \nu$ (for $\nu \le -\frac12$). The total energy of any 
hydrogen-like one-electron system equals $E_{\kappa} = - \frac{Q^{2}}{2 \; \kappa^{2}}$. This system 
is important to understand derivation of equations in the end of Section VI.


\begin{thebibliography}{99}
\bibitem{KS} The Kroneker symbol $\delta(a,b) = \delta_{ab}$ is defined as follows $\delta(a,b) = \delta_{ab} 
= 1$, if $b = a$, and $\delta(a,b) = \delta_{ab} = 0$, if $b \ne a$.
\bibitem{Lapl} The conservation of this vector in celestial mechanics was shown by P.-S. Laplace in 1790's. 
(see, e.g., the first volume of his famous book ``Trait\'{e} \; de \; m\'{e}canique \; c\'{e}leste'' 
published by Duprat, in Paris in 1799).
\end{thebibliography}
\end{document}